\documentclass[twoside,12pt]{article}
\usepackage{xcolor}
\usepackage{graphicx}
\usepackage{tikz}
\usepackage{amsmath}
\usepackage{amssymb}
\usepackage{amsfonts}
\usepackage{amsthm}
\usepackage{fixmath}
\usepackage{braket}
\usepackage{slashed}
\usepackage{fontenc}
\usepackage{bm}
\usepackage[colorlinks=true,linkcolor=blue,filecolor=blue,urlcolor=blue,citecolor=blue,pdftex,plainpages=false]{hyperref}
\usepackage[left]{lineno}

\usepackage{units}

\def\Journal#1#2#3#4{{#1} {#2} (#4) #3 }

\def\NPA{{\em Nucl. Phys.} A}

\def\PLB{{\em Phys. Lett.} B}

\def\PRL{\em Phys. Rev. Lett.}

\def\PREP{\em Phys. Rep.}

\def\PRD{{\em Phys. Rev.} D}
\def\PRC{{\em Phys. Rev.} C}

\newcommand{\be}{\begin{equation}}
\newcommand{\ee}{\end{equation}}
\newcommand{\bea}{\begin{eqnarray}}
\newcommand{\eea}{\end{eqnarray}}
\newcommand{\nn}{\nonumber}

\newcommand{\nnnl}{\nonumber\\}

\newcommand{\al}[1]{\vskip-3ex\begin{align}#1\end{align}}
\newcommand{\Eqref}[1]{\mbox{Eq.}~(\ref{#1})}
\newcommand{\Eqsref}[1]{\mbox{Eqs.}~(\ref{#1})}
\newcommand{\Figref}[1]{\mbox{Fig.}~\ref{#1}}
\newcommand{\ie}{\mbox{i.e.~}}
\newcommand{\eg}{\mbox{e.g.~}}

\renewcommand{\bm}{\mathbold}
\newcommand{\mr}{\mathrm}
\newcommand{\mb}{\mathbf}

\newcommand{\mf}{\mathfrak}
\newcommand{\mc}{\mathcal}

\def\ri{\mr{i}} 
\def\tr{\mr{tr}~}
\def\del{\partial} 
\def\nab{\nabla} 
\def\nabc{\nabla^{\ast}} 
\def\ga{\gamma} 
\newcommand{\adj}{{^\dagger}}
\def\als{\alpha_s} 
\def\lMSb{\Lambda_{\rm \overline{MS}}}
\def\md{m_{D}}

\begin{document}


\title{Color Screening in Quantum Chromodynamics}
\author{Alexei Bazavov, Johannes H. Weber\\
Department of \\
Computational Mathematics, Science and Engineering and\\
Department of Physics and Astronomy,\\
Michigan State University, East Lansing, MI 48824, USA}
\maketitle

\begin{abstract}
	We review lattice studies of the color screening in the quark-gluon plasma.
	We put the phenomena related to the color screening into the context of
	similar aspects of other physical systems (electromagnetic plasma or
	cold nuclear matter).
	We discuss the onset of the color screening and its signature and
	significance in the QCD transition region, and elucidate at which
	temperature and to which extent the weak-coupling picture based on
	hard thermal loop expansion, potential nonrelativistic QCD, or
	dimensionally-reduced QCD quantitatively captures
	the key properties of the color screening.
	We discuss the different regimes pertaining to the color screening and
	thermal dissociation of the static quarks in depth for various spatial
	correlation functions that are studied on the lattice, and clarify the
	status of their asymptotic screening masses.
	We finally discuss the screening correlation functions of dynamical mesons
	with a wide range of flavor and spin content, and how they conform with
	expectations for low- and high-temperature behavior.
\end{abstract}


\eject
\tableofcontents


\section{Introduction}

Whenever the interactions of particulate matter are described in terms of a 
field theoretical approach, these interactions can be understood as the 
exchange of one or more excitations of the fields mediating 
these interactions. 
The contribution to the potential from the exchange of $ n $ field excitations 
with mass $ m $ is $ V \sim (e^{-mr}/r)^n $. 
If the leading contribution is due to the exchange of a single massless, 
classical field excitation, then the dominant contribution takes the form 
of the classical Coulomb potential, \ie $ V_\mr{C} \sim 1/r $ and the 
force is $ F_\mr{C} \sim 1/r^2 $.  
If the leading contribution is due to the exchange of a single massive, 
classical field excitation, then the dominant contribution takes the form 
of the famous Yukawa potential~\cite{Yukawa:1935xg}, \ie $ V_\mr{Y} \sim e^{-mr}/r $, and the 
force is $ F_\mr{Y} \sim (e^{-mr}+mr)/r^2 = F_\mr{C}~(1+ \mc{O}(rm)^2) $. 
The Yukawa potential is the textbook example of a screened interaction 
with the Debye-H\"uckel or screening length $ \lambda = 1/m $~\cite{DebyeHuckel1923}. 
The basic features of the Coulomb or Yukawa forces are present in the 
classical limit of all known fundamental gauge forces of the Standard 
Model of Particle Physics, they are present in the classical field theory 
of gravitation, and they are present in most of the many-body effective 
field theories, too, \eg in the chiral effective theory that describes 
the formation of nuclei from individual nucleons, or in many condensed 
matter applications. 
\vskip1ex
We generally speak of \emph{screening} whenever the presence of mobile 
charges is the cause for the falloff of the leading contribution to a 
potential being larger than the power law $ V \sim 1/r^n $ that would be  
expected for the exchange of $ n $ massless, classical field excitations. 
This applies to the case of the quantum corrections due to the vacuum 
polarization, and it applies to the case of the thermal screening inside 
of a plasma. 
We take a first look at the more simple case of the electromagnetism and 
contrast it later with the more complicated case of the strong interactions.
\vskip1ex
Electromagnetism is an Abelian $ \mr{U}(1) $ gauge theory. 
For this reason the electromagnetic fields themselves carry no 
electromagnetic charge, and they couple to each other only indirectly 
by coupling to the same electromagnetically charged matter. 
The Coulomb potential $ V \sim \alpha_\mr{em}/r $ in (classical) 
electromagnetism is a long-range interaction. 
Nevertheless, at large but still finite distances the electromagnetic 
potential exhibits in many systems a dominant power-law behavior $ 1/r^n $, 
where $ n \geq 2 $. 
This may be understood in a many-body picture by applying Gauss's law 
to a case where multiple opposite charges reside in the same local volume. 
They compensate each other when summed up to the total charge, 
which may exactly cancel the leading Coulomb contributions to the force 
felt by another observing charge distribution at a far distance. 
However, subleading contributions due to the higher moments of this local 
charge distribution are associated with higher powers $ 1/r^n $, and 
are still relevant at large distances, if the observing charge distribution 
may accommodate higher moments of the same kind. 
We think of the multipole expansion and the dipole radiation as well-known 
classical examples of such a scenario.  
Since the dominant contribution exhibits a clean power-law falloff in such 
scenarios, this is not an example of screening.  
\vskip1ex
In a quantum field theory such as Quantum Electrodynamics (QED), the leading 
contribution to the Coulomb force between an electron-positron pair at rest 
is mediated by the exchange of a single \emph{electric $ A_0 $ photon}. 
However, an effect reminiscent to the aforementioned many-body picture plays 
out through the vacuum polarization, where each individual electron (or 
positron) creates a polarized charge distribution of virtual electron-positron 
pairs from the quantum fluctuations of the vacuum~\cite{Dyson:1949bp, PhysRev.82.664}. 
These lead to a larger charge $ \alpha_\mr{em}(1/r) $ being felt by an 
observer at smaller distances $ r $, where less quantum fluctuations 
can contribute to the total charge~\cite{PhysRev.48.55}. 
The dependence of $ \alpha_\mr{em}(1/r) $ is logarithmic at distances much 
smaller than the electron's Compton wave length $ \lambda_e = 1/m_e $, 
such that the QED potential behaves as 
$ V \sim \alpha_\mr{em}(1/r)/r \sim \log(r)/r $ for 
$ r \ll \lambda_e $, \ie $ V $ does not exhibit the power-law 
falloff of the classical Coulomb force!  
The same effect applies also to exchanges mediated by the emission of 
multiple (electric or magnetic) photons. 
Although this effect of the QED vacuum polarization is a screening mechanism, 
it is quite different from the mechanisms of the thermal screening inside of 
an electromagnetic plasma. 
\vskip1ex
A surrounding electromagnetic plasma breaks the Poincar\'e symmetry of the 
local interactions and the energies of all of the thermally equilibrated 
fields are discretized. 
On the one hand, the \emph{electric $ A_0 $ photons} are subject to the 
thermal modification and acquire an effective thermal mass -- the Debye 
mass -- that amounts to $ m = \mr{e}T/\sqrt{3} $ (for a theory with only 
one massless fermion with charge $ \mr{e} =\sqrt{4\pi \alpha_\mr{em}}$) 
at leading order~\cite{Kislinger:1975uy}. 
Hence, inside the plasma the electric fields are screened with the 
screening length $ \lambda = 1/m $. 
On the other hand, the magnetic fields are not screened at all, 
and the \emph{electric $A_0$ photons} can couple to magnetic photons through fermion loops.  
Hence, at large enough distances $ r \gg \lambda $, the leading contribution 
to the potential between two electromagnetic charges is not of the (screened) 
Coulomb form $ V \sim \alpha_\mr{em} e^{-mr}/r $ anymore, but instead it is 
dominated by the exchange of multiple \emph{magnetic photons}, \ie it is of 
the form $ V \sim \alpha_\mr{em}^6/r^6 $ for the magnetic van-der-Waals 
interaction. 
Due to the smallness of the electromagnetic coupling $ \alpha_\mr{em} $, 
there is an intermediate range, where the dominant contribution to the 
electromagnetic potential is still of the form 
$ V \sim \alpha_\mr{em}e^{-mr}/r $, such that the electromagnetic 
Debye mass $ m $ can be determined straightforwardly. 
However, at larger distances the van-der-Waals contribution from the exchange 
of two \emph{magnetic photons} is dominant, and thus the electromagnetic 
potential inside a plasma exhibits the power-law behavior at large enough distances, see \eg~Ref.~\cite{Gross:1980br}. 
\vskip1ex
Regarding the strong interaction, the picture is very different. 
In the many-body field theoretical description of nuclei and hypernuclei, 
the excitation and exchange of a single pion, kaon or eta meson is mediating 
the leading contribution to the potential between the octet baryons 
$ V \sim e^{-m r}/r $. 
Given the pion, kaon and eta masses of $ m_\pi = 135\,\mr{MeV} $, 
$ m_K = 496\,\mr{MeV} $, and $ m_\eta = 548\,\mr{MeV} $ the screening 
lengths are different for different processes and are of the order of 
$ \lambda \sim 1/m_\eta \sim 0.36\,\mr{fm} $ to $ 1/m_\pi \sim 1.5\,\mr{fm} $, 
which is about twice the charge radius of a charged octet baryon. 
The existence of different screening lengths in different interaction 
channels is a feature of QCD that is also found for the screening inside 
of a quark-gluon plasma. 
\vskip1ex
At much shorter distances this picture of baryons and mesons is not 
appropriate, and the strong interactions have to be described in terms 
of quarks and gluons in the Quantum Chromodynamics (QCD), which is a 
non-Abelian $ \mr{SU}(3) $ gauge theory. 
Due to the asymptotic freedom, the leading contribution to the force 
between two color charges at short enough distances is again the exchange 
of a single massless gluon, and the potential has the Coulomb form 
$ V \sim \alpha_s/r $ (with quantum corrections). 
In contrast to the QED, each individual quark or gluon creates a polarized 
charge distribution of virtual quark-antiquark pairs and virtual gluons from 
quantum fluctuations of the vacuum. 
While the former have a screening effect as in the QED, the latter have an 
even stronger anti-screening effect~\cite{Gross:1973ju,Politzer:1973fx}. 
These lead to a larger charge $ \alpha_s(1/r) $ being felt by an 
observer at larger distances $ r $, where more quantum fluctuations 
contribute to the total charge. 
The dependence of $ \alpha_s(1/r) $ is logarithmic such that the QCD 
potential behaves as $ V \sim \alpha_s(1/r)/r \sim 1/(r \log(r)) $, 
\ie $ V $ does not exhibit the classical power-law falloff~\cite{Appelquist:1977tw}.  
\vskip1ex
At distances $ r \sim 1/\lMSb $, where $ \lMSb $ is the intrinsic scale of 
the QCD (in the $\mr{\overline{MS}}$ scheme), the charge $ \alpha_s(1/r) $ 
would actually diverge, indicating the 
breakdown of a description in terms of individual quarks and gluons. 
At such distances the force approaches a constant, the QCD string tension, 
and the energy of this QCD string grows linearly with the distance. 
This is the color confinement of the QCD~\cite{Wilson:1974sk}. 
Due to the presence of sea quarks this string eventually breaks apart at 
a string-breaking distance $ \lambda_\mr{sb} $, where the total energy 
$ E(\lambda_\mr{sb}) $ of the QCD string becomes sufficiently 
large for creating a quark-antiquark pair $ q^\prime\bar q^\prime$ from 
the vacuum, see \eg~Refs.~\cite{Drummond:1998ar, Philipsen:1998de}. 
Two separate bound states with masses 
$ m_{q\bar q^\prime} = m_{q^\prime\bar q} \sim m_q+1/\lambda_\mr{sb} $
are formed by consuming this pair. 
For distances not too different from $ \lambda_\mr{sb} $ the energy of the 
quark-antiquark pair exhibits some characteristic features of the 
exponential screening of color charges, although the underlying unscreened 
QCD potential includes the contribution from the QCD string as well. 
\vskip1ex
Neither of these two screening mechanisms is the one of the 
color screening in the  quark-gluon plasma at high temperatures.  
As in the case of the electromagnetic plasma, the surrounding quark-gluon 
plasma breaks the Poincar\'e symmetry of the local interactions and the 
energies of all of the thermally equilibrated fields are discretized. 
On the one hand, inside such a quark-gluon plasma the 
\emph{electric $ A_0 $ gluons} are subject to the thermal modification and 
acquire a Debye mass -- that amounts to $ \md \sim gT $ at leading order, 
where $ g =\sqrt{4\pi \alpha_s}$. 
Hence, inside the quark-gluon plasma the \emph{electric fields} are screened 
with the screening length $ \lambda = 1/\md $. 
However, in contrast to the QED, one has to deal in the definition of the 
Debye mass with subtleties due to the gauge dependence of the vacuum 
polarization~\cite{Arnold:1995bh}. 
On the other hand, the \emph{magnetic fields} that are not screened at all 
are subject to the confining three-dimensional Yang-Mills theory. 
Hence, the nonperturbative interactions among the \emph{magnetic gluons} or 
with the \emph{electric $ A_0 $ gluons} lead to bound states with masses at 
the associated confining scale $ g^2T $~\cite{Gross:1980br,Nadkarni:1986cz}, which can be related to 
various inverse screening lengths of in-medium correlation functions in 
the corresponding channels. 
Moreover, since the \emph{electric $A_0$ gluons} can couple directly to 
the magnetic gluons due to the nontrivial structure of the gauge group 
$\mr{SU}(3)$, the dominant contribution from the magnetic gluons is larger 
than in QED and does not require the presence of fermion loops at all~\cite{Arnold:1995bh}.
\vskip1ex
At this point it is already evident that there are fundamental differences 
between the mechanisms of \emph{electric screening} in electromagnetic or 
\emph{color screening} in quark-gluon plasma. 
The key aspects of the \emph{color screening} can be understood only within 
a proper quantum field theoretical framework. 
We give a brief overview of the quantum field theoretical foundations 
of QCD and of the related thermal field theory in Section~\ref{sec:lattice}. 
In Section~\ref{sec:static} we discuss the screening of the static charges 
and of bound states of static charges in QCD. 
Later on, in Section~\ref{sec:dynamic} we discuss the interplay between screening and dissociation for 
static and non-static mesons in QCD. 
Finally, we conclude with a concise summary in Section~\ref{sec:summary}.

\section{Field theoretical foundations}\label{sec:lattice}


\subsection{Partition function and Lagrangian}

Nuclear matter and the strong interactions are realized on the fundamental 
level in the Standard Model in terms of the Quantum Chromodynamics (QCD). 
The hadron spectrum, hadron structure and hadron reactions are -- up to 
effects from the electroweak sector -- completely described by the QCD 
partition function (in any finite or infinite volume $ V $)
\al{\label{eq:ZT=0} 
 Z(V) 
 &= 
 \int \prod_{\mu} \mc D A_\mu \prod_{f} \mc D \bar\psi_f \mc D \psi_f 
 e^{~\ri\int_{-\infty}^{+\infty} dt \int_V d^{d-1}x  
 ~\mc L[A_\mu,~ \bar\psi_f,~ \psi_f] },  
}
which can be expressed in terms of a path integral over the 
$ N_f $ quark (and antiquark) fields $ ( \bar\psi_f,~ \psi_f ) $ with 
$ f= u,~ d,~ s,~ c,~ b,~ t$ (up, down, strange, charm, bottom, and top) 
and the gluon fields $ A_\mu $. 
This partition function implicitly depends on the strong coupling $ g $ 
(or equivalently $ \als=\tfrac{g^2}{4\pi} $) and the masses 
$ \bm m = ( m_u, m_d, m_s, \ldots ) $~of the quark flavors, which are the 
only parameters of the classical QCD Lagrangian $ \mc L[A_\mu,~ \bar\psi_f,~ \psi_f] $. 
Since the top quark mass $ m_t $ is significantly larger than the electroweak 
scale $ M_W $, it decouples in most respects from nuclear matter at 
any lower scales and will be omitted in the following. 
Observables can be calculated in QCD by expressing them through the fundamental 
fields and evaluating the path integral 
\al{\label{eq:Obs}
 \braket{O}
 &= \frac{1}{Z}
 \int \prod_{\mu} \mc D A_\mu \prod_{f} \mc D \bar\psi_f \mc D \psi_f 
 O[A_\mu,~ \bar\psi_f,~ \psi_f] 
 e^{~\ri S[A_\mu,~ \bar\psi_f,~ \psi_f] }, 
}
where we use the QCD action $ S = \int d^dx~\mc L $ and omit any explicit 
reference to the volume. 
\vskip1ex
All interactions among these fundamental fields are encoded into the monomials of the QCD 
Lagrangian, which satisfies a local symmetry under the gauge group $ \mr{SU}(N_c) $, 
where $ N_c=3 $ for QCD. 
Namely, the quark flavors $ ( \bar\psi_f^a,~ \psi_f^a ) $ exist in $ N_c $ copies 
(color indices $ a = 1,~ \ldots,~ N_c $) that transform in the fundamental representation 
of the gauge group, and the gluons
\begin{equation}
\label{eq:Amu}
A_\mu^{c} \equiv A_\mu^{ab} t_{ab}^c,
\end{equation}
in $ N_c^2-1 $ 
copies (color indices $ c = 1,~ \ldots,~ (N_c^2-1) $, $ a,~b = 1,~ \ldots,~ N_c $) that 
transform in the adjoint representation of the gauge group. 
Here, $ t_{ab}^c $ are the generators of the Lie algebra $ \mf{su}(N_c) $ normalized as 
$ \tr{( t^c t^d )} = \tfrac{1}{2}\delta^{cd} $. 
The classical QCD Lagrangian has a gauge part and a matter part,
\al{\label{eq:LagQCD}
 \mc L[A_\mu,~\bar\psi_f,~\psi_f]  
 &= \mc L_{\rm gauge}[A_\mu] + \mc L_{\rm matter}[A_\mu,~\bar\psi_f,~\psi_f] \\
 &= -\frac{1}{4g_0^2} F^{\mu\nu}_{c}(x) F_{\mu\nu}^{c}(x) 
  -\sum\limits_{f} \bar\psi_f^{\alpha\,a}(x) 
  \left\{ \ri \slashed{D}_{\alpha\beta}^{ab} - m_{0f} \delta_{\alpha\beta}\delta_{ab}  \right\} 
  \psi_f^{\beta\,b},
}
where all indices are understood to be summed over. 
The Dirac or spin indices $ \alpha = 1,~\ldots,~d $ are indicated with greek 
characters. 
In the notation used in the following, color, flavor, and spin indices, and the 
space-time arguments $ x $ will be generally omitted whenever this is appropriate.
The matter part is restricted to the $ N_f $ quark flavors that are considered 
as dynamical degrees of freedom. 
As such, the QCD Lagrangian explicitly depends on the bare gauge coupling $ g_0 $, 
the number $ N_f $ of dynamical quark flavors, and the $ N_f $ respective bare 
quark masses $ m_{0f} $ as its only parameters. 
\vskip1ex
The massless Dirac operator $ \slashed{D}_{\alpha\beta}^{ab} $ is 
given in terms of the covariant derivative  $ {D}_{\mu}^{ab} $ as 
\al{\label{eq:Dslash}
 \slashed{D}_{\alpha\beta}^{ab} 
 &= \sum\limits_\mu \ga_\mu^{\alpha\beta} {D}_{\mu}^{ab} 
 ,&
 {D}_{\mu}^{ab} 
 &= \ri \left(\delta^{ab}\del_\mu -\ri t_c^{ab} A_\mu^{c} \right).  
}
The matter part of the QCD Lagrangian is locally gauge invariant under a transform 
\al{\label{eq:gaugeinv}
 &\psi(x) \to \Omega(x) \psi(x),\qquad
 \bar\psi(x) \to \bar\psi(x)\Omega\adj(x),\nnnl& 
 A_\mu(x) \to \Omega(x) \{ A_\mu(x) +\ri \Omega\adj(x) \del_\mu \Omega(x) \} \Omega\adj(x)
}
for any $ \Omega(x) \in \mr{SU}(N_c) $. 
The field strength tensor $ F_{\mu\nu}^{a}(x) $ can be expressed in terms 
of the structure constants $ f_{abc} $ of the Lie algebra $ \mf{su}(N_c) $, 
or in terms of commutators of $ A_\mu $ or of covariant derivatives $ {D}_{\mu} $,
\al{\label{eq:Fmunu}
 F_{\mu\nu}^{c}(x) 
 &= \del_\mu A_\nu^{c} - \del_\nu A_\mu^{c} + f_{abc} A_\mu^{a} A_\nu^{b}\nonumber\\
 &= t^c_{ab} \left\{ \del_\mu A_\nu^{ab} - \del_\nu A_\mu^{ab} 
   -\ri [ A_\mu, A_\nu ]^{ab} \right \}
 = -\ri t^c_{ab} [ D_\mu, D_\nu ]^{ab}.
} 
The invariance of the gauge part follows immediately from \Eqref{eq:gaugeinv}.
The gauge part on its own is also referred to as the Yang-Mills or pure gauge 
Lagrangian.  
For mostly technical reasons this is still of particular relevance to 
lattice gauge theory as will be explicitly addressed later on. 

\begin{figure*}\center
\includegraphics[width=8.6cm]{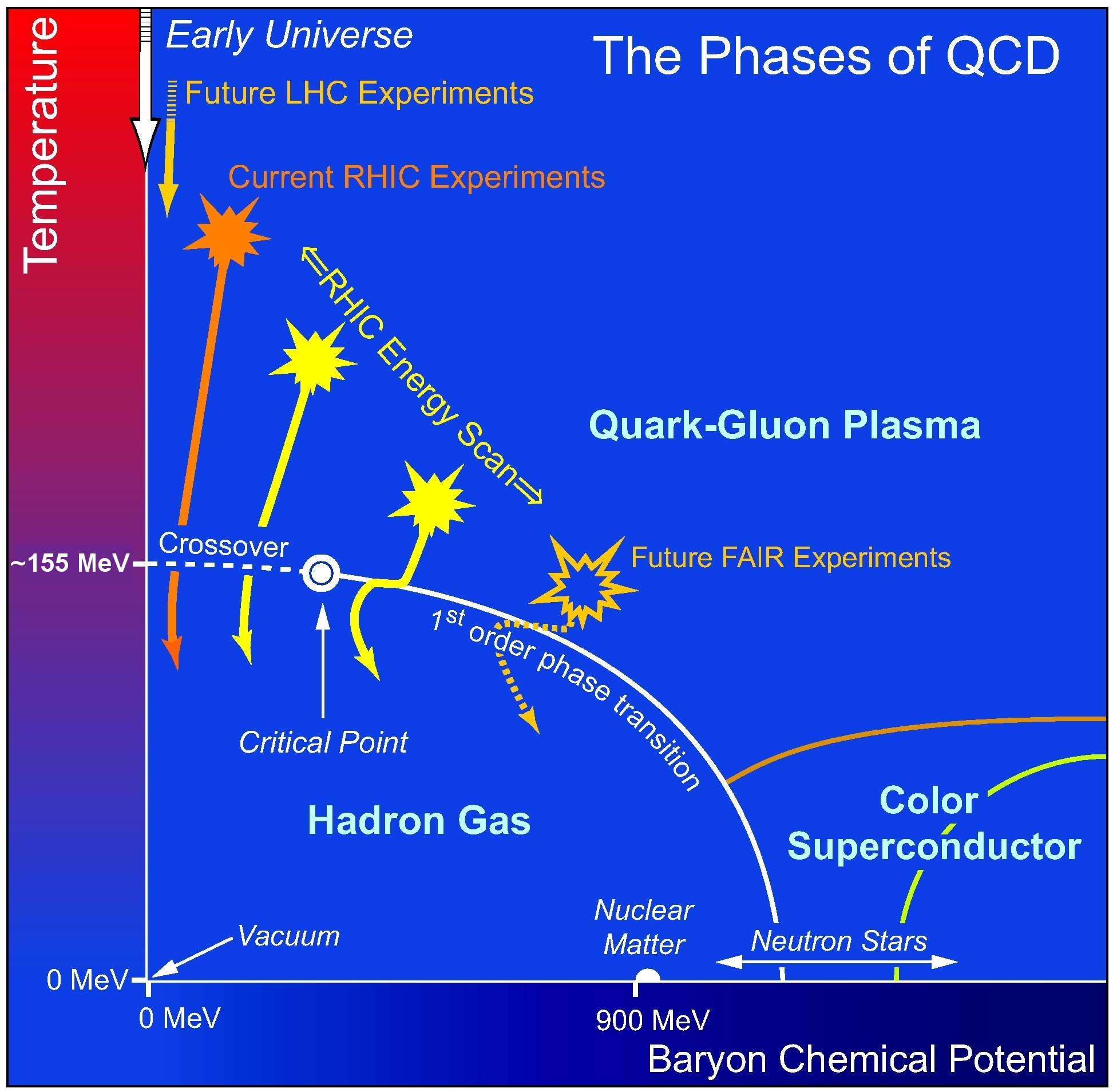}
\caption{
The conjectured QCD phase diagram in the plane of two external parameters:
temperature and the baryon chemical potential. Possible reach of the current
and future heavy-ion collision experiments is indicated with orange and yellow
arrows and symbols.
}
\label{fig:phases}
\end{figure*}

QCD has multiple inherently nonperturbative aspects such as the confinement, 
the spontaneous chiral symmetry breaking, and the axial anomaly, all of which 
are directly reflected in the low-energy hadron spectrum and control the key 
properties of nuclear matter. 
These key properties change quite dramatically as the external parameters such 
as the temperature $ T $, the volume $ V $, or the quark flavor chemical 
potentials $ \bm \mu = ( \mu_u,~ \mu_d,~ \mu_s,~ \ldots) $ are modified. 
Quark flavor chemical potentials can be introduced by adding a source term 
\al{\label{eq:muN}
 \mc L_{\bm \mu \cdot \bm n} 
 &= \bm \mu \cdot \mc N[\bar\psi_f,~\psi_f] 
 = \sum\limits_f \mu_f~\bar\psi_f \ga_0 \psi_f, 
} 
to the QCD Lagrangian of \Eqref{eq:LagQCD}, 
where $ \mc N_f = \bar\psi_f \ga_0 \psi_f $ is the quark number density. 
A finite temperature can be introduced through a compact imaginary time 
direction with periodic boundary conditions for the bosons (gluons) and 
antiperiodic boundary conditions for the fermions (quarks). 
In practice this can be achieved by replacing the Minkowski space-time 
through a Wick rotation $ t \to -\ri \tau $ with a Euclidean space-time 
with (anti-)periodic boundaries (in time). 
Thermally equilibrated strongly interacting matter is then described by the grand canonical 
QCD partition function $ Z(T,~ V,~ \bm \mu) $, 
\al{\label{eq:ZQCD}
 Z(T,~V,~\bm \mu) 
 &= 
 \int \prod_{\mu} \mc D A_\mu \prod_{f} \mc D \bar\psi_f \mc D \psi_f 
 e^{~-\int_{~0}^{\tfrac{1}{T}} d\tau \int_V d^{d-1}x  
 \mc L_E[A_\mu,~ \bar\psi_f,~ \psi_f] }, 
}
where $ \mc L_E $ is the classical QCD Lagrangian in Euclidean space-time. 
The definition of observables in a background of thermally equilibrated 
nuclear matter follows from applying the same steps to \Eqref{eq:Obs}. 
Since the matter part of the QCD Lagrangian in \Eqsref{eq:LagQCD} and 
\eqref{eq:muN} is bilinear in the quark and antiquark fields, the quark 
degrees of freedom can be integrated out in the path integral to obtain 
 \al{\label{eq:Zdet}
 Z(T,~V,~\bm \mu) 
 &= 
 \int \prod_{\mu} \mc D A_\mu 
 ~\prod\limits_{f} \det \{D[A_\mu](m_{0f},\mu_f) \}
 ~e^{~-\int_{~0}^{\tfrac{1}{T}} d\tau \int_V d^{d-1}x  
 \mc L_{\rm gauge}[A_\mu] }.
}
\vskip1ex
Strongly interacting matter described by \Eqref{eq:ZQCD} has a rich phase structure, see 
\Figref{fig:phases}, with many similarities to well-studied condensed matter systems. 
In particular, QCD has a well-established hadron gas phase for small values 
of the temperature $ T $ and the chemical potentials $ \bm \mu $, and another 
well-established quark-gluon plasma phase for large values of the temperature $ T $. 
In the limit of massless quarks and vanishing chemical potentials, there is a 
sharp chiral phase transition that takes place at the critical temperature 
$T_c^0=132^{+3}_{-6}\,\mr{MeV}$~\cite{Ding:2019prx}.
For small enough quark masses and small enough chemical potentials $ \bm \mu $, 
this transition is turned into a smooth crossover at a somewhat higher 
pseudo-critical temperature~\cite{Aoki:2009sc, Bazavov:2011nk, Bazavov:2018mes, 
Borsanyi:2020fev}, \ie $T_{c}=156.5(1.5)\,\mr{MeV}$ for physical quark masses, 
while being still sensitive to the sharp phase transition.
Recently, there have been hints~\cite{Rohrhofer:2019qwq} of a 
\emph{stringy-fluid} phase with a spin-chiral symmetry at values of the 
temperature slightly above this transition, although not much 
is known about this state so far. 
For a long time, there have been speculations about a critical endpoint and phase 
transition line for larger values of the chemical potentials, but these prove 
elusive despite major efforts in experimental or theoretical studies to date. 
For small temperatures and densities around the average nuclear matter density, 
one encounters the larger nuclei of the elements in the periodic table, and 
-- for much larger densities -- eventually neutron stars. 
While this regime is well-studied experimentally, theory calculations are still 
challenging due to a sign problem 
(i.e. breakdown of stochastic importance sampling and/or exponentially decreasing
signal-to-noise ratio)
in the partition function of \Eqref{eq:ZQCD}. 
\vskip1ex
\subsection{Finite temperature field theory}

The restriction of a quantum field theory to finite temperature has profound 
consequences that play out in QCD quite similar as in most other quantum field 
theories. 
The energies of all fields become quantized and are restricted to the 
Matsubara modes $ \omega_n $, which are fixed for any bosons as 
$ \omega_n= 2n~\pi T$ and for any fermions as $ \omega_n= (2n+1)~\pi T $, 
where $n=0,~1,~\ldots $. 
Hence, the lowest bosonic Matsubara mode $ \omega_0 = 0 $ is the 
\emph{Matsubara zero mode} or \emph{static Matsubara mode}, and is 
well-separated from all higher (bosonic or fermionic) Matsubara modes. 
\vskip1ex
This scale separation entails an approximate decoupling of the 
\emph{static modes} of the fields from their higher Matsubara modes. 
Namely, the interactions between these different modes take place at the 
temperature scale, and the coupling depends on the temperature. 
The most important interactions between the \emph{static modes} 
and the higher Matsubara modes induce (in the massless limit divergent) 
contributions that modify the large-distance behavior of the 
\emph{static modes}.
Namely, the correlation lengths of the \emph{static modes} may be temperature 
dependent and finite in the interacting theory, despite being infinite at the 
tree level. 
The inverse of such a screening length can be rephrased in terms of a 
screening mass that vanishes in the non-interacting or zero-temperature 
limits. 
While a similar screening mechanism affects the higher Matsubara modes as well, 
it is generally subleading compared to the non-zero Matsubara frequencies 
$ \omega_n =(2n)~\pi T $ for $ n>0 $. 
Since the lowest fermionic Matsubara mode is $ \omega_0 = \pi T $,
contributions from fermions are generally suppressed similarly to the higher Matsubara modes, too.  

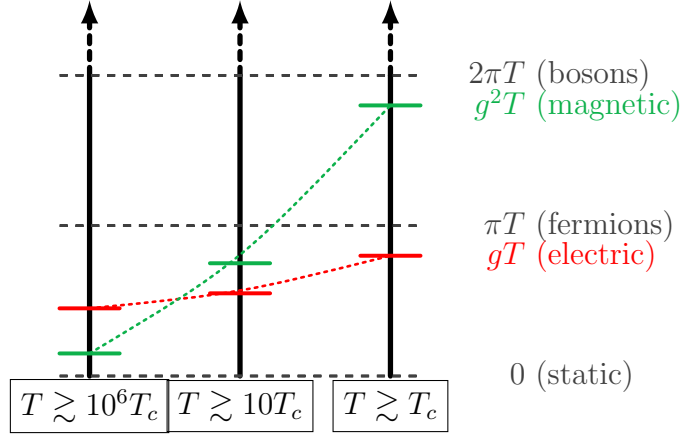
\begin{figure*}\center

\definecolor{spartandarkgreen} {RGB}{24, 69, 59}
\definecolor{spartanlightgreen} {RGB}{13, 177, 75}
\begin{tikzpicture}
\draw [>=latex,line width=2pt, dashed, color=black,line cap=round,->] (0,4) -- (0,5);
\draw [>=latex,line width=2pt, dashed, color=black,line cap=round,->] (2,4) -- (2,5);
\draw [>=latex,line width=2pt, dashed, color=black,line cap=round,->] (4,4) -- (4,5);

\draw [>=latex,line width=2pt,color=black,line cap=round] (0,-0) -- (0,4);
\draw [>=latex,line width=2pt,color=black,line cap=round] (2,-0) -- (2,4);
\draw [>=latex,line width=2pt,color=black,line cap=round] (4,-0) -- (4,4);

\node [draw] at (0,-0.4) {$T \gtrsim 10^6 T_c$};
\node [draw] at (2,-0.4) {$T \gtrsim 10 T_c$};
\node [draw] at (4,-0.4) {$T \gtrsim T_c$};

\draw [darkgray,line width=1pt, dashed] (-0.4,-0.00) rectangle (4.4,-0.00) node[anchor=west] {~~$\phantom{2\pi T} 0$ (static)};
\draw [darkgray,line width=1pt, dashed] (-0.4,+2.00) rectangle (4.4,+2.00) node[anchor=west] {~~$\phantom{20}\pi T$ (fermions)};
\draw [darkgray,line width=1pt, dashed] (-0.4,+4.00) rectangle (4.4,+4.00) node[anchor=west] {~~$\phantom{0} 2\pi T$ (bosons)};

\draw [>=latex,line width=1.5pt, color=red, line cap=round] (-0.4,+0.90) -- (0.4,+0.90);
\draw [>=latex,line width=1.5pt, color=red, line cap=round] (+1.6,+1.10) -- (2.4,+1.10) ;
\draw [>=latex,line width=1.5pt, color=red, line cap=round] (+3.6,+1.60) -- (4.4,+1.60) node[anchor=west] {~~~~$\phantom{^2}gT$ (electric)} ;
\draw [>=latex,line width=1.0pt, color=red, dotted, line cap=round] (-0.0,+0.90) .. controls (+2.0,+1.10) .. (+4.0,+1.60);

\draw [>=latex,line width=1.5pt, color=spartanlightgreen, line cap=round] (-0.4,+.30) -- (0.4,+0.30) ;
\draw [>=latex,line width=1.5pt, color=spartanlightgreen, line cap=round] (+1.6,+1.50) -- (2.4,+1.50) ;
\draw [>=latex,line width=1.5pt, color=spartanlightgreen, line cap=round] (+3.6,+3.60) -- (4.4,+3.60) node[anchor=west] {~~~~$g^2T$ (magnetic)}; 
\draw [>=latex,line width=1.0pt, color=spartanlightgreen, dotted, line cap=round] (-0.0,+0.30) .. controls (+2.0,+1.50) .. (+4.0,+3.60);
  
\end{tikzpicture}
\caption{
Thermal QCD hierarchies for very high temperatures (left), and 
high (center) or low (right) temperatures in the phenomenologically 
interesting range. 
The naive hierarchy $ g^2 T \ll gT $ between magnetic and electric scales 
is inverted for phenomenologically interesting temperatures. 
Eventually, it is not even clear if the magnetic or even electric scales 
can be distinguished from the lowest non-static scales at all, if the 
coupling $ g(T) $ becomes too large. 
}
\label{fig:thermal hierarchy}
\end{figure*}

In particular, the interactions of these \emph{static modes} can be 
understood in terms of the three-dimensional effective field 
theory~\cite{Braaten:1995cm, Braaten:1995jr}, see~\cite{Moller:2012zz} 
for a concise review. 
The higher Matsubara modes and the fermions are integrated out and are 
manifest in the Wilson coefficients of suppressed higher order operators. 
In the case of QCD, there are a few key differences between the vacuum 
field theory and the thermal field theory.
First, the symmetry between the \emph{magnetic gluons} and the 
\emph{electric $ A_0 $ gluons} is broken in the thermal theory.
Second, there are contributions at odd powers of the gauge coupling $ g(T) $, 
which are due to the screening of the \emph{electric $ A_0 $ gluons}. 
Third, perturbation theory has an infrared cutoff at order $ \sim g^2T $, 
which leads to a complete breakdown of perturbation theory at order 
$ g^6T^4 $ for the thermodynamic potentials~\cite{Linde:1980ts}. 
These scales -- $ T $ (or rather $ 2\pi T $ or $ \pi T $), $ g(T)T $, 
and $ g^2(T)T $ -- are hierarchically ordered for very high temperatures. 
For phenomenologically interesting temperatures not too far above the QCD 
scale $ \lMSb $, this hierarchy might not be realized at all in practice, 
since the coupling $ g(T \gtrsim \lMSb) $ is large, 
see \Figref{fig:thermal hierarchy}.
Under these circumstances, the predictive power of the weak-coupling 
approach may strongly vary between different quantities. 
\vskip1ex
On the one hand, the interactions among the \emph{magnetic gluons} can be 
described by a three-dimensional, \ie confining $ \mr{SU}(N_c) $ pure gauge 
theory with the coupling $ g_M^2 \sim g^2T $ and the confinement 
radius $ \sim 1/(g^2T) $~\cite{Gross:1980br}. 
This effective field theory is called the \emph{magnetostatic QCD} (MQCD). 
The \emph{magnetic gluons} themselves are not affected by the thermal 
screening, although they have finite correlation lengths due the 
confinement radius $ \sim 1/(g^2T) $ of the pure gauge theory. 
Accordingly, these \emph{magnetic gluons} combine to the same glue-ball 
spectrum as in the three-dimensional Yang-Mills theory at zero temperature. 
The lowest-lying glue-ball masses $ m \sim g^2 T $ correspond (for 
$ m \ll 2\pi T $) to certain inverse correlation lengths of long-range 
correlations between the \emph{static modes} of the thermal QCD medium. 
These \emph{magnetostatic QCD} contributions to the correlation lengths in 
the thermal QCD medium cannot be determined using perturbation theory.

\begin{figure*}\center
\includegraphics[width=8.6cm]{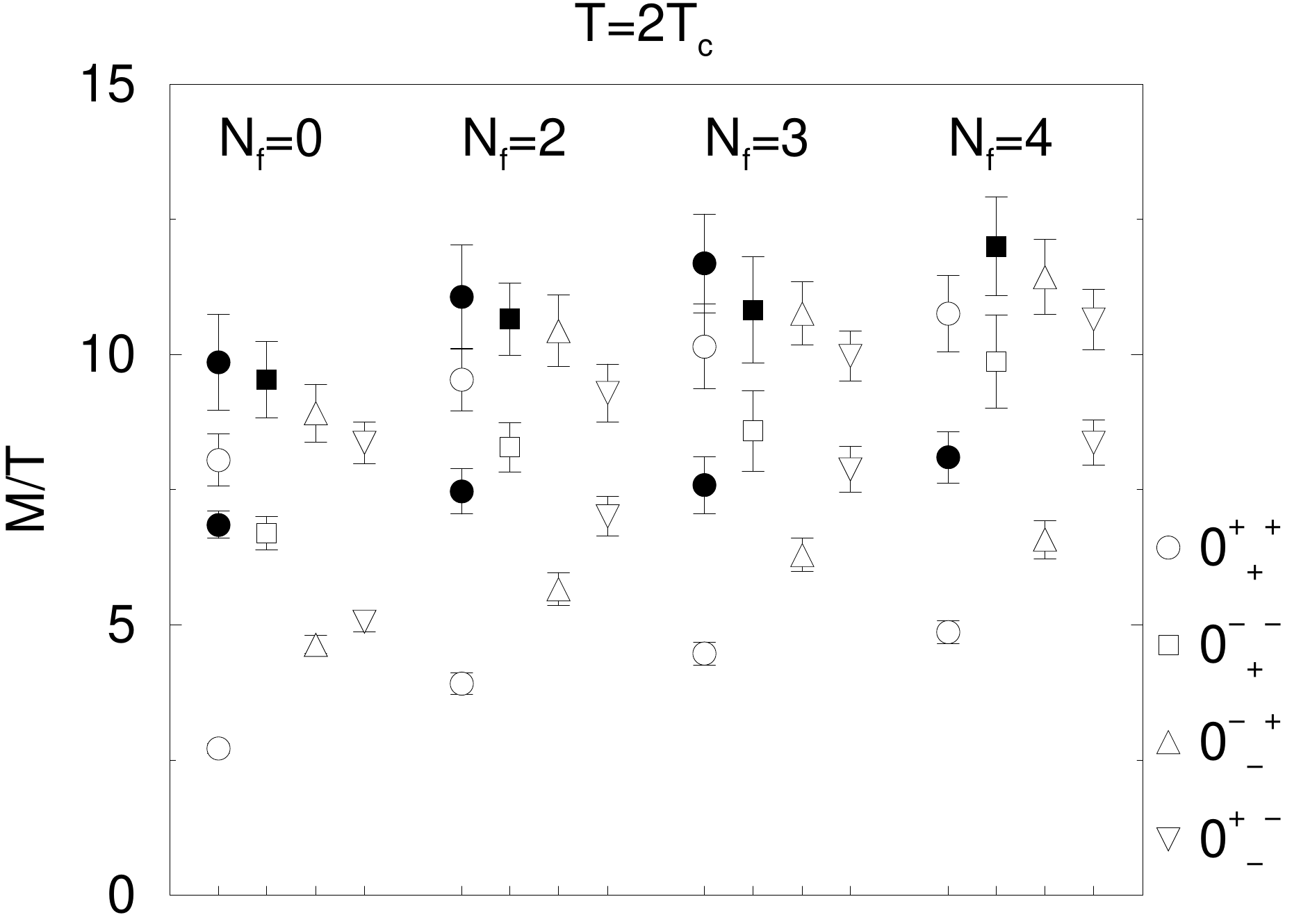}
\caption{
The spectrum of electrostatic QCD as obtained in nonperturbative 
lattice simulations~\cite{Hart:2000ha}. 
Open symbols correspond to states containing \emph{electric $ A_0$ gluons}, 
filled symbols correspond to genuine magnetic glue-balls. 
The naive hierarchy is inverted in all channels. 
From Ref.~\cite{Hart:2000ha}.
}
\label{fig:EQCD spectrum}
\end{figure*}

On the other hand, the large-distance correlations of the 
\emph{electric $ A_0 $ gluons} are severely impacted by the thermal 
modifications.  
These modifications can be accounted for in terms of a thermal 
\emph{Debye mass}, which is of the (leading) order $ \md \sim gT $ and 
affects only the \emph{electric $ A_0 $ gluons}. 
To the leading order the \emph{electric $ A_0 $ gluons} may be treated as a 
massive adjoint scalar field with $ A_0^4 $ self-interactions (with the 
coupling at the order $ \lambda_E \sim g^4T $), and with the minimal coupling 
to the \emph{magnetic gluons} using the charge $ g_E^2 \sim g^2T $. 
This effective field theory is called the \emph{electrostatic QCD} (EQCD). 
Its spectrum also contains nonperturbative bound states of multiple 
\emph{electric $ A_0 $ gluons} or of \emph{electric $ A_0 $ gluons} with 
\emph{magnetic gluons}. 
These bound states, which are formed at the confinement scale $ \sim g^2T $ of 
the \emph{magnetostatic QCD}, can be understood as a manifestation of the Linde 
problem~\cite{Linde:1980ts}. 
Whether the lightest states with any given quantum numbers are purely 
magnetic glue-balls or bound states involving \emph{electric $ A_0 $ gluons} 
depends on the value of the coupling $ g(T) $, and therefore on the 
temperature, see \Figref{fig:EQCD spectrum} for a glance on the spectrum of 
the \emph{electrostatic QCD}~\cite{Hart:2000ha}. 
\vskip1ex
In the weak-coupling approach a \emph{Debye mass} can be defined 
order-by-order in terms of the pole position of the \emph{static mode} of the 
thermal gluon propagator.  
The higher Matsubara modes or the fermions contribute order-by-order only at 
odd powers of the coupling, \ie $ \md/T \sim c_1 g + c_2 g^3 + \ldots $. 
However, while this pole position is gauge invariant order-by-order in 
perturbation theory, one cannot avoid using the gauge-dependent gluon 
propagator for its definition, and hence it is dependent on a gauge-fixed 
approach. 
Moreover, contributions from the \emph{magnetic gluons} yield a 
nonperturbative contribution to $\md$ that is of order $ \sim g^2 T $. 
This is yet another manifestation of the Linde problem~\cite{Linde:1980ts}. 
The perturbative expansion breaks down as soon as scales of 
order $ g^2T $ contribute. 
For these reasons it is not obvious to which extent the weak-coupling 
approach may yield a physically meaningful description of the screening of 
the \emph{electric $ A_0 $ gluons} at all~\cite{Arnold:1995bh}.  
Answers to this problem can be provided through the comparison between the 
results obtained in the weak-coupling approach or in a direct nonperturbative 
calculation. 
We return to this point in \mbox{Section}~\ref{sec:plcor}. 
\vskip1ex
\subsection{Lattice regularization}

Quantum effects are inherent to the properties and interactions of
strongly interacting matter. 
As such the classical QCD Lagrangian of \Eqref{eq:LagQCD} has physical 
significance only through its role inside of the path integral definition of the 
partition function of \Eqsref{eq:ZT=0} or \eqref{eq:ZQCD}. 
Alas, this definition is not well-defined, but fraught with divergences.
These are both of the infrared kind due to the propagation of massless 
(the gluons) or almost massless modes (some of the quark flavors), or of the 
ultraviolet kind due to the presence of quantum loops with infinite 
momenta. 
The former are only present in an infinite space-time volume and eventually 
regulated by the nonperturbative phenomena at the QCD scale $\lMSb$, 
the latter originate in the contributions from infinitesimally separated fields. 
Hence, \Eqsref{eq:ZQCD} only permits the calculation of QCD 
amplitudes if all of the divergences are regulated accordingly. 
\vskip1ex
Any such regularization scheme introduces its particular variety of unphysical 
properties at intermediate stages of a calculation. 
Eventually, physical predictions for the strong interactions are recovered 
only after the removal of this regulator. 
A reformulation of QCD on a finite (hybercubic) lattice of $ N_\sigma^{d-1} \times N_\tau $ 
sites with the lattice spacing $ a $ is such a regularization scheme that removes 
all of the divergences and explicitly enforces the local gauge symmetry. 
Here, $ N_\tau $ is the number of points along the time axis and $ N_\sigma $ the number 
of points along each spatial axis. 
Usually, periodic boundary conditions are used for all axes (antiperiodic for 
fermions in the time direction), although use of open boundary conditions 
in one of the directions has become more common in recent 
years\footnote{The reasons why open boundary conditions may be more beneficial
are related to sampling of the topology of the gauge fields but are too technical
to be discussed in this review.}. 
The path integral in lattice gauge theory with (anti-)periodic boundary conditions 
in the time direction and a Euclidean metric automatically samples the partition 
function of thermally equilibrated matter at temperature $ T = \tfrac{1}{aN_\tau} $. 
Yet as long as this temperature is significantly smaller than the most infrared scale, 
\mbox{i.e.} the pion mass, this partition function is practically indistinguishable 
from a true $ T = 0 $ result. 
These periodic boundaries of the spatial directions imply that the lattice is a 
finite volume regulator with the smallest nonzero spatial momentum being 
$ k_\mu = \tfrac{\pi}{aN_\sigma} $. 
The finite lattice spacing enforces $ k_\mu = \tfrac{\pi}{a} $ as the largest 
accessible four-momenta, and thus provides an ultraviolet cutoff. 
\vskip1ex
Any lattice regularization entails (at least) the explicit breaking of the rotational 
symmetry from $ O(d) $ to a discrete symmetry of lattice rotations (typically a 
hypercubic symmetry $ W_d $). 
Since there is no derivative operator on a lattice, it has to be 
approximated through finite difference operators for any lattice regularization, 
\mbox{i.e.}, in the simplest symmetric form 
\al{
 \del_\mu 
 &= \frac{\nab_\mu+\nabc_\mu}{2} + \mathcal{O}(a^2),&\nonumber\\
 & \nab_\mu \psi(x) = \frac{\psi(x+a\hat{\mu})-\psi(x)}{a},&
 & \nabc_\mu \psi(x) = \frac{\psi(x)-\psi(x-a\hat{\mu})}{a}.
}
Here and in the following $ a\hat{\mu}$ denotes the vector of one lattice spacing 
$ a $ in the $\mu$ direction.
Due to this lack of an exact derivative operator, \Eqref{eq:gaugeinv} 
cannot be realized on the lattice for arbitrary gauge transforms $ \Omega $. 
However, QCD can be rephrased on a (hypercubic) lattice by substituting the gauge 
fields $ A_\mu(x) $, which are elements of the Lie algebra $ \mf{su}(N_c) $, with 
the gauge links 
\al{\label{eq:links}
 & U_{\mu,x} = \exp{[iaA_\mu(x+\tfrac{a}{2}\hat{\mu})]},&
 & U_{\mu,x} \to \Omega_x U_{\mu,x} \Omega_{x+a\hat{\mu}}^\dag, 
}
that transform as elements of the Lie group $ \mr{SU}(N_c) $. 
Whereas the quark and antiquark fields are defined on the sites of the lattice, 
the gauge fields on the lattice are placed on the links between the sites. 
Wilson lines are constructed by attaching the gauge links as path-ordered 
segments of a continuous path, and the gauge invariance of the trace of any 
closed contour follows from \Eqref{eq:links}. 
There are two general classes of closed contours -- the topologically trivial 
Wilson loops, which can be deformed into a point (\mbox{i.e.}, do not wrap 
around a boundary) and the topologically nontrivial Wilson loops, which 
can be deformed into a line wrapping around the lattice. 
The smallest possible Wilson loop wrapping around one elementary square is 
called the plaquette. 

The covariant derivative can be expressed with the gauge links 
\al{\label{eq:covder}
 D_\mu \psi(x)
 &= \frac{U_{\mu,x}\psi(x+a\hat{\mu}) - U_{\mu,x-a\hat{\mu}} \psi(x-a\hat{\mu})}{2} 
 + \mathcal{O}(a^2).
}
From \Eqref{eq:covder} the gauge invariance of the matter part 
of the QCD Lagrangian is evident, too. 
It follows that Wilson lines with a quark and an antiquark on their ends are 
also gauge invariant.
\vskip1ex
There are different possibilities for the discretization of the field monomials of 
the continuum QCD Lagrangian of \Eqsref{eq:LagQCD} and~\eqref{eq:muN} 
that differ only in terms of the discretization errors, \mbox{i.e.}, unphysical 
effects of order $ a,~a^2,~\ldots $, which vanish as the regulator is removed in 
the \emph{continuum limit}. 
Various formulations of lattice QCD present different regularization schemes that are 
distinguished by the type and the magnitude of the unphysical effects that are 
introduced by the regulator. 
These effects can always be systematically diminished in so-called 
\emph{improved actions}, 
but never fully eliminated without taking the continuum limit.
\vskip1ex
\subsection{Renormalization and weak coupling}

The fundamental charge $ g(\nu) = \sqrt{ 4\pi \als(\nu) } $ associated 
with the local gauge symmetry explicitly depends on the scale $ \nu $, at which 
the respective interactions transpire. 
The evolution of the charge $ \als(\nu) $ with the scale $ \nu $ is controlled 
by the QCD beta function~\cite{Gross:1973ju, Politzer:1973fx}  
\al{\label{eq:betafun}
 \frac{d\,\als(\nu)}{d\ln \nu}
&=
\als~\beta(\als)
= -\frac{\als^2}{2\pi}\sum\limits_{n=0}^\infty \left( \frac{\als}{4\pi} \right)^n \beta_n
=-2\alpha_s\left[\beta_0\frac{\alpha_s}{4\pi}+\beta_1\left(\frac{\alpha_s}{4\pi}\right)^2+\cdots\right],
}
Only its first two coefficients are universal (independent of the 
regularization scheme) 
\al{
 &\beta_0 =  \frac{11}{3}C_A-\frac{4}{3}T_FN_f,&
 \beta_1 & =  \frac{34}{3}C_A^2-\frac{20}{3} C_A N_f T_F-4 C_F N_f T_F, 
}
with the color factors being 
\al{
C_F  =\frac{N_c^2-1}{2N_c},\qquad C_A=N_c,\qquad T_F=\frac{1}{2},
}
where $ N_c $ is the number of colors.
$ N_f $ has to be understood as the number of (approximately massless) 
quark flavors contributing up to the scale $ \nu $. 
As such, the multiplicative factor $ 1/g_0^2 $ that scales the gauge Lagrangian
$ \mc L_{\rm gauge} $ has to be understood as a bare gauge coupling that is 
related to the physical charge $ g(\nu) $ at the scale $ \nu $ through the QCD 
beta function of \Eqref{eq:betafun}. 
In a similar way, the bare quark masses $ m_{0f} $ are related to renormalized quark 
masses $ m_f(\nu) $ at the scale $ \nu $ through the renormalization group flow. 
\vskip1ex
In the lattice formulation, the lattice spacing is not an explicit input parameter. 
Instead it is related to the bare gauge coupling $ g_0 $ by the QCD beta function as 
\al{\label{eq:latbfun}
 a \Lambda_{\rm lat} 
 &= \left(\frac{1}{\beta_0 g_0^2}\right)^{\frac{\beta_1}{2\beta_0^2}} 
 e^{-\frac{1}{2\beta_0 g_0^2}},
}
where $ \Lambda_{\rm lat} \sim \lMSb $ (parametrically) is a representation of 
the QCD Lambda parameter that is specific to the details of each lattice 
regularization scheme. 
Any quantities with nontrivial mass dimension are given in units of the lattice 
spacing. Namely, the bare quark masses are given as $ a m_{0f} $. 
The lattice spacing is determined a posteriori by \emph{setting the scale} using 
an observable with nontrivial mass dimension that can be easily computed with 
high precision and has only a small sensitivity to the discretization errors, 
\mbox{i.e.} higher powers in $ a $. 
Keeping physical observables constant while approaching the continuum limit 
requires that the bare gauge coupling $ g_0 $ as well as the bare quark masses 
$ a m_{0f} $ are adjusted to smaller values along lines of constant physics.  
\vskip1ex
Due to the adjoint charge of the gluons ($ A_\mu^{ab} $) the gauge Lagrangian
$ \mc L_{\rm gauge} $ contains anti-screening self-interactions, which express 
themselves in the first, \mbox{i.e.}, positive contributions to 
$ \beta_0 $ and $ \beta_1 $. 
These self-interactions give rise to the emergent scale of QCD, $ \lMSb $, at 
which \Eqref{eq:betafun} has a Landau pole, where the strong coupling 
$ \als(\nu=\lMSb) $ diverges. 
In fact, the confinement property of QCD at low values of the temperature implies 
that the energy density in strongly interacting matter grows linearly with the separation.
Hence, fields in any nontrivial representation of the gauge group must be dressed 
in clouds of nuclear matter with typical energies $ E \sim \lMSb $, 
and may only propagate over very small distances $ r \ll 1/\lMSb $. 
On longer times scales $ t \gtrsim 1/\lMSb $ these fields have to aggregate into 
bound states that transform in the trivial representation. 
For temperatures as high as $ T \gtrsim \lMSb $, the confinement property is lifted 
due to the color screening of long-range forces in the quark-gluon plasma. 
\vskip1ex
At much higher scales $ \nu \gg \lMSb $ the strong coupling constant becomes small, 
and this asymptotic freedom explicitly permits a weak-coupling expansion of 
the strong interactions. 
In an equivalent scheme that is particularly convenient for a weak-coupling 
expansion the gauge fields are defined as $ A_\mu = g_0 \mc A_\mu $, \mbox{i.e.}, 
the bare gauge coupling $ g_0 $ explicitly multiplies all three-point functions 
and $ g_0^2 $ multiplies all four-point functions. 
\vskip1ex
Processes at scales in the vicinity of $ \lMSb $ are inherently nonperturbative, 
\mbox{i.e.} a perturbative expansion in powers of $ g $ (or $ \als $) cannot 
describe such interactions at all. 
The lattice formulation lends itself to a nonperturbative calculation of the QCD 
partition function. 
This is accomplished by performing a Wick rotation $ t \to -\ri \tau $ that 
trades Minkowski space-time for Euclidean space-time and transforms the action 
as $ \ri S_M \to - S_E $. 
Thus, the exponential factors in \Eqsref{eq:ZQCD} and~\eqref{eq:Obs} become real, 
and importance sampling can be applied for evaluating the path integral in a 
Markov Chain Monte Carlo simulation in practice. 
Under these conditions the weak-coupling expansion is not necessary anymore and 
the nonperturbative calculation is explicitly feasible.
\vskip1ex
The different quark flavors can be generally grouped into the two subsets of 
\emph{light} or \emph{heavy} quark flavors, which are distinguished by the 
ordering of the respective quark masses $ m_f $ and $ \lMSb $. 
Namely, the \emph{light} quark flavors $ f=u,~d,~s$ have $ m_f \lesssim \lMSb $, 
while the \emph{heavy} quark flavors $ f=c,~b $ have $ m_h \equiv m_f \gg \lMSb $. 
These two different ordered hierarchies may permit the use of the effective field 
theory (EFT) approach for QCD. 
The low-energy limit of QCD is rephrased in terms of specific sets of low-energy 
degrees of freedom, and the high-energy modes above some matching scale $ \nu $ are 
integrated out and absorbed into the Wilson coefficients of the EFT. 
Suitable EFT approaches for the \emph{light} quark flavors are the chiral perturbation 
theory (for a review see, \ie~\mbox{Refs.}~\cite{Scherer:2002tk, Scherer:2012xha}) 
and chiral effective field theory (for a review see, 
\ie~\mbox{Refs.}~\cite{Epelbaum:2008ga, Machleidt:2011zz}), whose 
convergence is restricted to $ p \ll \lMSb $ for all external momenta. 
The corresponding Wilson coefficients are calculated at the scale $ \nu \sim \lMSb $. 
Suitable EFT approaches for the \emph{heavy} quark flavors take the nonrelativistic 
limit of QCD, whose convergence is restricted to $ p \ll m_h $ for all external momenta. 
The corresponding Wilson coefficients are calculated at the scale $ \nu \sim m_h $. 
Thus, whereas the interactions of the former necessitate nonperturbative 
approaches, the latter can be addressed by means of the weak-coupling expansion. 
\vskip1ex

\subsection{Light quarks}

On the one hand, since $ m_f/\lMSb $ is a small quantity for the \emph{light} 
quark flavors, these behave as almost massless in QCD, and it is permissible to 
expand QCD amplitudes involving the \emph{light} quark flavors in terms of 
$ m_f/\lMSb $. 
This introduces a power counting even for large values of the coupling. 
The matter Lagrangian has an accidental, chiral symmetry in the massless limit. 
This chiral symmetry is in part broken by the axial anomaly of the fermion measure 
(flavor singlet chiral symmetry), and in part spontaneously broken for all states 
of the hadron spectrum due to the formation of a scalar quark-antiquark condensate 
(\emph{chiral condensate}) at low temperatures (all flavor nonsinglet chiral 
symmetries), and, finally, explicitly broken by the nonzero \emph{light} quark masses. 
As a consequence of the interplay between the nonzero \emph{light} quark masses and 
the spontaneous breaking of the chiral symmetry the associated Pseudo-Goldstone 
bosons (pion, kaon, eta) acquire masses significantly below one unit of $ \lMSb $ 
for each contributing quark (or antiquark). 
These are the lightest modes that transform in the trivial representation, and thus 
propagate over large distances in the hadronic phase and induce the strongest 
sensitivity to effects of the finite spatial volume. 
A further consequence of the spontaneously broken chiral symmetry is the 
absence of parity doubling in the hadron spectrum.  
\vskip1ex
If the temperature or the density of a nuclear matter system is increased 
beyond a certain delimiting region in the phase diagram, the description in 
terms of hadronic degrees of freedom breaks down. 
Instead, the bulk properties of the strongly interacting matter system have to be 
understood in terms of the partonic (quark and gluon) degrees of freedom, 
or in terms of a mix between hadronic and partonic degrees of freedom. 
The corresponding thermal expectation values undergo violent thermal 
fluctuations when the system changes between the low and high temperature 
phases. 
The details of these fluctuations and of the transition strongly depend on 
the number of \emph{light} quark degrees of freedom and on the values of the 
parameters (\emph{light} quark masses and \emph{light} quark chemical 
potentials) in the QCD partition function of \Eqref{eq:ZQCD}. 
\vskip1ex
In the limit of pure gauge theory, where the matter part of \Eqref{eq:LagQCD} 
is absent (or, equivalently, where all quarks are infinitely heavy), the QCD partition 
function has a $ Z(N_c) $ center symmetry that is manifest in the hadronic 
phase due to the frequent tunneling between the $ N_c $ different sectors of 
the $ Z(N_c) $ center symmetry. 
In the quark-gluon plasma phase this $ Z(N_c) $ center symmetry is broken
spontaneously and the system becomes stuck in one of the 
$ N_c $ different sectors of the $ Z(N_c) $ center symmetry for 
arbitrarily long times. 
For four-dimensional $ \mr{SU}(N_c) $ pure gauge theory with $ N_c=3 $, 
the transition between the phases is of first order. 
The associated non-local \textit{order parameter} is the Polyakov loop, 
namely, the most simplest case of a topologically nontrivial Wilson 
loop wrapping around the periodic Euclidean time direction, whose 
fluctuations diverge at the phase transition, 
see \mbox{Section}~\ref{sec:ploop}. 
Above the transition, the Polyakov loop assumes a non-zero expectation value 
(in the infinite volume limit) due to the color screening. 
\vskip1ex
In full QCD with dynamical \emph{light} quark flavors this $ Z(N_c) $ center 
symmetry is explicitly broken by the presence of the \emph{light} quark 
flavors. 
Namely, the associated \emph{chiral condensate} couples to the Polyakov loop 
and acts as a $ Z(N_c) $ center symmetry breaking field already in the vacuum. 
Therefore, the Polyakov loop does not play the role of an order parameter in 
full QCD with dynamical \emph{light} quark flavors and is not related to the
critical behavior in the chiral limit.
Instead, the \emph{chiral condensate} is the order parameter of full QCD, and 
the partition function becomes singular at the \emph{chiral phase transition} 
in the limit of vanishing masses of the \emph{light} quark flavors. 
The role of the symmetry breaking field in full QCD is thus played by the 
finite masses of the \emph{light} quark flavors. 
At higher temperatures, the thermal fluctuations of the 
\emph{chiral condensate} eventually become too large, and the thermal 
expectation value of the \emph{chiral condensate} drops to zero. 
Depending on the number of \emph{light} quark flavors and their masses, 
the universality class and the details of the transition vary. 
For the physical values of the quark masses, there is no sharp phase 
transition, but a smooth crossover. 
Nevertheless, the physical \emph{light} quark masses are small enough that 
the real world is quite sensitive to the \emph{chiral phase 
transition}.
The fluctuations of the \emph{chiral condensate} diverge in the 
massless limit at the critical temperature 
$T_c^0=132^{+3}_{-6}\,\mr{MeV}$~\cite{Ding:2019prx}, and reach a pronounced 
maximum for physical \emph{light} quark masses at the pseudo-critical 
temperature $T_{c}=156.5(1.5)\,\mr{MeV}$~\cite{Aoki:2009sc, Bazavov:2011nk, 
Bazavov:2018mes}.
\vskip1ex
In the quark-gluon plasma phase at temperatures sufficiently above the 
\emph{chiral phase transition}, namely for temperatures at or above 
$T \gtrsim 2T_c \sim \lMSb$, the \emph{chiral condensate} has already melted, all 
flavor nonsinglet chiral symmetries are restored, and there are 
no associated Pseudo-Goldstone bosons accordingly. 
The masses of the thermalized \emph{light} quark flavors are much smaller 
than the lowest fermionic Matsubara frequency $ \omega_0 = \pi T $.
Hence, the bulk properties of the quark-gluon plasma depend only mildly on 
the details of the dynamical \emph{light} quark flavors for such 
temperatures. 
Nevertheless, thermalized \emph{heavy} quark flavors may become relevant at 
such high temperatures. 
\vskip1ex
\subsection{Heavy quarks}


\begin{figure*}\center
\includegraphics[width=5.4cm]{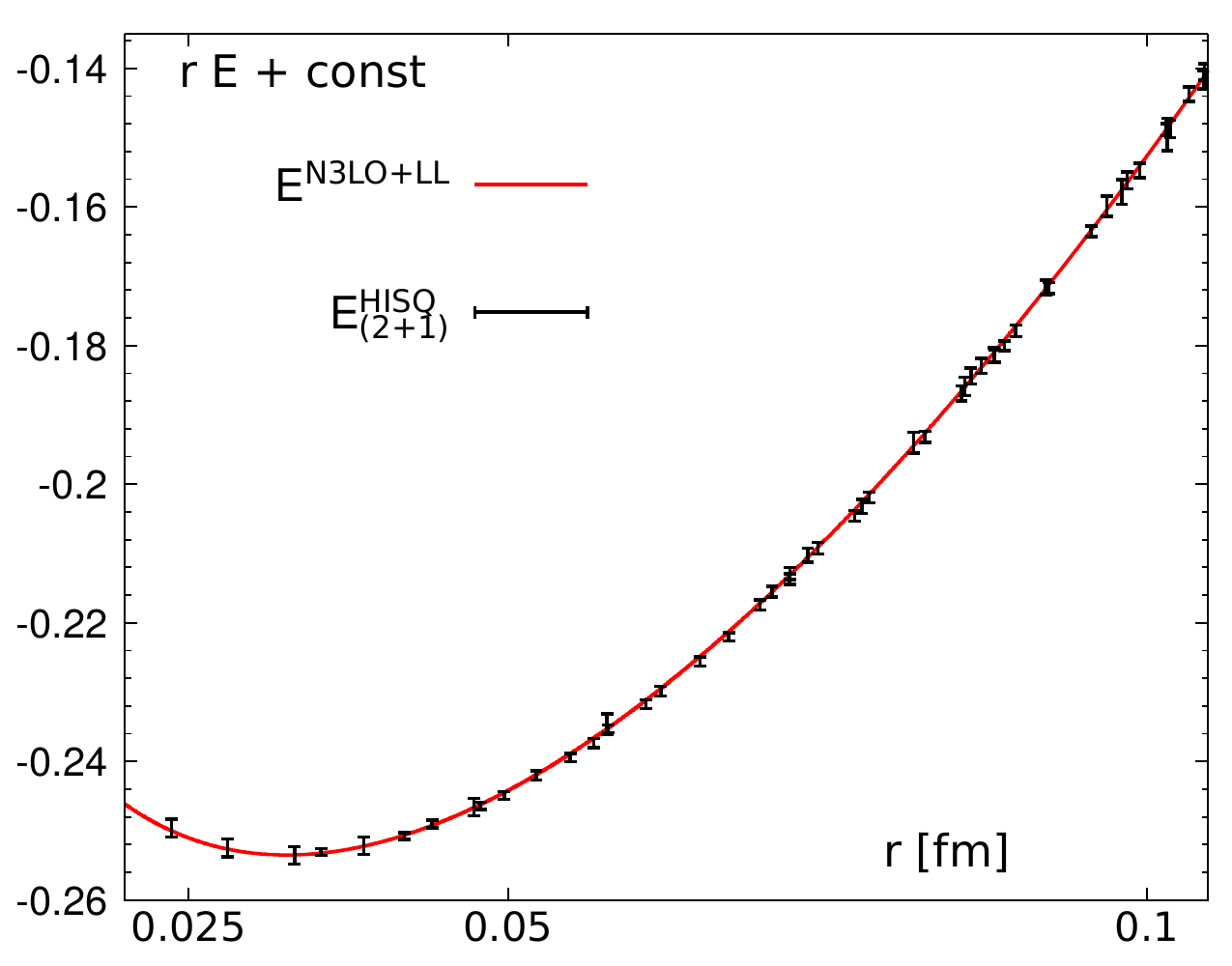}
\includegraphics[width=6.6cm]{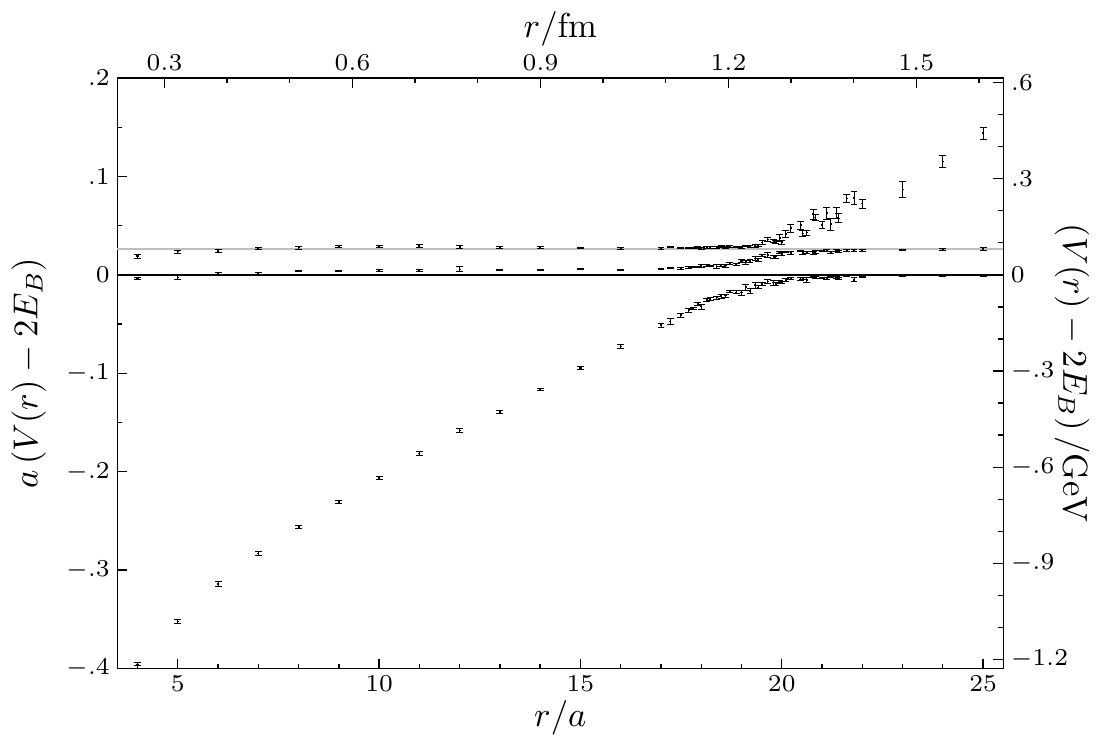}
\caption{
The static energy in (2+1)-flavor QCD on the lattice. 
(Left) After removing the discretization errors at small distances 
from the lattice result (with the HISQ action), the result is well-described 
by the $\mr{N^3LO}$ result for $\als(M_Z)=0.1167$ (with resummation of 
leading ultrasoft logarithms, $\mr{LL}$) up to 
$r \sim 0.10\,\mr{fm}$~\cite{Bazavov:2019qoo}. 
(Right) In full QCD (with improved Wilson fermions) the string-breaking 
disrupts the QCD string~\cite{Bulava:2019iut}. 
There are two avoided level crossings due to the two non-degenerate 
quark masses of the sea, where the horizontal lines correspond to the 
twice static-light (lower) or static-strange (higher) meson masses.
From \mbox{Ref.}~\cite{Bulava:2019iut}.
\label{fig:static energy}
}
\end{figure*}

On the other hand, since $ \lMSb/m_h $ is a small quantity for the \emph{heavy} 
quark flavors, these behave as almost infinitely heavy in QCD, and it is permissible 
to expand QCD amplitudes for the \emph{heavy} quark flavors in terms of $ \lMSb/m_h $. 
In the static limit the \emph{heavy} quark flavors are treated as infinitely heavy 
(\emph{static quarks}) and do not propagate in space. 
Such immobile \emph{static quarks} are test charges in the fundamental representation 
of $ \mr{SU}(N_c) $, and one is able to define a potential, or rather the 
quark-antiquark static energy $ E $, which can be calculated on the lattice and in 
the weak-coupling approach, reaching good agreement at small 
distances, see \Figref{fig:static energy} (left). 
The typical scale of the gluons that contribute to this energy is 
$ p \sim 1/r $. 
Due to the asymptotic freedom this energy has a Coulombic core $ E \sim \als/r $ 
at short distances $ r \ll 1/\lMSb $. 
Due to the confinement the energy also exhibits a linearly rising contribution 
$ E \sim \sigma r $ at larger distances $ r \sim 1/\lMSb $, the QCD string or 
QCD flux tube, where  $ \sigma \sim \lMSb^2 $.
The most simple potential model that features these two aspects is the Cornell 
potential $ V_{\rm Cornell} = -\alpha/r + \sigma r $. 
At even larger distances, $ r \sim 1/m_\pi $, the potential picture breaks down and 
the QCD string rips apart and a \emph{light} quark-antiquark pair is generated from 
the vacuum for the formation of two \emph{heavy-light} mesons, 
\ie a pair of static $ D $ and $D^\ast $ mesons, 
see \Figref{fig:static energy} (right). 
This string-breaking process can occur only if \emph{light} quark flavors are 
present in the sea, \mbox{i.e.}, this mechanism is impossible with the pure gauge 
Lagrangian. 
Further terms (relativistic corrections, Darwin-term, spin-orbit or spin-spin coupling, 
etc.) are included in the sophisticated nonrelativistic quark models, which describe 
the spectra and transitions in \emph{heavy-heavy} quark-antiquark bound states quite 
successfully.  

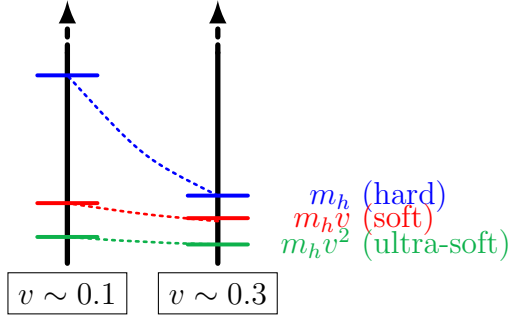
\begin{figure*}\center

\definecolor{spartandarkgreen} {RGB}{24, 69, 59}
\definecolor{spartanlightgreen} {RGB}{13, 177, 75}
\begin{tikzpicture}
\draw [>=latex,line width=2pt, dashed, color=black,line cap=round,->] (0,2.8) -- (0,3.5);
\draw [>=latex,line width=2pt, dashed, color=black,line cap=round,->] (2,2.8) -- (2,3.5);

\draw [>=latex,line width=2pt,color=black,line cap=round] (0,-0) -- (0,2.8);
\draw [>=latex,line width=2pt,color=black,line cap=round] (2,-0) -- (2,2.8);


\node [draw] at (0,-0.4) {$v \sim 0.1$};
\node [draw] at (2,-0.4) {$v \sim 0.3$};

\draw [>=latex,line width=1.5pt, color=blue, line cap=round] (+1.6,+0.90) -- (2.4,+0.90) node[anchor=west] {~~$\phantom{v^2}m_h$ (hard)} ;
\draw [>=latex,line width=1.5pt, color=blue, line cap=round] (-0.4,+2.50) -- (0.4,+2.50);
\draw [>=latex,line width=1.0pt, color=blue, dotted, line cap=round] (+2.0,+0.90) .. controls (+1.0,+1.40) .. (+0.0,+2.50);

\draw [>=latex,line width=1.5pt, color=red, line cap=round] (+1.6,+0.60) -- (2.4,+0.60) node[anchor=west] {~~$\phantom{^2}m_hv$ (soft)} ;
\draw [>=latex,line width=1.5pt, color=red, line cap=round] (-0.4,+0.80) -- (0.4,+0.80);
\draw [>=latex,line width=1.0pt, color=red, dotted, line cap=round] (+2.0,+0.56) .. controls (+1.0,+0.65) .. (+0.0,+0.80);

\draw [>=latex,line width=1.5pt, color=spartanlightgreen, line cap=round] (+1.6,+.25) -- (2.4,+0.25) node[anchor=west] {~~$m_hv^2$ (ultra-soft)} ;
\draw [>=latex,line width=1.5pt, color=spartanlightgreen, line cap=round] (-0.4,+0.35) -- (0.4,+0.35); 
\draw [>=latex,line width=1.0pt, color=spartanlightgreen, dotted, line cap=round] (+2.0,+0.25) .. controls (+1.0,+0.28) .. (+0.0,+0.35);
  
\end{tikzpicture}
\caption{
Nonrelativistic QCD hierarchies for the bottom quark (left), and 
the charm quark (right). 
}
\label{fig:nr hierarchy}
\end{figure*}

Upon expansion about the static limit one obtains the nonrelativistic QCD 
(NRQCD)~\cite{Caswell:1985ui, Bodwin:1994jh} for the \emph{heavy} quark flavors. 
The nonrelativistic \emph{heavy} quarks and antiquarks decouple to leading 
order, namely, pair creation or annihilation are suppressed, since typical gluon 
momenta are $ p \sim \lMSb \ll m_h $.  
The symmetry of the couplings between nonrelativistic quarks (and antiquarks) 
to chromoelectric and -magnetic gluons is broken by powers of $ \lMSb/m_h $. 
Eventually, \emph{heavy-heavy} bound states of quarks and antiquarks are small 
and compact objects, where the typical binding energies $ E \sim m v^2/2 $ are 
even smaller than the typical momenta $ p \sim m v $ (with $ v \ll c $). 
Thus, there is another layer of the 
hierarchy $ E \sim \als/r \ll p \sim 1/r \ll m_h $, which permits the multipole 
expansion of such amplitudes, see \Figref{fig:nr hierarchy}. 
The EFT obtained in this manner is called potential nonrelativistic QCD 
(pNRQCD)~\cite{Brambilla:1999qa, Brambilla:1999xf}. 
pNRQCD has non-local Wilson coefficients, which have the meaning of various 
potential or non-potential terms similar to those used in nonrelativistic quark 
models. 
In pNRQCD, such potentials can be derived from first principles. 
For a review of NRQCD and pNRQCD see \mbox{Refs.}~\cite{Brambilla:2004jw, Pineda:2011dg}.
\vskip1ex
\subsection{Implementation of QCD on the lattice}

In lattice gauge theory, generally only the three \emph{light} quark flavors, or 
additionally the charm quark, are considered as thermalized degrees of freedom. 
Since the up or down quark masses $ m_u $ or $ m_d $ are much smaller than the 
strange quark mass $ m_s $, QCD is very often approximated in the isospin limit, 
where instead of up and down quarks two degenerate light quarks with an average 
light quark mass $ m_l = (m_u+m_d)/2 $ are considered. 
For this scenario the terms (2+1)- or (2+1+1)-flavor QCD are commonly used. 
\vskip1ex
The most simple lattice action for the gauge fields is the Wilson plaquette action, 
\mbox{i.e.} a sum of the trace of all elementary $ 1 \times 1 $ Wilson loops, 
which has leading discretization errors of order~$ a^2 $.
The breaking of rotational symmetry is less pronounced for the improved gauge actions, 
namely the tree-level (or one-loop) Symanzik gauge action, which includes all elementary 
$ 1 \times 2 $ (or also $ 1 \times 1 \times 1 $) Wilson loops, too, and has 
discretization errors of order~$ a^4 $ and~$ \als a^2 $ (or~$\als^2 a^2 $).  
\vskip1ex
Local link distributions in lattice simulations tend to suffer from 
violent UV fluctuations at short separations, which can be ameliorated 
through the use of a variety of iterative link smoothing techniques. 
Typical link smoothing algorithms are the APE smearing~\cite{Albanese:1987ds}, 
the HYP (hypercubic) smearing~\cite{Hasenfratz:2001hp}, 
stout smearing~\cite{Morningstar:2003gk} or the 
Wilson flow~\cite{Luscher:2010iy,Luscher:2011bx}. 
These techniques can be applied to subsets of the links, \ie typically 
only to spatial links or to all links, and with or without mixing links 
at different times. 
Any of these link smoothing techniques tend to distort the physics at 
small temporal and/or spatial separations, while suppressing the UV 
divergences of lattice operators. 
For this reason, the appropriate amount and type of smoothing is problem 
specific in lattice simulations. 
\vskip1ex
Quarks on the lattice are more complicated. 
The naive lattice derivative of \Eqref{eq:covder} has discretization errors of 
order~$ a^2 $, but produces $ 2^d $ degenerate lattice fermions. 
Hence, it is not directly suitable for QCD, since the continuum limit of such 
a discretization does not produce a theory with correct number of dynamical 
degrees of freedom. There are two commonly used approaches\footnote{
The more recent, overlap~\cite{Neuberger:1997fp} and 
domain-wall~\cite{Kaplan:1992bt} fermions realize an exact or almost exact 
chiral symmetry with the correct axial anomaly on the lattice and also have 
discretization errors of order~$ a^2 $, but are substantially more 
computationally expensive. 
As such, we will not discuss these formulations here.}
to remedy this problem: Wilson and staggered.
\vskip1ex
The Wilson fermion approach~\cite{Wilson:1974sk} 
 includes a momentum-dependent mass term at order~$ a $ 
-- the Wilson term -- which breaks chiral symmetry explicitly even in the limit 
$ m_{0f}\to 0 $, and introduces the additive mass renormalization (due to the 
different renormalization of the two parts of the Wilson term). 
An improved formulation of Wilson fermions introduces the so-called Clover term 
to achieve discretization errors of order~$ a^2 $~\cite{Sheikholeslami:1985ij}, 
but all individual quark bilinears using (improved) Wilson fermions require 
additional corrections for the order~$ a $ errors. 
\vskip1ex
The staggered fermion approach~\cite{Kogut:1974ag}, which is derived from the 
naive discretization, has the same discretization errors of order $ a^2 $. 
For even number of dimensions $d$ the corresponding $ 2^{d/2} $~spinor 
components are decoupled through a unitary transform and $ 2^{d/2}-1 $ of these 
are omitted. 
The staggered discretization still has a reduced doubling problem with $ 2^{d/2} $ 
degenerate fermions that can only be accounted for by taking the $ (2^{d/2}) $th 
root of the staggered quark determinant. 
The components of the fermions become distributed over $ 2^d $ lattice sites and 
the $ 2^d $ hyperquadrants of the Brillouin zone. 
Due to couplings between the $ 2^d $ hyperquadrants in the interacting field 
theory, the discretization errors are numerically large. 
Various formulations of improved staggered fermions have reduced discretization 
errors of order $ a^4 $ and $ \als a^2 $, and suppress the coupling between the 
hyperquadrants, namely, the AsqTad~\cite{Orginos:1999cr} and HISQ (highly improved 
staggered quarks)~\cite{Follana:2006rc} 
formulations, whereas the stout formulation~\cite{Morningstar:2003gk} simply 
reduces the $ a^2 $ and $ a^4 $ errors substantially.
Staggered fermions are the most popular quark discretization for QCD at finite temperature.
\vskip1ex
In practice the evaluation of the quark determinant $ \det \{D[A_\mu](m_{0f},\mu_f) \} $ 
(using stochastic estimators) is the most expensive part of the Markov 
Chain Monte Carlo algorithm, in particular, if the sea quark masses are small. 
At each lattice site $ \det \{D[A_\mu](m_{0f},\mu_f) \} $ in \Eqref{eq:Zdet} is sensitive 
to the gauge fields on all links of the entire lattice, whereas the gauge part of the QCD 
Lagrangian contains only small Wilson loops, and thus, very localized couplings.  
For this reason, lattice QCD simulations are often performed with a \emph{light} quark mass 
that is larger than the physical value, or even without sea quarks at all, in the so-called 
quenched approximation, where the quark determinant is approximated by $ 1 $ 
and purely local updating can be used. 
It needs to be stressed that the Markov Chain that is generated in the quenched 
approximation samples the partition function of $ \mr{SU}(N_c) $ pure gauge theory. 
\vskip1ex

\section{Screening of static charges}
\label{sec:static}


Before introducing the observables related to the properties of the
medium and screening effects in a non-Abelian gauge theory,
such as QCD, let us recall some of the important features of the well-studied,
both on classical and quantum level, Abelian gauge theory --
Quantum Electrodynamics.

One of the simplest questions one can ask when encountering a new
force of nature is what is the force between two static point-like
objects capable of interacting through that force. For 
macroscopic electromagnetism
the answer to this question has been experimentally established
in 18th century with the discovery of Coulomb's law.
As is well known today, the force between two static probe charges
$q_1$ and $q_2$ 
is long-range and falls of as $F~\sim q_1 q_2/r^2$ with the distance between
the charges, which corresponds to the Coulomb-type potential
$V~\sim\alpha_\mr{em}/r$.
If the test charges are embedded in an electromagnetically
interacting medium, such as an electromagnetic plasma, this long-range
force gets \emph{screened} and the range of interaction (excluding 
the van-der-Waals force that arises due to the polarization of the medium) becomes
finite, governed by the screening length. The potential becomes
of a Yukawa type $e^{-mr}/r$, where $r_D=1/m$ -- is the
Debye-H\"{u}ckel length~\cite{DebyeHuckel1923}.
Thus, studying the force between static
probe charges, one learns about the properties of the medium, and
the nonperturbative lattice QCD approach follows a very similar strategy
as discussed in the following sections.

Another ingredient that helps to set up the stage for QCD is
the Aharonov-Bohm effect~\cite{1959PhRv..115..485A} in electrodynamics.
Namely, while propagation of a charged particle on the classical
level is completely determined by the electric and magnetic
fields, on the quantum level it is sensitive to the phase induced
by the electromagnetic potential $A_\mu$, even if the particle
is propagating through the regions of space where the electric
and magnetic fields are zero. This phase, acquired by the
particle propagating in the gauge field background plays
a fundamental role in the discussion of the thermodynamic properties
of the strongly interacting medium.

\subsection{The Polyakov loop and related quantities}
\label{sec:ploop}

\subsubsection{Wilson line and the Polyakov loop}

For a non-Abelian gauge theory the analog of the 
path-dependent phase in the
Aharonov-Bohm effect is the \emph{gauge connection} or
\emph{Wilson line}:
\begin{equation}
\label{eq:Wyx}
W(y,x)=P \exp\left\{
i\int_x^y A_\mu(z)dz^\mu
\right\}.
\end{equation}
where the integral
is understood as a line integral along the path connecting
$y$ and $x$ and $P$ represents \emph{path ordering}.
The latter plays
an important role for a non-Abelian case, since local $A_\mu$ fields at different points generally do not commute, 
as clear from Eq.~(\ref{eq:Amu}).
The discussion is still in the continuum, but one can easily
recognize that the gauge link variables introduced in
Eq.~(\ref{eq:links}) are the shortest possible Wilson lines
that can be resolved on a lattice with spacing $a$.

A \emph{Wilson loop} is a Wilson line over a closed path.
Such a construction is gauge invariant, and lattice observables
built of only gauge fields need to be always formulated in terms
of Wilson loops traced over the color indices.
A Wilson line running in the temporal direction represents the
phase acquired by a static probe charge 
whose position in space remains fixed.
A flat Wilson loop of size, \emph{e.g.} $L_x\times L_t$ in
$(x,t)$ plane represents a static quark-antiquark pair
(since the temporal Wilson lines have to run in the opposite direction
when going over the loop, thus, naturally representing a
particle and anti-particle). Wilson loops 
or correlators of spatially separated temporal Wilson lines
are therefore natural
objects to study the force between static probe charges.

While planar loops that do not wind around the lattice can be contracted to a point,
there are also non-trivial Wilson loops that are closed paths
due to the periodic boundary conditions.
A special kind of a loop represented by a temporal Wilson line
that forms a closed path by connecting to itself through the
periodic boundary in the temporal direction is called \emph{the Polyakov loop}:
\begin{equation}
\label{eq:ploop}
L(\bm{r}) = \frac{1}{N_c}\tr P \exp\left\{
i\int_0^\beta A_0(t,\bm{r})dt
\right\}.
\end{equation}
We refer to this object as a \emph{thermal Wilson line} if the trace is not taken.
On the lattice Eq.~(\ref{eq:ploop}) is particularly simple, it
amounts to taking the trace of a product of gauge links going in
the temporal direction
to obtain the bare Polyakov loop:
\begin{equation}
L^\mr{bare}(\bm x) = \frac{1}{N_c} \tr \prod\limits_{\tau=a}^{aN_\tau} U_{0,(\tau,~\bm x)}.
\end{equation}
It is customary in lattice QCD calculations to take advantage of the 
translational invariance and define quantities summed over the whole lattice.
This self-averaging often significantly improves the signal-to-noise ratio, even
though the nearby points are strongly correlated.
The lattice-averaged Polyakov loop is
\begin{equation}
L = \frac{1}{V} \sum\limits_{\bm x}L(\bm x).
\end{equation}
The expectation value of the Polyakov loop in the sense of
Eq.~(\ref{eq:Obs}) is related to the difference in free energy
between the medium and the medium with a single static charge inserted into it:
\begin{equation}
\label{eq:Fq}
\langle L\rangle=\exp\left\{-\frac{F_q}{T}\right\}.
\end{equation}
On the one hand, in a pure gauge theory (\emph{i.e.} with dynamical gauge fields
but no dynamical fermions) the Polyakov loop is a (non-local)
order parameter for the confinement-deconfinement transition,
similar to the magnetization in spin systems. 
As discussed in Ref.~\cite{Svetitsky:1982gs}, the Polyakov loop is associated
with the center symmetry, $Z(N_c)$. In the symmetric, confined phase
its expectation value is zero, which is interpreted as an
infinite free energy $F_q$ associated with insertion of a static
probe color charge in the fundamental representation.
This also hints that a concept of an isolated color charged object is not
consistent with the confinement property of non-Abelian gauge theories. 
Namely, it is rigorously impossible to combine any number of gluons, which transform in the adjoint representation of the group, and one quark that transforms in the fundamental representation into an object that transforms in the trivial representation that would be allowed by the confinement. On the contrary, some higher representations of the Polyakov loop do not vanish in the confined phase, and may be related to the \emph{glue-balls}. 
The Polyakov loop has been studied in higher representations in 
the $\mr{SU}(3)$ pure gauge theory~\cite{Gupta:2007ax} and in the (2+1)-flavor 
QCD~\cite{Petreczky:2015yta}. 
Some of its higher representations can be defined (with $L_3 \equiv L$) as 
\begin{equation}
\label{eq:Lrep}
\langle L_6 \rangle = \frac{1}{6} (\langle (3L_3) ^2\rangle-\langle 3L_3^\ast \rangle), \quad
\langle L_8 \rangle = \frac{1}{8} (\langle |3L_3|^2 \rangle-1), \ldots~,
\end{equation}
where $\langle L_8 \rangle$ is the expectation value of the Polyakov loop in the adjoint representation. 
In the broken, deconfined phase $\langle L\rangle$ has
a non-zero value, corresponding to finite $F_q$. Viewed from this perspective
the high-temperature phase of a $\mr{SU}(N_c)$ pure gauge theory is 
a phase with spontaneously broken center symmetry, similar to e.g. a 
low-temperature phase of a ferromagnet. 
On the other hand, in a theory with dynamical fermions the quark condensate 
acts as a symmetry-breaking field, and the expectation value of the Polyakov 
loop in the fundamental representation does not go to zero in the low 
temperature phase.

Since the physics in the low-temperature phase of QCD and around the
confinement-deconfinement transition is nonperturbative, a fully
nonperturbative approach, such as lattice QCD is necessary to study
the behavior of the Polyakov loop with full theoretical control.
Indeed, the Polyakov loop has been a subject of study on the lattice
since the earliest days of lattice gauge theory~\cite{McLerran:1981pb,Engels:1981qx}
until today~\cite{Bazavov:2016uvm}.

For computational reasons simulations with dynamical fermions were
hardly affordable in 1980s and 1990s and studies of the qualitative features of
QCD were performed in SU(2) and SU(3) pure gauge theory. The
finite temperature phase transition in SU(2) is of the second order and in
SU(3) is of the first order. 
In the pure gauge theory the Polyakov loop susceptibility
\begin{equation}
\chi_L=VT^3\left(
\langle |L|^2\rangle-\langle|L|\rangle^2\right)
 =VT^3\left( \frac{1}{9} + \frac{8}{9} \langle L_8 \rangle-\langle L_3\rangle^2 \right)
\end{equation}
can be used to unambiguously define the transition temperature in the
continuum and thermodynamic limit\footnote{Given that numerical
simulations are always done with a finite number of degrees of freedom,
proper finite-size scaling techniques are employed to extrapolate to the thermodynamic
limit and rigorously determine the order of the phase transition.}. 
In full QCD with fermions this is not such a simple issue, since the 
Polyakov loop susceptibility mixes different representations.

\subsubsection{The static quark potential and the renormalized Polyakov loop}

While the temperature $T_\chi$ where $\chi_L$ becomes infinite defines the confinement-deconfinement
phase transition temperature in pure gauge theory, in full QCD the transition is
a smooth crossover and one needs to understand the quantitative behavior
of the Polyakov loop and the associated free energy across the transition and
in the deconfined phase.

\begin{figure*}\center
	\includegraphics[width=8.6cm]{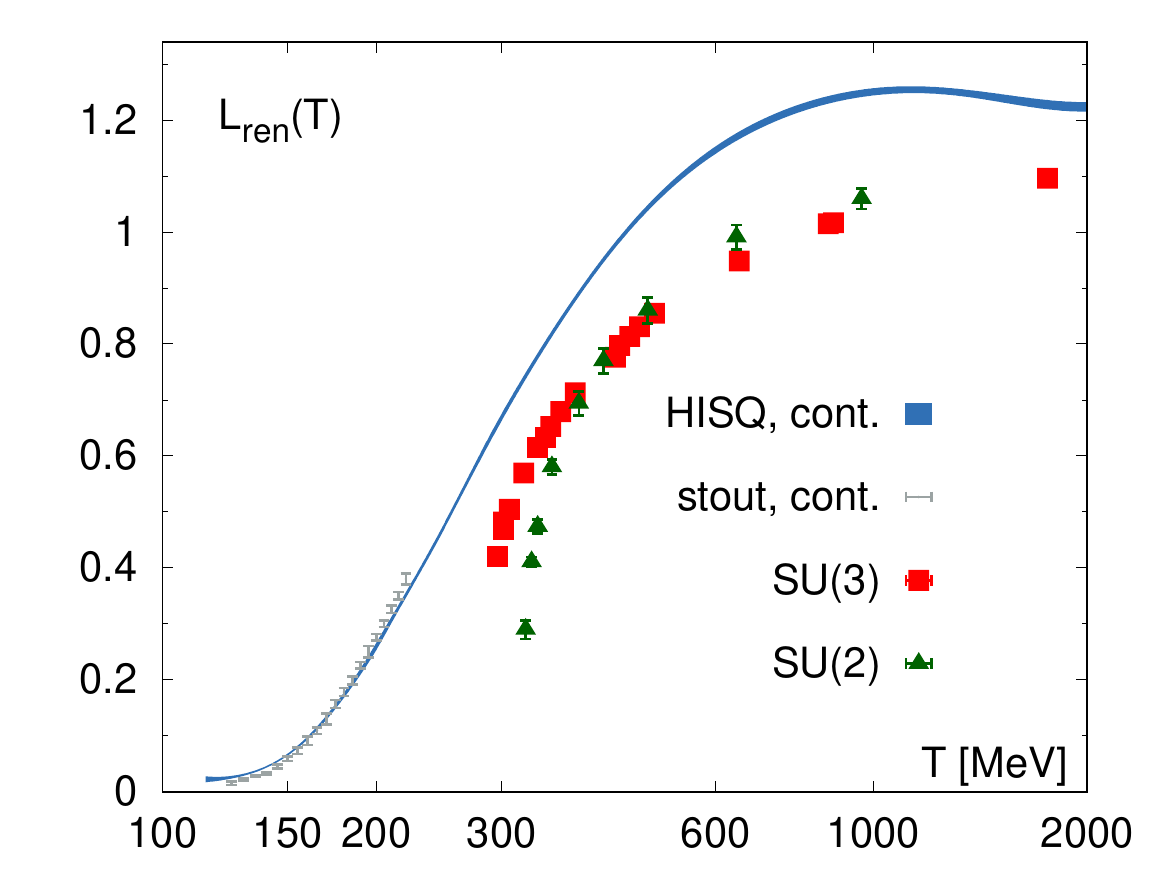}
	\caption{
		The renormalized Polyakov loop in SU(2) and SU(3) pure gauge theory
		(zero values in the confined phase are not shown) and in 2+1 flavor QCD with
		the HISQ and stout action. The chiral crossover temperature in QCD is
		$T_c=155$~MeV.
		\label{fig:Lren}
	}
\end{figure*}

The bare free energy as calculated on the lattice contains a linear divergence
which makes the bare Polyakov loop vanish in the continuum limit.
Renormalization of non-local operators such as Wilson loops 
was considered first in perturbation theory~\cite{Gervais:1979fv,Polyakov:1980ca}
and later nonperturbatively on the
lattice~\cite{Kaczmarek:2002mc,Dumitru:2003hp}.
A common procedure to renormalize the Polyakov loop is to first consider the free
energy of a static quark-antiquark pair.
The ultraviolet divergence in this quantity 
is temperature independent and is already present at zero temperature where the free
energy coincides with the static quark-antiquark energy.


On the lattice the zero-temperature static quark-antiquark energy $E(r)$ can 
be extracted from the expectation value of the temporal Wilson loop of size 
$r\times \tau$ or a correlator of temporal Wilson lines of length $\tau$ 
separated by a distance $r$ in a fixed gauge (the Coulomb gauge is an 
especially convenient choice)

\begin{align}
 \label{eq:E vacuum}
 W^\mr{bare}(\tau,\bm{r};g_0^2) 
 &= \frac{1}{N_c} \Bigg\langle~\sum\limits_{\bm{x}}~\tr \Big[
 {W}(\tau+\tau_0, \bm{x}; \tau, \bm{x}) 
 {W}(\tau, \bm{x}+\bm{r}; \tau+\tau_0, \bm{x}+\bm{r}) 
 ~ \Big]\Bigg\rangle^\mr{gf} \nn\\
 &=\exp{[-E^\mr{bare}(r;g_0^2)\tau]} 
 \left( A_0(r) + \sum\limits_{i=1}^\infty A_i(r) \exp{[-\Delta_i(r)\tau]} \right),
\end{align}
where $\Delta_i(r) \sim \lMSb$ are the positive energy differences between the 
ground state $E^\mr{bare}(r)$ and the excited states.
The bare gauge coupling $g_0$ is an input parameter
of lattice QCD simulations and here we explicitly emphasize that we refer to a quantity
evaluated at non-zero lattice spacing.
Non-perturbative renormalization of the static quark-antiquark energy 
then amounts to choosing a renormalization condition that the
renormalized energy
\begin{equation}
\label{eq:Eren}
E^\mr{ren}(r)=E^\mr{bare}(r;g_0^2)-C(g_0^2)
\end{equation}
is equal to a prescribed value at some fixed distance $\tilde r$, for instance,
\begin{equation}
\label{eq:Econd}
E^\mr{ren}(\tilde r)=0.
\end{equation}
The chosen renormalization condition, \eg\Eqref{eq:Econd},
determines the additive shift $C(g_0^2)$, which is twice the
self-energy of a static quark
in the lattice scheme.

The renormalized Polyakov loop can then be defined as
\begin{equation}
\label{eq:Lren}
L^\mr{ren}(T)=L^\mr{bare}(T;g_0^2)e^{N_\tau C(g_0^2)/2}.
\end{equation}
An alternative renormalization procedure for the Polyakov loop is
possible~\cite{Petreczky:2015yta}
with using recently introduced 
\emph{gradient flow}~\cite{Luscher:2010iy}. In any case, the renormalized
Polyakov loop has a well defined continuum  limit $g_0\to0$, or equivalently,
$a\to0$.

The behavior of the renormalized Polyakov loop $L^\mr{ren}(T)$
 in SU(2), SU(3) pure gauge theory
and in 2+1 flavor QCD is shown in Fig.~\ref{fig:Lren}. One can observe that in
full QCD the renormalized Polyakov loop is smooth across the transition.
It may be tempting~\cite{Aoki:2006we,Aoki:2009sc}
to interpret the inflection point of the renormalized Polyakov loop
as a location of the confinement-deconfinement transition, however,
Eq.~(\ref{eq:Lren}) shows that the exact details of the temperature dependence of
$L^\mr{ren}$ depend on $C(g_0^2)$ and thus on the renormalization prescription.
Such an approach typically yields a transition temperature above 170~MeV,
about 15~MeV  higher than the chiral crossover temperature~at $T_{c}=156.5$~MeV. 
See Sec.~\ref{sec:lattice} for brief remarks on the chiral crossover.

The renormalization of the Polyakov loops in the higher representations 
$\langle L_\bm{R} \rangle$ is a somewhat different 
case. 
The renormalization constant can be obtained from the static energy only in 
the fundamental representation ($\langle L_3 \rangle \equiv \langle L \rangle$). 
For all other representations, one has to rely on other procedures, such 
as the direct renormalization~\cite{Gupta:2007ax,Bazavov:2016uvm} or the 
gradient flow renormalization~\cite{Petreczky:2015yta}. 
For temperatures below about twice the chiral crossover temperature, 
\ie $T < T_{c}=300$~MeV in (2+1)-flavor QCD, the Polyakov loop exhibits a 
large violation of the Casimir scaling between the different representations, 
which begin at the four-loop order in the weak-coupling 
expansion~\cite{Berwein:2015ayt}.

\subsubsection{Static quark free energy and entropy}

\begin{figure*}\center
	\includegraphics[width=8.6cm]{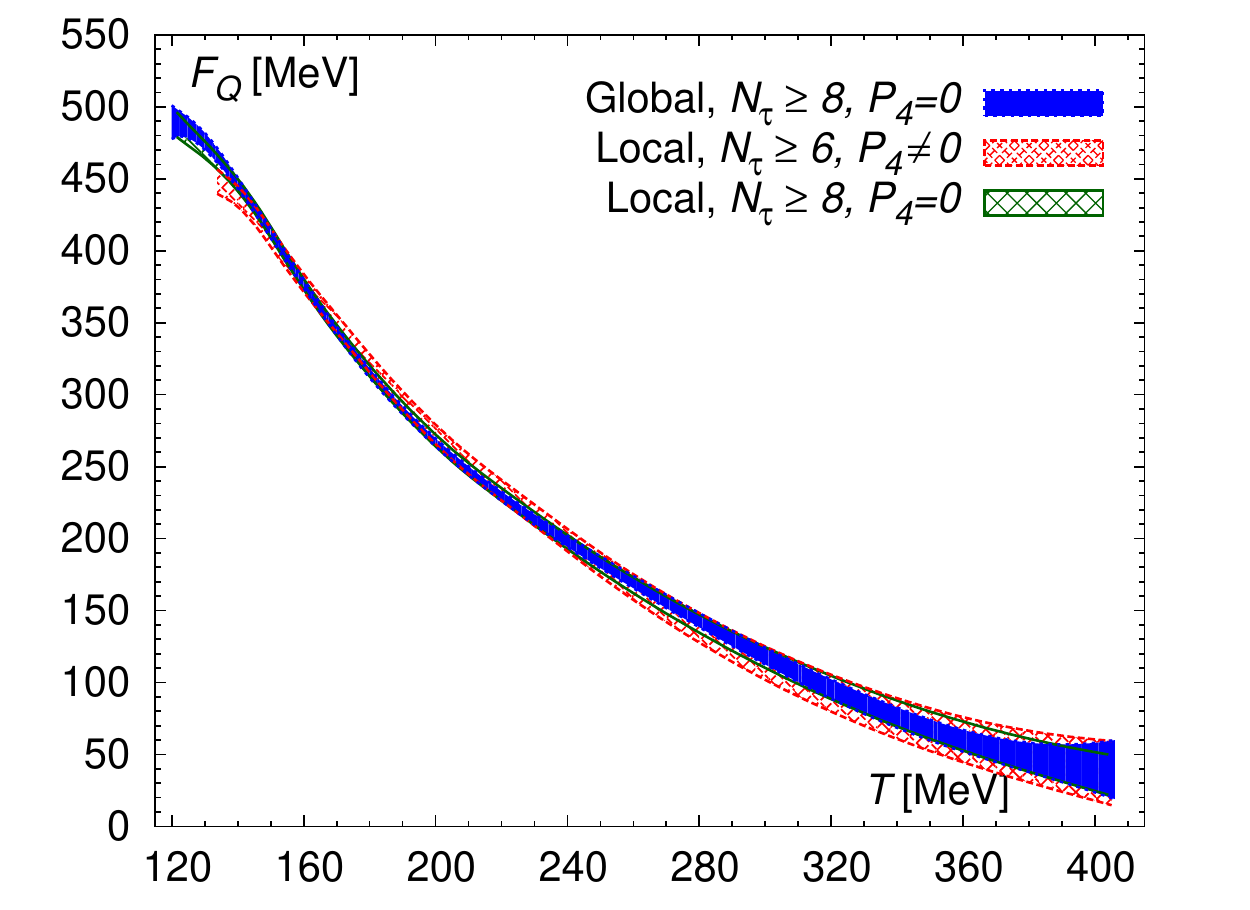}
	\caption{
		The renormalized static quark free energy $F_q$ in 2+1 flavor
		QCD extrapolated to the
		continuum limit. The details of the extrapolations are given
		in Ref.~\cite{Bazavov:2016uvm}.
		\label{fig:Fq}
	}
\end{figure*}

In contrast to the Polyakov loop, 
the renormalization of the static quark free energy $F_q$, Eq.~(\ref{eq:Fq}),
is additive like the static quark potential, and it is therefore
a more robust observable to describe the transition region.
The temperature dependence of the continuum extrapolated renormalized
free energy in 2+1 flavor QCD is shown in Fig.~\ref{fig:Fq}.
At the lowest temperature in the confined phase $T=120$~MeV accessible 
in the lattice calculation it is about 500~MeV, rapidly decreasing to 400~MeV at
the chiral crossover at the pseudo-critical temperature $T_{c}=156.5$~MeV 
and then gradually dropping by an order of magnitude in the deconfined phase in 
the temperature range from 155 to 400~MeV, or $T_c$ -- $2.5T_c$.
The color screening effects of the medium in QCD kick in slowly which is quite
different from the rapid drop of the static quark free energy associated with the 
first order phase transition in SU(3) pure gauge theory.

\begin{figure*}\center
	\includegraphics[width=8.6cm]{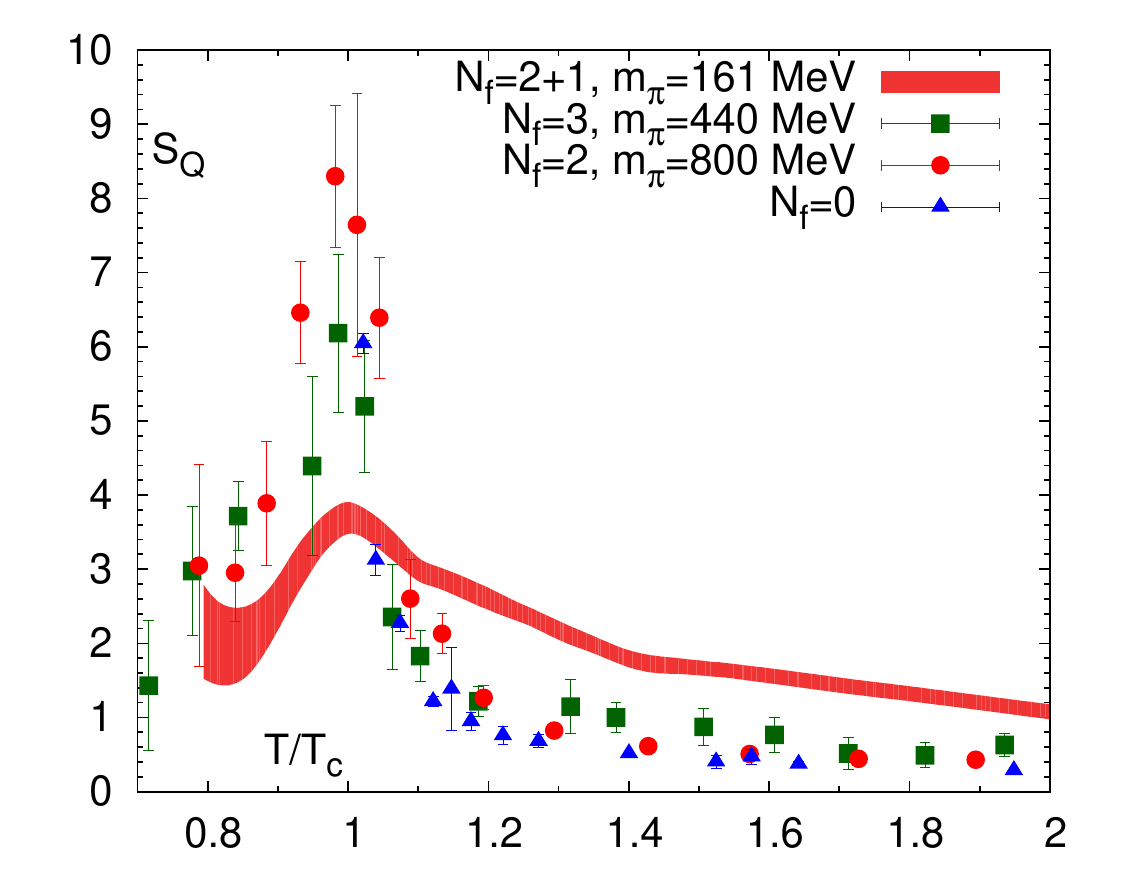}
	\caption{
		The static quark entropy $S_q$ in 2+1 flavor
		QCD extrapolated to the
		continuum limit (solid red curve) compared with earlier
		calculations at finite cutoff and heavier than physical light quark masses
		and and in pure gauge theory. Because of the different particle content
		the numerical value of the transition temperature $T_c$ and thus the temperature
		scale is presented in units of $T_c$ corresponding to $T_S=153$ for 2+1 flavor QCD
		(see text), $T_\chi=193$~MeV and $T_\chi=200$~MeV for the  $N_f=3$ and $N_f=2$
		results and $T_L=270$~MeV for the pure gauge ($N_f=0$) case.
		\label{fig:Sq}
	}
\end{figure*}

Nevertheless, it is still meaningful to consider the inflection point in the static quark
free energy as one of the quantitative measures of the crossover into the
deconfined phase.
For this purpose we first define the 
entropy shift due adding to the medium
a static quark~\cite{Kaczmarek:2005ui,Kaczmarek:2005gi} as
\begin{equation}
\label{eq:Sq}
S_q(T) = -\frac{\partial F_q(T)}{\partial T}.
\end{equation}
The inflection point of $F_q(T)$ corresponds to the location of the peak in the
static entropy
\begin{equation}
\frac{\partial S_q}{\partial T}=0.
\end{equation}
The static quark entropy in the transition region for 2+1 flavor QCD at the almost
physical light quark mass\footnote{Although the pion mass in calculations of
Ref.~\cite{Bazavov:2016uvm} is about 160~MeV, this deviation of about 20~MeV 
from the physical value plays no role within the statistical precision that can
be reached on the thermodynamic observables.} and at 
larger
light quark 
masses~\cite{Petreczky:2004pz,Kaczmarek:2005gi}
is shown in Fig.~\ref{fig:Sq}. Again, one can observe that the
temperature dependence of the static entropy in the real-world QCD is smoother
in the transition region than in the theories with 
larger quark masses.
The location of the maximum of $S_q$ can be interpreted
as the deconfinement transition temperature.
The numerical result of Ref.~\cite{Bazavov:2016uvm} is
$T_S=153^{+6.5}_{-5}\ {\rm MeV}$ which matches within the uncertainties
the latest determination of the chiral crossover temperature 
$T_c=156.5\pm1.5$~MeV~\cite{Bazavov:2018mes,Borsanyi:2020fev}.
One has to bear in mind that while the latter is related to the critical behavior of QCD
in the two-flavor chiral limit (masses of the up- and down-quark set to 0), the former
comes from the observables that are not related to any singular behavior in the
chiral limit. It may be that this agreement is incidental for the real-world QCD\footnote{
It is worth pointing out that ratios of properly renormalized Polyakov loop susceptibilities as suggested in \mbox{Ref.}~\cite{Lo:2013etb}, \ie 
\begin{equation}
\label{eq:R_T}
R_T = \frac{\langle (\mr{Im} L)^2 \rangle}{\langle |L|^2 \rangle-\langle L \rangle^2}
\quad\text{due to}\quad \langle \mr{Im} L \rangle = 0,
\end{equation}
show smooth crossover behavior in full QCD~\cite{Bazavov:2016uvm}. 
Due to the mixing between different representations, these ratios can only 
be calculated with controlled uncertainties in the gradient flow renormalization at large enough flow time. 
}.
In the chiral limit the chiral crossover turns into an actual phase transition whose
temperature $T^0_c=132^{+3}_{-6}$~MeV~\cite{Ding:2019prx}
is about 20~MeV below the chiral crossover
temperature at the physical light quark masses.
One may wonder whether or not the transition temperature defined from the static 
quark entropy $T_S$ would decrease further when going towards the chiral limit, 
since, on the one hand, it has little obvious sensitivity to the chiral critical 
behavior, but, on the other hand, has been found to follow the chiral crossover 
temperature $T_\chi$ closely in terms of the quark mass dependence and the 
discretization effects. 
This open question however requires further theoretical investigation.

\begin{figure*}\center
	\includegraphics[width=8.6cm]{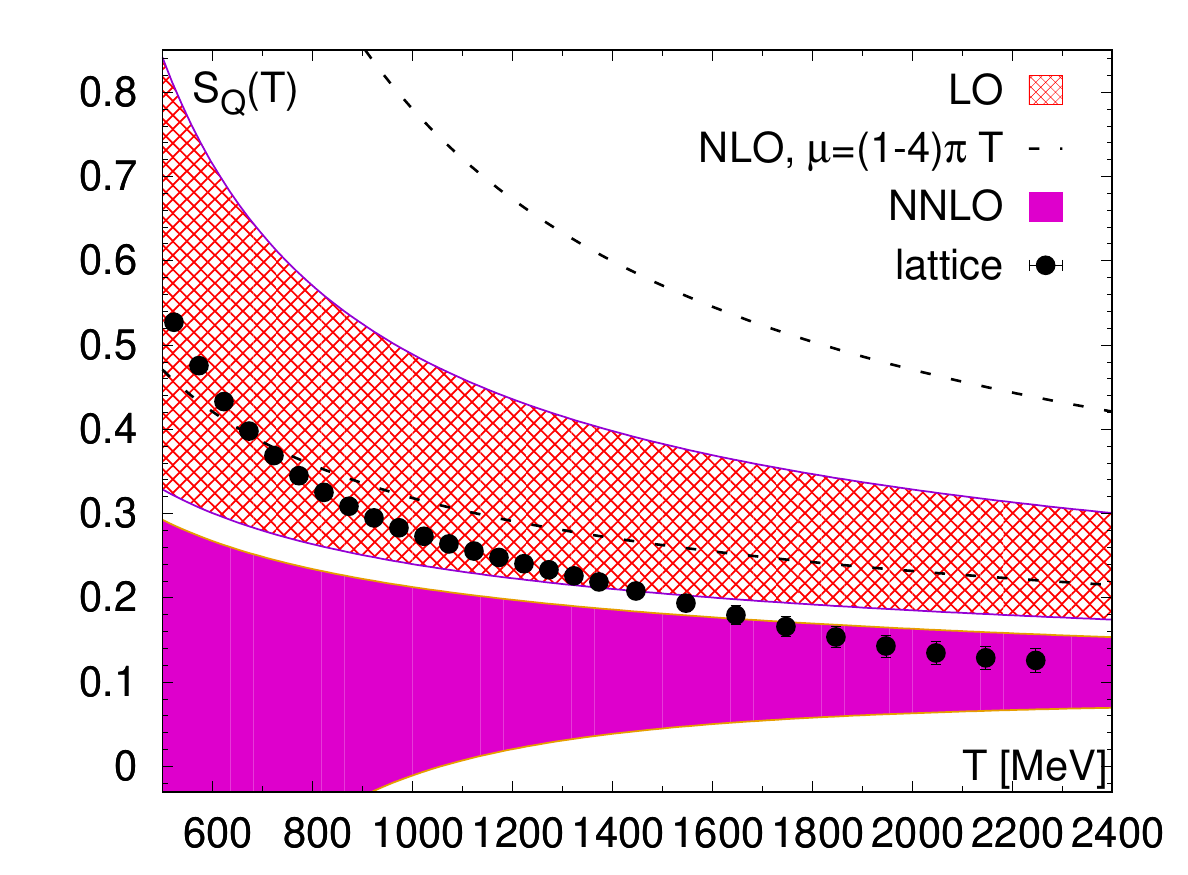}
	\caption{
		The static quark entropy $S_q$ calculated on the lattice~\cite{Bazavov:2018wmo}
		and in a weak-coupling expansion~\cite{Berwein:2015ayt} up
		to the next-to-next-to-leading order.
		\label{fig:Sqhigh}
	}
\end{figure*}

While the physics of the strongly interacting medium is nonperturbative in the
vicinity of the transition, one expects that at high temperatures weak-coupling
expansions may be reliable. The static quark free energy was recently
calculated  to next-to-next-to leading order (NNLO) in
Ref.~\cite{Berwein:2015ayt}. Direct comparison of free energies between
the lattice and weak-coupling calculations is complicated because the
calculations are performed in different renormalization schemes.
They can be related by matching the temperature independent shift,
e.g. $C(g_0^2)$ arising from renormalizing the static 
quark-antiquark energy, 
Eq.~(\ref{eq:Eren}). However,
the most straightforward way to perform the comparison is
to consider the entropy where this shift is eliminated by the derivative with
respect to temperature, Eq.~(\ref{eq:Sq}). A challenge for lattice QCD calculations
is however to reach very high temperatures, since this requires very fine lattices
which are computationally expensive. 
Ref.~\cite{Bazavov:2018wmo} calculated $S_q$ in the continuum limit up to 
temperature $T\sim 2.2$~GeV, shown in Fig.~\ref{fig:Sqhigh}.
The bands shown in the figure correspond to scale variation from
$\mu=\pi T$ to $4\pi T$. At the highest temperature the lattice results 
and the NNLO results agree within the uncertainties.
At temperatures below 1.5~GeV the lattice results are closer to the LO
weak-coupling results. 
The rather poor convergence at the level of the NLO calculation 
and its apparent inconsistency with the lattice result can be understood from 
the observation that this is an expansion in $g$ as required for the static 
Matsubara modes, which still misses the leading correction from an expansion 
in $\als=g^2/(4\pi)$ for the non-static Matsubara modes and the fermions.

\subsection{Polyakov loop correlators}
\label{sec:plcor}

It was noted quite early in the considerations concerning thermal field 
theories that a single, isolated probe charge cannot be considered as a 
physically meaningful concept in a confining theory such as the non-Abelian 
$ \mr{SU}(N_c) $ pure gauge theory~\cite{Polyakov:1978vu}. 
In other words the physical meaning of the Polyakov loop is not evident in 
the confined phase of Yang-Mills theory. 
In fact, only after combining the charges into even representations of 
the gauge group one may obtain states that are consistent with confinement. 
If the probe charges transform in the (anti-)fundamental representation like 
the static (anti-)quarks, then confinement demands that the most simple states 
are obtained by combining fields in the fundamental and anti-fundamental 
representations as 
$ \mb{N_c} \times \mb{\overline{N_c}} = \mb{1} + \mb{(N_c^2-1)} $ into states 
in either the trivial or the adjoint representations. 
Given the importance of $ \mr{SU}(3) $ as the gauge group of QCD, it is 
customary to speak of the \emph{color singlet} and \emph{color octet} 
independent of the actual $ N_c $. 
Nevertheless, in the case of full QCD, a single, isolated static quark can be 
complemented in the confined phase by drawing an antiquark out of the vacuum 
and combining the two charges into an even representation. 

\begin{figure*}\center
\includegraphics[width=8.6cm]{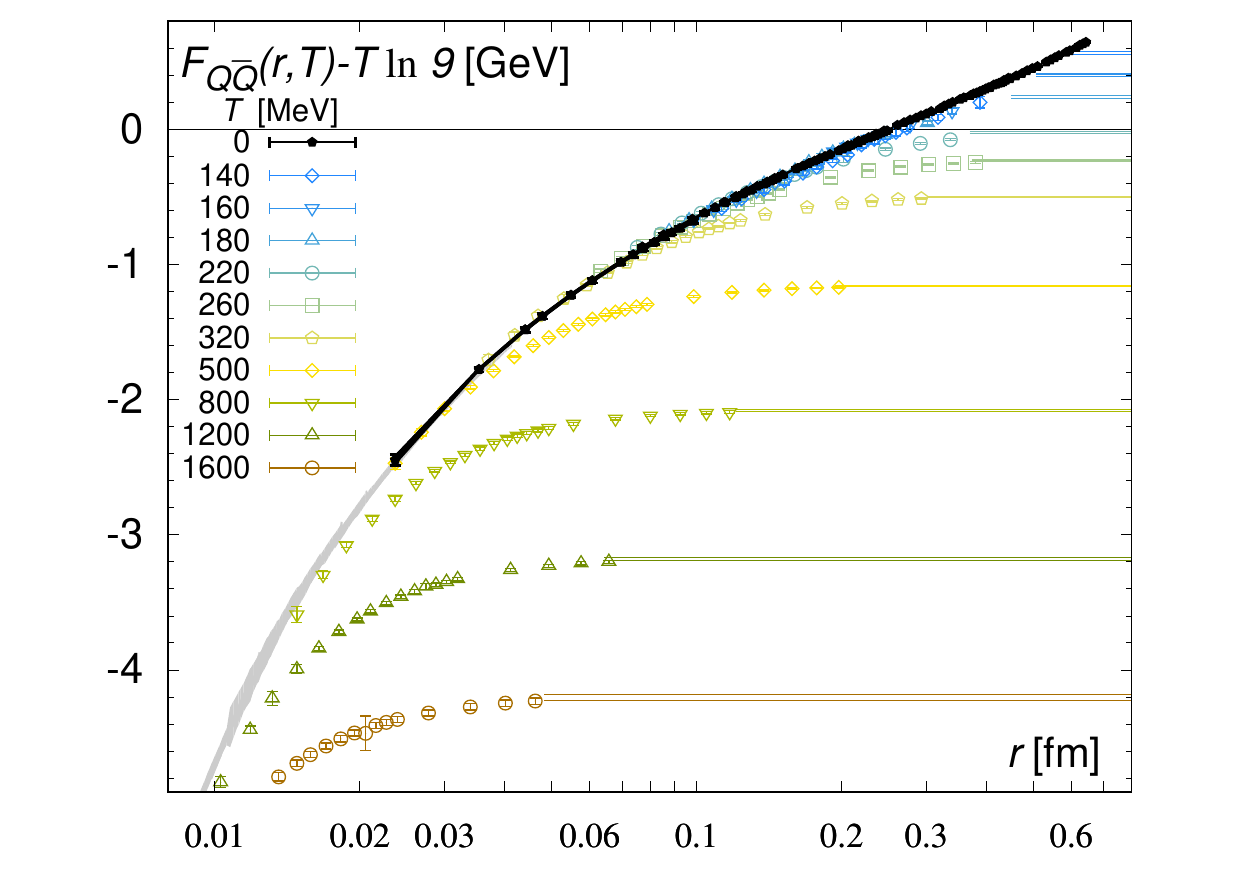}
\caption{
The continuum limit of the free energy $F_{q\bar q}$. 
The black band and symbols show the $T=0$ QCD static energy \(E\) and the 
light gray band shows the singlet free energy \(F_S\) at high temperatures 
and very short distances, where finite temperature effects are smaller than the 
statistical errors. 
The subtraction of \(T\ln 9\) is required for matching to the static energy 
at short distances due to the normalization convention used for \(F_{q\bar q}\). 
The horizontal bands outline the $r \rightarrow \infty$ limit of 
$F_{q\bar q}$, \ie, $2 F_q-T\ln 9$.
\label{fig:Fqq}
}
\end{figure*}

Hence, correlation functions of paired static quarks and antiquarks were 
considered already in the earliest numerical studies~\cite{Kuti:1980gh, 
McLerran:1981pb}, \mbox{i.e.} in the simplest case the correlation function 
of a single static quark-antiquark pair in $ \mr{SU}(2) $ pure gauge theory. 
Representing each static quark by a Polyakov loop as in \Eqref{eq:ploop}, one 
arrives at the Polyakov loop correlator, which is related to the difference 
in free energy between a system with and without the static quark-antiquark pair,  
\al{\label{eq:CL}
 C_L^\mr{bare}(T,r) 
 &= \frac{1}{N_c^2} \Braket{~\sum\limits_{\bm{x}}~L(\bm{x}) L\adj(\bm{x}+\bm{r})~}
  = \exp{\left\{-\frac{F_{q\bar q}^\mr{bare}(T,r)}{T}\right\}}. 
}
For infinite separation $ C_L $ approaches the limit of $ \braket{ L }^2 $, \ie the 
static quark and antiquark decouple from each other, either due to the string breaking 
in the vacuum or due to the color screening in the high temperature phase. 
For this reason it is clear that the multiplicative renormalization factor of $ C_L^\mr{bare} $ 
is just the square of the renormalization factor of $ \braket{ L^\mr{bare} } $. 
Hence, the ratio $ C_L/\braket{L}^2 $ does not require renormalization and defines a 
subtracted free energy $ F_{q\bar q}^{\rm sub}(T,r) = F_{q\bar q}(T,r) -2F_q(T) $ that 
vanishes for infinite separation. 
Results for the free energy $ F_{q\bar q} \equiv F_{q\bar q}^\mr{ren} $ with 
the renormalization scheme of \Eqref{eq:Lren} from a recent lattice 
calculation are shown in \Figref{fig:Fqq}.
\vskip1ex
Before embarking on the further discussion of the properties of $ C_L $ 
in a non-Abelian gauge theory let us begin again with a brief look at the 
Abelian case, namely Quantum Electrodynamics. 
We consider the thermal correlation function of an electron-positron pair, 
which can only combine to a state in the trivial representation. 
Hence, the associated free energy and potential are both Coulomb-like. 
For infinite separation only two isolated charges are left due to the cluster 
decomposition, thus giving physical meaning to the notion of a single, 
isolated electron. 
Although the electric field between the two probe charges is screened inside 
of an electromagnetic plasma, there are unscreened contributions from the 
magnetic fields that contribute at higher loop orders and eventually lead to 
the takeover by a power-law falloff instead~\cite{Arnold:1995bh}.
\vskip1ex
\subsubsection{Singlet and octet free energies}

Now let us juxtapose this to the case of the non-Abelian pure gauge theory. 
The $ \mr{SU}(N_c) $ Fierz identity 
permits rewriting \Eqref{eq:CL} in terms of the sum of the 
color singlet and octet contributions 
\al{\label{eq:fso}
 C_L(T,r) 
 &= \exp{\left\{-\frac{F_{q\bar q}(T,r)}{T}\right\}}\nonumber\\
 &= \frac{1}{N_c^2} \exp{\left\{-\frac{F_{S}(T,r)}{T}\right\}}
  + \frac{N_c^2-1}{N_c^2} \exp{\left\{-\frac{F_{O}(T,r)}{T}\right\}}. 
}
For infinite separation, both $ F_S $ and $ F_O $ share the same limit 
$ 2 F_q $, since the two color charges are decoupled entirely. 
For this reason it is convenient to define subtracted singlet and octet 
free energies  $ F_{S,O}^\mr{sub} \equiv F_{S,O} -2F_q$ just as in the 
case for the Polyakov loop correlator. 

\begin{figure*}\center
\includegraphics[width=0.5\textwidth]{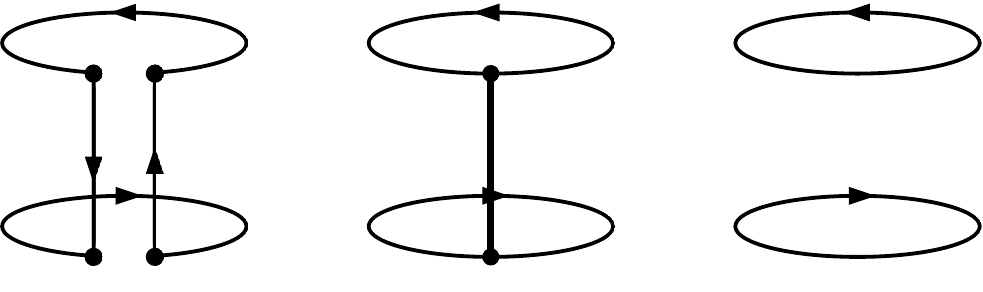}
\caption{
The thermal expectation values of cyclic Wilson loops (center) are obtained 
from the thermal expectation values of non-cyclic Wilson loops (left) by 
collapsing the two spatial Wilson lines to the same time Euclidean time 
across the periodic temporal boundary.  
Cyclic Wilson loops mix with the Polyakov loop correlator (right). 
From \mbox{Ref.}~\cite{Berwein:2012mw}).
\label{fig:CWL}
}
\end{figure*}

A direct operator representation for the singlet or octet correlation 
functions or free energies is less straightforward than for the Polyakov 
loop correlator. 
The former two are mixed under renormalization, and only \Eqref{eq:fso} or 
the difference 
\al{\label{eq:Wc-Cl}
 \exp{\left\{-F_{S}/T\right\}} - \exp{\left\{-F_{O}/T\right\}}
 = \frac{N_c^2}{N_c^2-1} \left( \exp{\left\{-F_{S}/T\right\}} - C_L \right)
 } 
are renormalized multiplicatively~\cite{Berwein:2017thy}. 
The difference in \Eqref{eq:Wc-Cl} is multiplicatively renormalizable at 
short distances, where the weak-coupling expansion is reliable.  
\vskip1ex
The singlet or octet free energies require two color charges fixed 
with respect to each other at a finite distance between them. 
Any thermal correlation function 
$ C_{S,O} \equiv \exp{\left\{-F_{S,O}/T\right\}} $ 
involving the hermitian conjugate for one of its two thermal Wilson 
lines defines a scheme for one among the free energies $ F_{S,O} $. 
Its counterpart in the same scheme is automatically given by \Eqref{eq:Wc-Cl}. 
We note that any choice of the scheme only reshuffles the individual 
contributions between the color singlet and octet configurations, but 
cannot modify the energies and hierarchies for any of these states. 
\vskip1ex
Such a scheme may be realized in a manifestly gauge-invariant manner in the 
form of a single closed contour --- a thermal or cyclic Wilson loop $ W_C $, 
see \Figref{fig:CWL} --- which introduces multiple types of UV divergences 
that are not contained in the Polyakov loop. 
These divergences have to be suppressed through suitable link-smoothing 
techniques in lattice simulations (see \mbox{Sec.}~\ref{sec:lattice} for 
details), \ie the spatial Wilson lines $ \widetilde{W} $ 
making up the cyclic Wilson loops
\al{\label{eq:cycWL}
 W_C^\mr{bare}(T,r) 
 &= \frac{1}{N_c} \Bigg\langle~\sum\limits_{\bm{x}}~\tr \Big[
 {W}(aN_\tau, \bm{x}; 0, \bm{x}) 
 \widetilde{W}(aN_\tau, \bm{x}+\bm{r}; aN_\tau, \bm{x}) 
 \nn\\&\phantom{=~\frac{1}{N_c} \Bigg\langle~\sum\limits_{\bm{x}}~\tr \Big[}
 {W}(0, \bm{x}+\bm{r}; aN_\tau, \bm{x}+\bm{r}) 
 \widetilde{W}(0, \bm{x}; 0, \bm{x}+\bm{r})  ~ \Big]\Bigg\rangle
} 
have to be drawn from a modified thermal distribution. 

\begin{figure*}\center
\includegraphics[width=8.6cm]{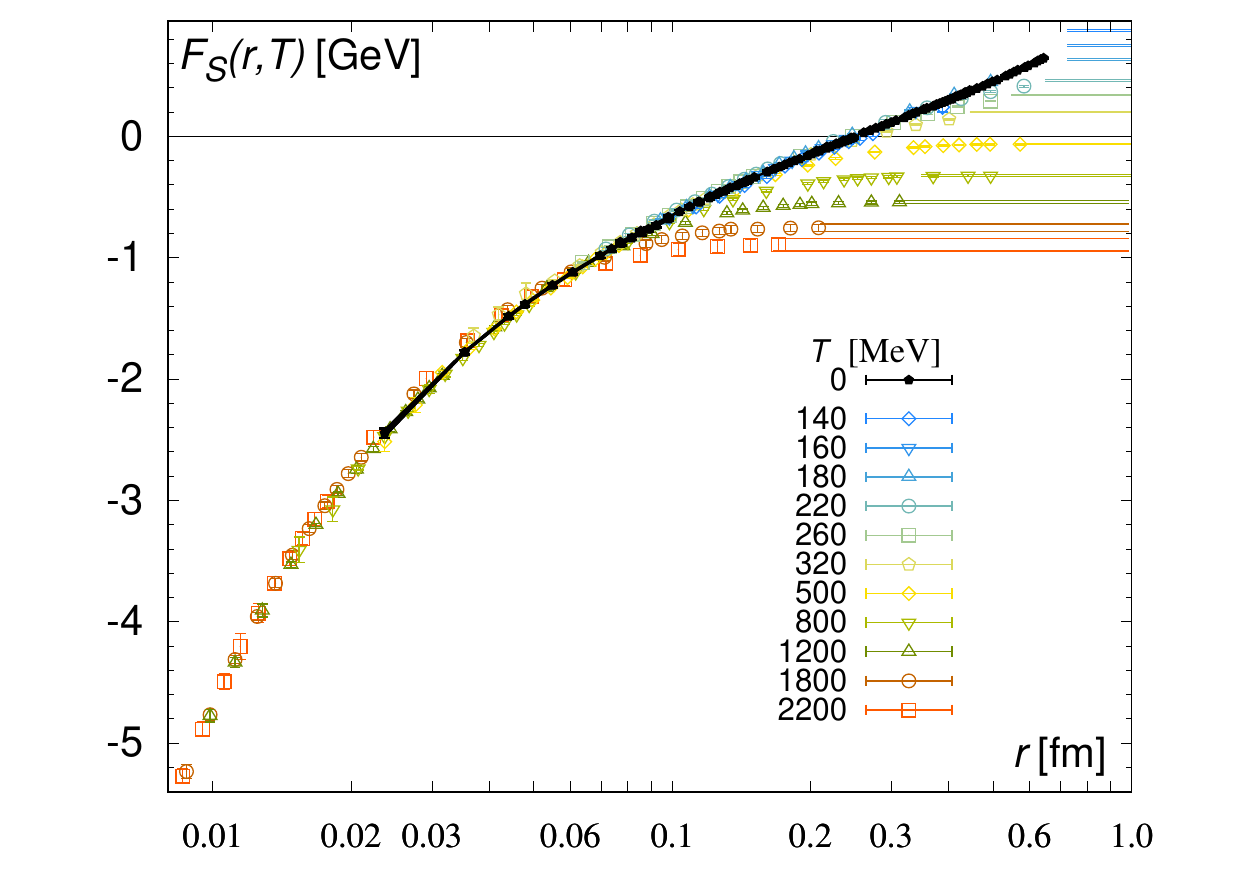}
\caption{
The continuum limit of the singlet free energy $F_S$ defined with Wilson line 
correlators in Coulomb gauge. 
The black band and symbols show the $T=0$ QCD static energy \(E\). 
The horizontal bands outline the $r \rightarrow \infty$ limit of 
$F_S$, \ie, $2 F_q$.
\label{fig:FS}
}
\end{figure*}

\begin{figure*}\center
\includegraphics[width=8.6cm]{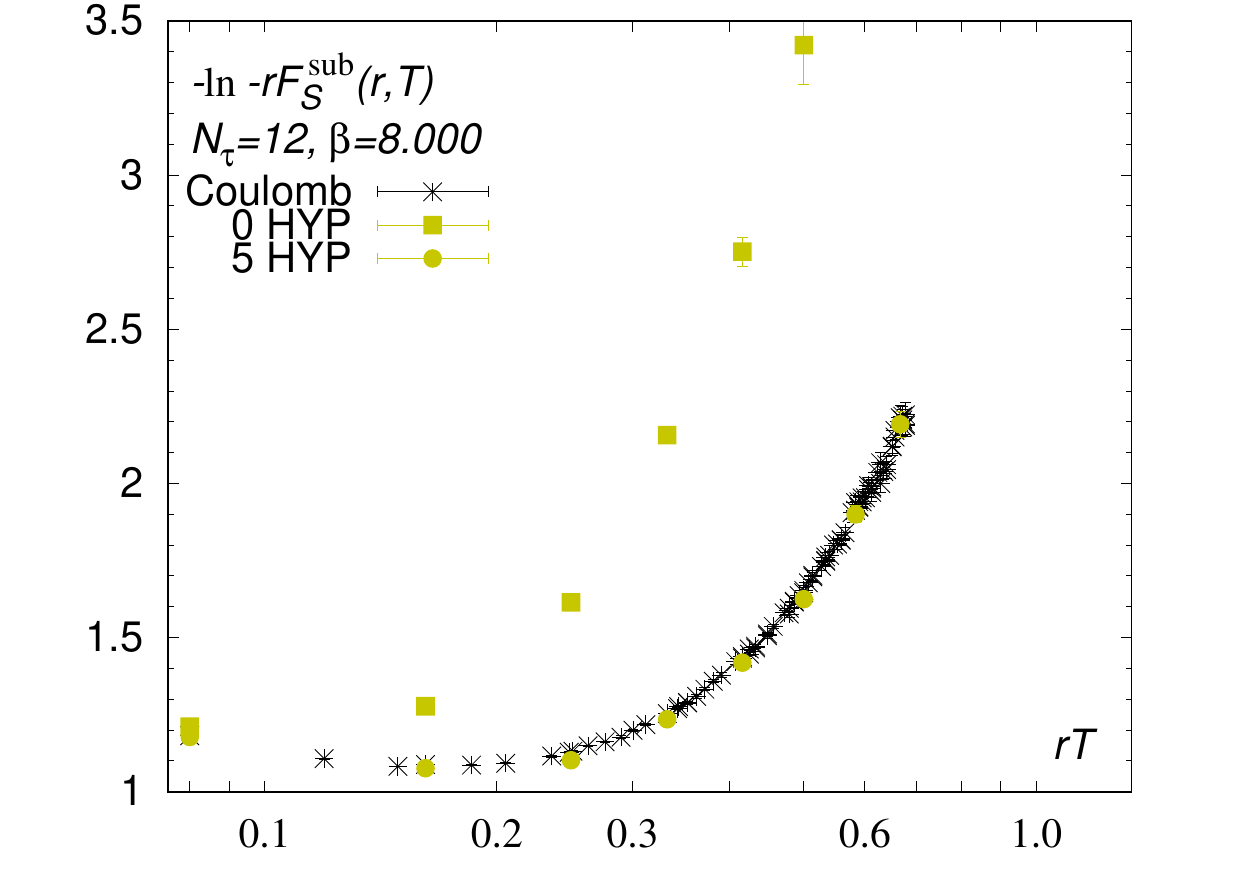}
\caption{
The logarithm of the singlet screening function $ -rF_S^\mr{sub} $ defined 
in terms of the Coulomb gauge Wilson line correlator or the gauge-invariant 
cyclic Wilson loop with appropriate amount of spatial hypercubic (HYP) 
link smearing show the quantitatively very similar screening 
behavior~\cite{Bazavov:2018wmo}. 
At shorter distances small differences can be seen. 
\label{fig:FW}
}
\end{figure*}

Alternatively, these singlet and octet free energies may be realized in terms of thermal
Wilson line correlation functions that are evaluated in a suitably fixed gauge 
such that the spatial Wilson lines can be omitted altogether, 
\al{\label{eq:CS}
 C_S^\mr{bare}(T,r) 
 &= \frac{1}{N_c} \Big\langle~\sum\limits_{\bm{x}}~\tr \Bigg[
 {W}(aN_\tau, \bm{x}; 0, \bm{x}) 
 {W}(0, \bm{x}+\bm{r}; aN_\tau, \bm{x}+\bm{r}) ~ \Big]\Bigg\rangle^\mr{gf}
 \nonumber\\
 &= \exp{\{-F_S^\mr{bare}/T\}}.
} 
We have to stress the key difference between the thermal Wilson line correlation function and the Polyakov loop correlator. 
Namely, the thermal Wilson lines are traced individually in \Eqref{eq:CL}, 
while the trace of the product of two separated Wilson lines is taken in \Eqref{eq:CS}, which requires the fixing of a gauge.
Coulomb gauge is particularly useful, since the free energies are finite 
in Coulomb gauge and the renormalization is particularly 
simple~\cite{Berwein:2017thy}, see \Figref{fig:FS}. 
While both alternative definitions of the singlet free energy exhibit 
different UV behavior~\cite{Berwein:2017thy}, they are found to be 
consistent at intermediate and large distances within statistical 
errors in direct (2+1)-flavor QCD lattice 
simulations~\cite{Bazavov:2016qod, Bazavov:2018wmo} using staggered 
(HISQ) quarks, see \Figref{fig:FW}.  

Lastly, using the weak-coupling approach and pNRQCD it is possible to define 
gauge-invariant thermal singlet and octet correlators. 
Their expectation values are the exponentiated pNRQCD singlet or octet free 
energies $ f_s $ and $ f_o $~\cite{Brambilla:2010xn}, whose linear combination 
formally recombines to $ C_L $ as in \Eqref{eq:fso}. 
Key relations between $ f_s $ and $ f_o $ that have been tested are found to 
be reproduced quite well by $ F_S $ and $ F_O $ defined in terms of the 
Coulomb gauge Wilson line correlation function on the lattice~\cite{Bazavov:2018wmo}. 
\vskip1ex

\subsubsection{Vacuum-like regime and vacuum physics} 

\begin{figure*}\center
\includegraphics[width=8.6cm]{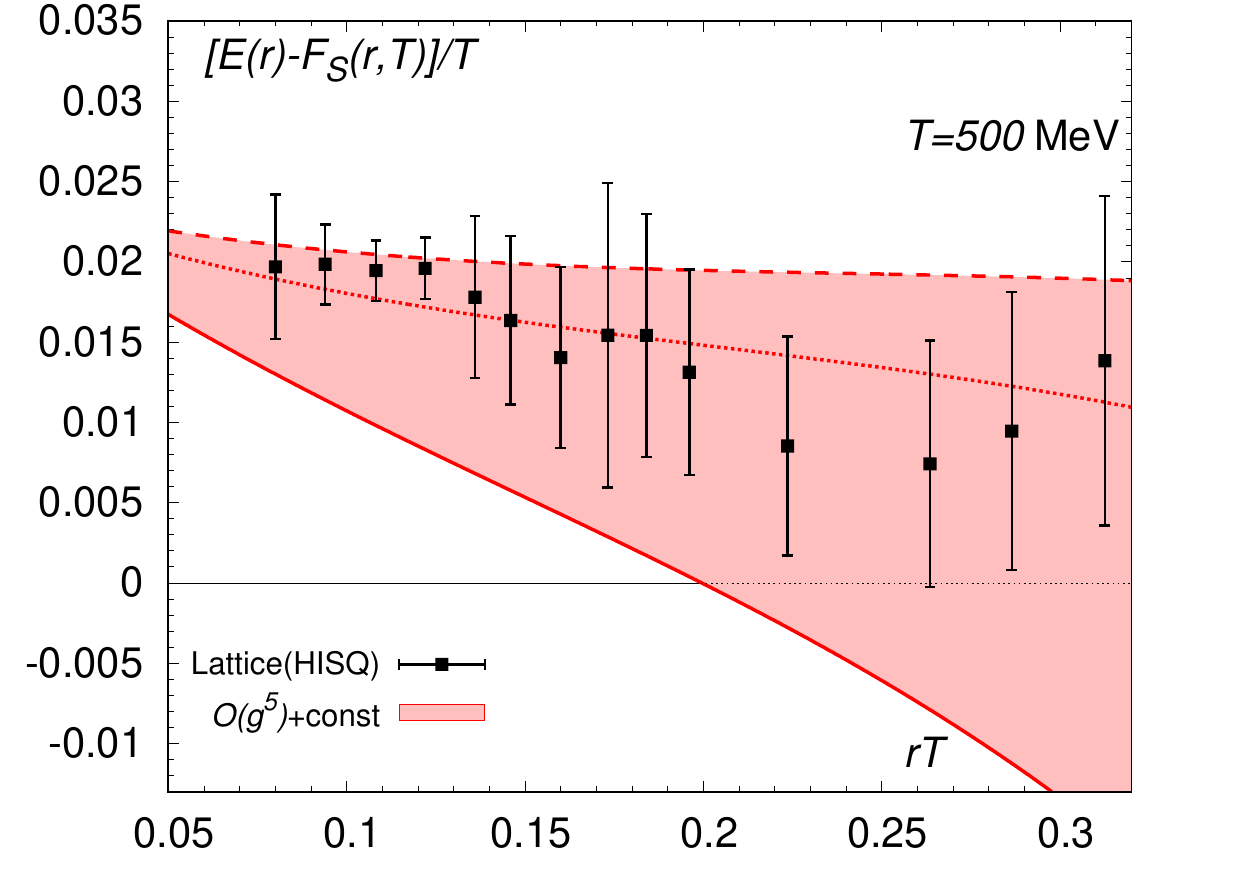}
\caption{
Lattice and weak-coupling results for $ E-F_S $~\cite{Bazavov:2018wmo}. 
The weak-coupling results are shifted by a small constant to match 
them to the lattice results at the shortest distance to account for an 
matching constant of order $ g^6 $. 
The dotted line corresponds to the renormalization scale $ \mu=2\pi T $, 
while the band corresponds to its variation from $ \pi T $ (solid) to 
$ 4\pi T $ (dashed).
\label{fig:E-Fs}
}
\end{figure*}

The singlet or octet free energies $ F_{S} $ or $ F_O $ are given to 
leading order $ g^2 $ by the corresponding zero temperature Coulomb-like 
potentials $ V_{s,o} = c_{s,o}~\als/r  $ with $ c_s = -C_F $ 
or $ c_{o} = +C_F/(N_c^2-1) $. 
At very short distances $ r \ll \als/T $, the running of the coupling 
$ \als(1/r) $ is controlled by the inverse distance as in the vacuum. 
The two Coulomb-like potentials with opposite signs become large, and thus the 
repulsive octet contribution can be neglected. 
In other words, at such short distances, the free energy behaves in the non-Abelian 
case up to a temperature dependent shift $ +T\ln(N_c^2) $ 
due to the color factors  and the different running coupling 
(\ie different beta function) just as in the Abelian case, 
namely $ F_{q\bar q} = F_S +T\ln(N_c^2) = V_s +T\ln(N_c^2) $ at leading order. 
We recall that the singlet potential $V_s$ and the static energy $E$ coincide at leading order.
The former relation (between $F_{q\bar q}$ and $F_S$ or $E$) has been observed for nonperturbative lattice calculations 
in (2+1)-flavor QCD using staggered (HISQ) quarks at shorter and shorter 
distances up to temperatures $ T \lesssim 0.5\,\mr{GeV} $~\cite{Bazavov:2018wmo}, 
while the latter relation (between $F_S$ and $E$) even 
holds at larger distances $ r \lesssim 0.3/T $ for much higher temperatures 
$ T \lesssim 2\,\mr{GeV} $~\cite{Bazavov:2019qoo}. 
The corresponding thermal corrections are small due to a partial cancellation 
between the contributions from non-static gluons or sea quarks, and from the 
static gluons, see \Figref{fig:E-Fs}. 
In particular, as the contributions from the former are restricted to even 
powers of the gauge coupling $g$ starting at $g^4$, while the latter are restricted 
to odd powers of the gauge coupling $g$ starting at $g^5$, it is clear that this 
partial cancellation is effective only in a limited temperature window which 
happens to coincide with phenomenologically interesting temperatures.

\begin{figure*}\center
\includegraphics[width=8.6cm]{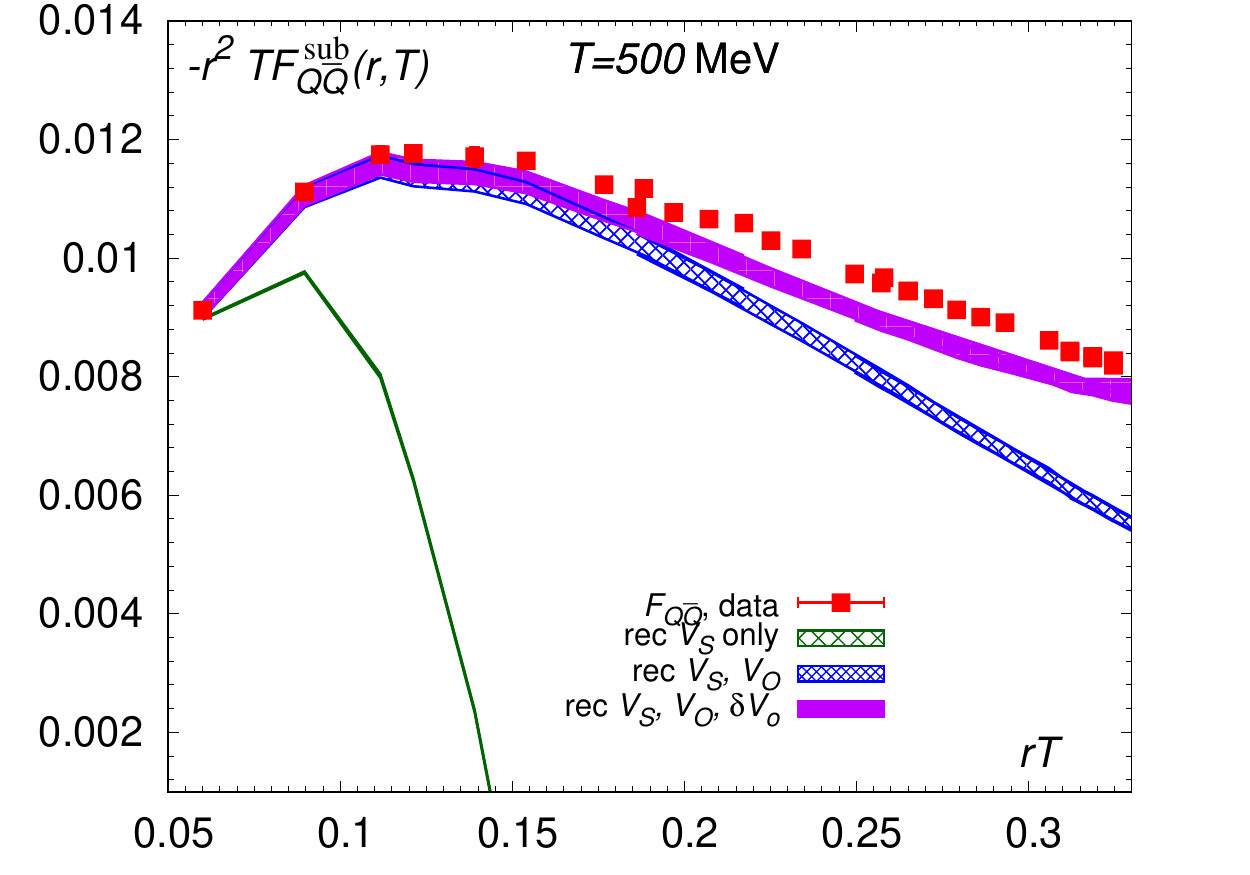}
\caption{
The subtracted free energy $ F_{q\bar q}^\mr{sub} $ multiplied by $-r^2 T$ 
calculated on an $ N_{\tau}=16 $ lattice (squares, using 1 step of 
4D hypercubic (HYP) smearing~\cite{Hasenfratz:2001hp} for 
$r \ge \sqrt{6}/a \approx 0.15/T$) and compared to the 
reconstruction based on pNRQCD (bands) at $ T=500\,\mr{MeV} $
(the fully reconstructed result is in magenta, while the blue band 
ignores the Casimir scaling violating contributions to the octet 
potential and the green one ignores the octet 
contribution).
\label{fig:Fqqrec}
}
\end{figure*}

\vskip1ex
In particular, the Polyakov loop correlator at distances up to $ r \lesssim 0.3/T $ 
is determined to good accuracy in terms of the singlet and octet zero temperature 
potentials and the adjoint Polyakov loop as predicted in 
pNRQCD~\cite{Brambilla:2010xn}, see \Figref{fig:Fqqrec}. 
\vskip1ex

For all larger distances we have to distinguish between the vacuum and the 
thermal medium at high temperatures. 
Let us first take a closer look at $ C_L $ in the vacuum, \ie in the confining 
phase.  
At larger distances $ r \sim 1/\lMSb $, the Coulomb-like interaction is not 
dominant anymore. 
Instead, the singlet and octet free energies are dictated to a large extent 
by the energies of the respective lowest states, \ie the QCD string or the 
excited QCD string, which both behave as $ \sim \sigma r \sim \lMSb $. 
As a consequence, the free energy behaves in pure $ \mr{SU}(N_c) $ gauge theory 
up to details of the normalization similar to the quark-antiquark static energy. 
This can be clearly resolved only at the shortest distances that are 
accessible in lattice gauge theory, see \Figref{fig:Fqq} for the case 
of full (2+1)-flavor QCD. 
In full QCD, the thermal Wilson lines can decouple via string breaking, since 
each thermal Wilson line -- being a closed contour -- can combine with a sea 
(anti-)quark into the trivial representation of the gauge group. 
For this reason, screening due to the string-breaking mechanism in the vacuum 
modifies the (singlet) free energy already in the vacuum phase of QCD. 
However, due to the severe signal-to-noise problem, the free energies cannot 
be studied in the string-breaking regime at low temperatures without extensive 
application of noise-reduction techniques, see 
\mbox{Ref.}~\cite{Steinbeisser:2018sde} for preliminary work in this direction. 
\vskip1ex
\subsubsection{Thermal dissociation and the free energy}

\begin{figure*}\center
\includegraphics[width=8.6cm]{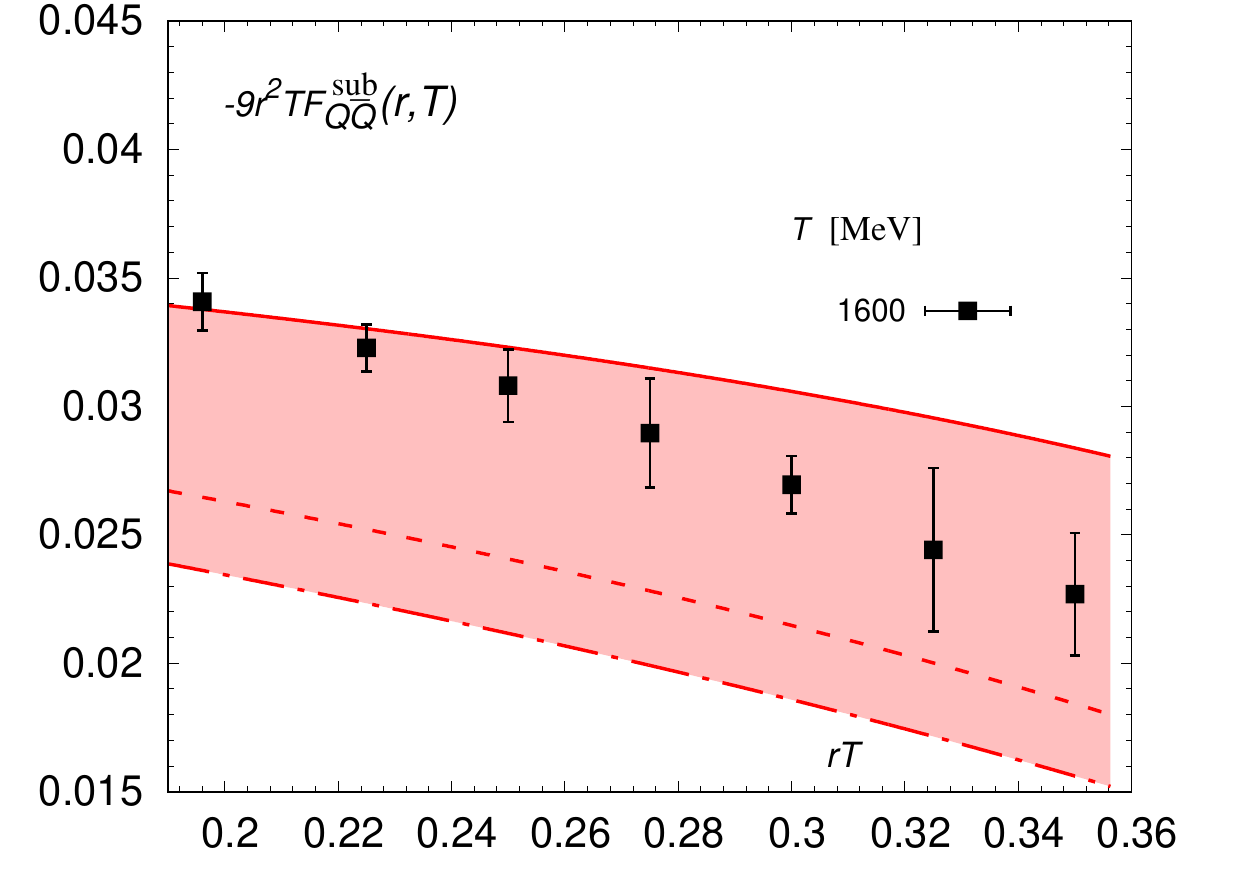}
\caption{
The subtracted free energy $ F_{q\bar q}^\mr{sub} $ multiplied by $-9r^2 T$ 
evaluated for $ T=1600\,\mr{MeV}$ and compared with the $ \mr{N^3}LO $ 
weak-coupling expression (lines)~\cite{Bazavov:2018wmo}.
The upper line corresponds to $\mu=\pi T$, the middle one to $\mu=2 \pi T$, 
and the lower line corresponds to $\mu=4 \pi T$.
\label{fig:Fqqsweak}
}
\end{figure*}

In the high temperature phase without confinement the free energy behaves rather 
differently. 
Let us consider distances $ r \ll 1/T $, for which the running of the coupling 
$\als(T) $ is controlled by the temperature $ T $, which is the lowest scale. 
If --- on top of that --- the hierarchy $ \als/T \ll r \ll 1/T $ is satisfied, then thermal 
gluons are sufficiently energetic to overcome the spectral gap between the singlet 
and octet configurations and the singlet states begin to dissociate and recombine 
again. 
Whereas the dissociation and recombination are dynamical processes that cannot 
be resolved directly in an imaginary time approach, the thermally equilibrated 
distribution that is their consequence is resolved by the Polyakov loop correlator! 
Due to $ \als/(rT) \ll 1 $, the exponential functions in \Eqref{eq:fso} can be 
expanded in the gauge coupling, and the color factors lead to a cancellation 
of the leading Coulomb-like terms between the singlet and octet contributions. 
In this hierarchy the subtracted free energy is given at leading order as 
\al{
 F_{q\bar q}^{\rm sub}(T,r) = - \frac{N_c^2-1}{8N_c^2} \left(\frac{\als}{r}\right)^2, 
}
with a non-Coulombic $ (\als/r)^2$ behavior. 
This behavior clearly indicates that the leading interaction between the two 
Polyakov loops at such distances is the emission of two 
\emph{electric $ A_0 $ gluons}, which have to be in a color singlet 
configuration. 
The Polyakov loop correlator has been calculated in the weak-coupling approach up 
to order $ g^7 $ in the small $ r $ expansion~\cite{Berwein:2017thy} and found to 
be compatible with the nonperturbative lattice calculation for (2+1)-flavor QCD 
using staggered (HISQ) quarks for $ 0.2/T \lesssim r \lesssim 0.3/T $ at 
temperatures $ T \gtrsim 1.5\,\mr{GeV} $~\cite{Bazavov:2018wmo}, see \Figref{fig:Fqqsweak}. 
\vskip1ex

\subsubsection{Chromoelectric screening}

At larger distances the quark-antiquark system enters the scale hierarchy 
$ r \sim 1/\md $, which is associated with the electric screening that 
is controlled by the mass parameter of the adjoint scalar field 
associated with \emph{electric $ A_0 $ gluons} in EQCD, which is often 
called the (perturbative) Debye mass.  
It is given to leading or next-to-leading order~\cite{Braaten:1995jr} as 
\al{\label{eq:mdlo}
 \md^2|_\mr{LO}\phantom{_\mr{N}}(\nu) 
 &= \frac{2N_c+N_f}{6}~g^2(\nu)T^2,\\
 \md^2|_\mr{NLO}(\nu) 
 &= \md^2|_\mr{LO}(\nu) 
  \nn\\&\phantom{=~}\times
 \frac{\als(\nu)}{4\pi}\Big[
 2 \beta_0 \left(\gamma_E+\ln\frac{\nu}{4 \pi T} \right)
+ \frac{5N_c}{3} 
+ \frac{2N_f}{3} (1-4\ln 2)
 \Big]\Big) \nn\\
 &\phantom{=~}-C_F N_f \als^2(\nu) T^2 
 \label{eq:mdnlo},
} 
where typical values of the scale $ \nu $ are of the order of $ 2\pi T $, 
\ie of the order of the lowest nonstatic Matsubara mode. 
The next-to-next-to-leading order contribution to the Debye mass has been 
calculated so far only in pure Yang-Mills theory~\cite{Ghisoiu:2015uza}. 
The $ \mr{N^2LO} $ Debye mass lies between the central values of $ \mr{NLO} $ 
and $ \mr{LO} $, but has a much smaller scale dependence. 
Since the effect of the $ N_f $ quarks on the Debye mass is rather mild 
$ \sim 20\% $ to $ 30\% $, similar results are to be expected for the full 
QCD calculation.  
The regime of electric screening is mixed up with the previous regime 
of thermal dissociation, since the hierarchies are not well separated 
(separate regimes would require $ \als(\nu)\md(\nu) \gtrsim T $, \ie $g(\nu) \gtrsim 2.5$). 
To leading order the free energy and singlet free energy are given in the 
regime of electric screening by 
\al{\label{eq:Fqqlo}
 F_{q\bar q}^\mr{sub}|_\mr{LO} (T,r,\nu) 
 &= -\frac{N_c^2-1}{8N_c^2} \left(\frac{\als(\nu)e^{-\md(\nu)r}}{r}\right)^2, \\ 
 \label{eq:Fslo}
 F_S^\mr{sub}|_\mr{LO} (T,r,\nu) 
 &= -C_F \frac{\als(\nu)e^{-\md(\nu)r}}{r}.   
}

\begin{figure*}\center
\includegraphics[width=6.6cm]{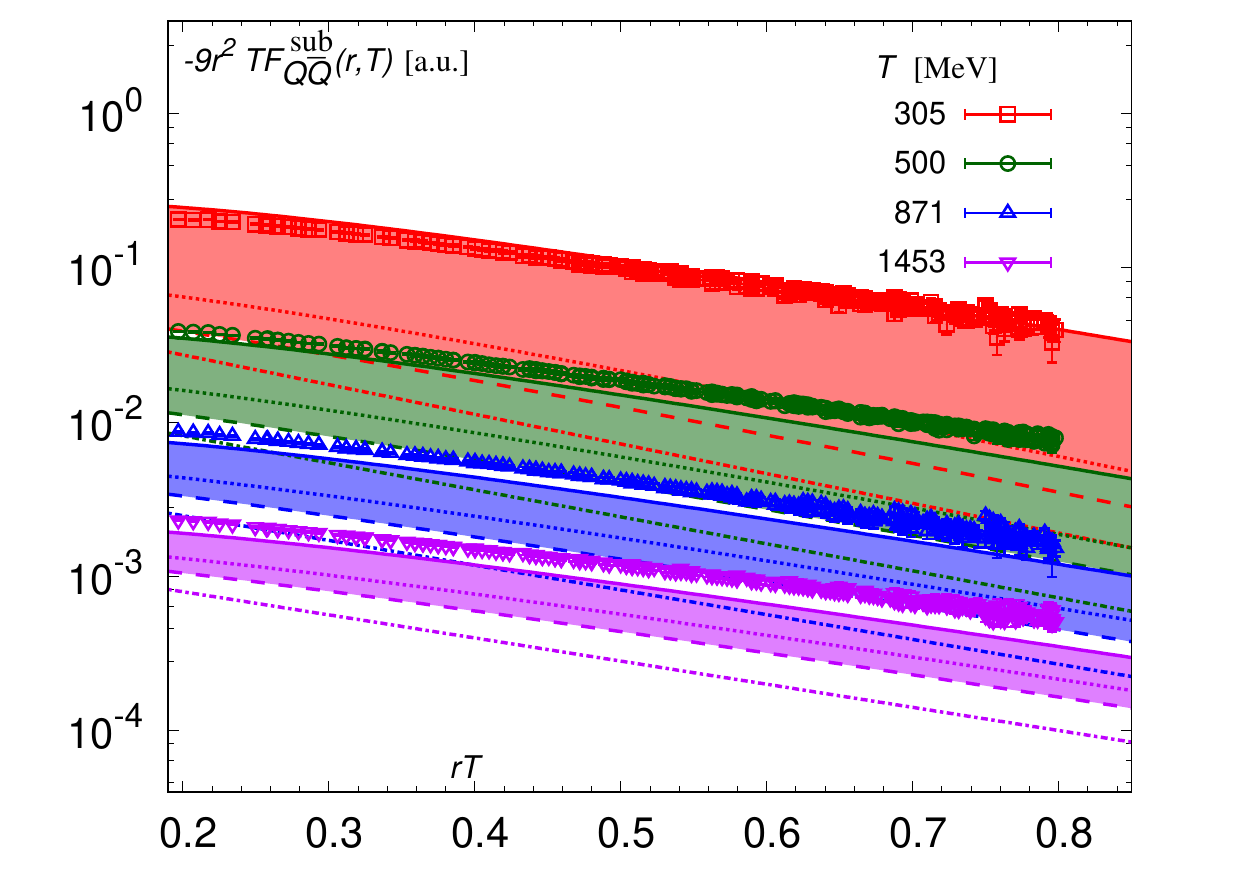}
\includegraphics[width=6.6cm]{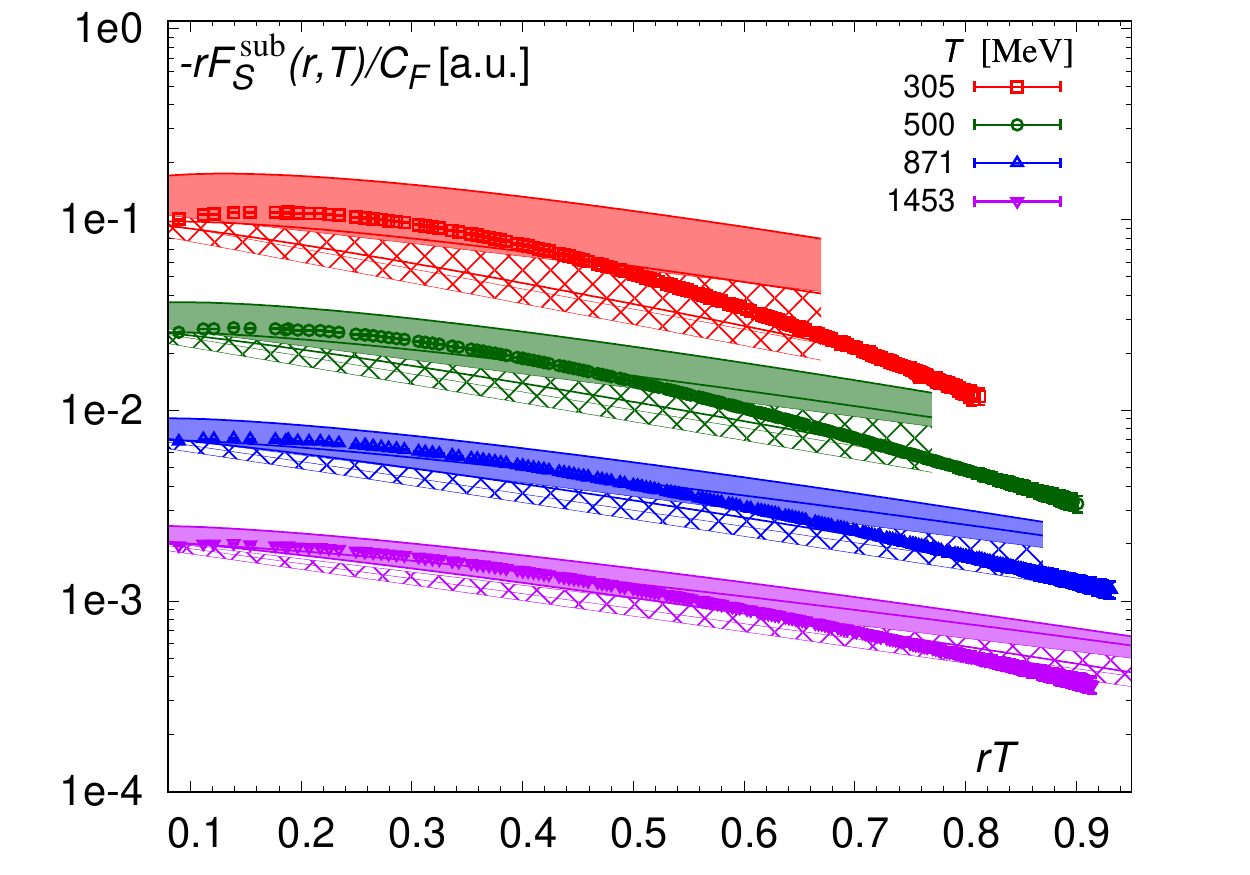}
\caption{
The screening functions of the subtracted free energy $ F_{q\bar q}^\mr{sub} $ 
or singlet free energy $ F_S^\mr{sub} $, \ie $ F_{q\bar q}^\mr{sub} $ is 
multiplied by $-9r^2 T$ or $ F_S^\mr{sub} $ is multiplied by 
$ -r/C_F $, see~\cite{Bazavov:2018wmo} for details of the calculation.
The bands represent the NLO (solid) results, where the scale has been varied as 
$\mu=\pi T,\,2\pi T$, and $4\pi T$ (solid, dotted, and dashed lines). 
Data are calculated on an $ N_{\tau}=16 $ lattice using 1 step of 
4D hypercubic (HYP) smearing~\cite{Hasenfratz:2001hp} for $r \ge \sqrt{6}/a \approx 0.15/T$.
(Left) 
The separate dash-dotted lines correspond to the LO result evaluated at 
$\mu=4 \pi T$. 
(Right) 
The hashed bands correspond to the LO result evaluated at $\mu=\pi T,\,2\pi T$, and 
$4\pi T$ (solid, dotted, and dashed lines).
\label{fig:electric}
}
\end{figure*}

After including the full next-to-leading order 
corrections~\cite{Berwein:2017thy} quantitative consistency with (2+1)-flavor 
QCD lattice simulations using staggered (HISQ) quarks could be shown for 
$ T > 300\,\mr{MeV} $ and 
$ 0.3/T \lesssim r \lesssim 0.6/T $~\cite{Bazavov:2018wmo}, see 
\Figref{fig:electric}. 
In particular, the interaction between Polyakov loops and adjoint Polyakov 
loops is predominantly mediated in the electric screening regime by 
the emission of two \emph{electric $ A_0 $ gluons} in a color singlet configuration 
with one unit of the perturbative Debye mass each. 
On the contrary, the interaction between thermal Wilson lines and adjoint 
thermal Wilson lines is predominantly mediated in the electric screening 
regime by the exchange of one \emph{electric $ A_0 $ gluon} with one unit of charge 
in the adjoint representation and one unit of the perturbative Debye mass. 
\vskip1ex
\subsubsection{Asymptotic screening} 

The color screening at asymptotically large distances cannot be understood 
in terms of the perturbative Debye mass defined in terms of the pole 
position of the \emph{electric $ A_0 $ gluon} propagator in EQCD. 
This is most easily understood in the dimensionally-reduced effective field 
theory (\emph{magnetostatic QCD}) picture, where the three-dimensional 
$ \mr{SU}(N_c) $ pure gauge theory of the \emph{magnetic gluons} gives rise 
to a confinement radius $ r \sim 1/(g^2T) $. 
At distances of the order of this magnetic confinement radius, the 
nonperturbative interaction between the \emph{electric $ A_0 $ gluons} and 
the magnetic gluons becomes too strong. 
The \emph{electric $ A_0 $ gluons} have to be dressed with compensating 
charges in order to obtain bound states that transform in the trivial 
representation of $ \mr{SU}(N_c) $ with binding energies $ \sim g^2T $.
Thus, the \emph{magnetic} confinement scale acts as the infrared cutoff for 
the \emph{electric $ A_0 $ gluons}. 
It is not evident whether a gauge-dependent \emph{electric $ A_0 $ gluon} 
propagator could even be employed for defining an order-by-order (odd 
powers of $ g $) gauge-independent pole mass in a nonperturbative framework, 
or whether the notion of a single \emph{electric $ A_0 $ gluon} at such scales 
is physically meaningful at all. 
On this basis the perturbative definition of the Debye mass in 
\Eqref{eq:mdnlo} is inadequate for describing the asymptotic screening. 
\vskip1ex
Instead, the interactions between the \emph{electric $ A_0 $ gluons} and the 
fluctuating \emph{magnetic} fields entail the additive, nonperturbative 
renormalization of the leading-order Debye mass $ \sim gT $ at the 
\emph{magnetic} confinement scale $ \sim g^2T $, and yields the expression 
for the nonperturbative Debye mass 
\al{\label{eq:mdnp}
 \md|_\mr{NP}(\nu) 
 &= \md|_\mr{LO}(\nu) 
 + \frac{N_c}{4\pi} {g^2(\nu)  T} \ln{\left(\frac{\md|_\mr{LO}(\nu) }{g^2(\nu)  T}\right)}
 + c_{N_c} {g^2(\nu)  T} +\mc{O}(g^3T).
} 
Here, $c_{N_c} $ is a constant that has to be determined numerically using 
lattice simulations\footnote{  
$c_{N_c}$ has been determined from the exponential falloff 
$ \sim \exp{[-m_Wr]} $ of an adjoint Wilson line in the lattice 
regularization (Wilson plaquette action) between a suitable discretization 
of the magnetic field strength tensor~\cite{Laine:1999hh}. 
Matching between the lattice regularization and the continuum result in 
dimensional regularization yields at one loop order~\cite{Arnold:1995bh} 
\al{\label{eq:mdnp-cnc}
 c_{N_c} 
 &= \frac{m_W}{g^2T} + \frac{N_c}{4\pi}\left(\ln{\left(g^2\right)}-1\right). 
} 
Specifically, for $ \mr{SU}(2) $ or $ \mr{SU}(3) $ the coefficients 
$ c_2=1.14(4)$, or $ c_3 = 1.65(6) $ were calculated using three-dimensional 
$ \mr{SU}(N_c) $ + adjoint Higgs theory~\cite{Laine:1999hh}.
}, 
while the leading logarithmic contribution can be determined in one-loop 
resummed perturbation theory~\cite{Rebhan:1994mx,Arnold:1995bh}. 
When counting the second term into the \emph{perturbative 
contribution} (\ie that is calculable with perturbative methods), then the 
\emph{nonperturbative contribution} due to $ c_{N_c} $ (only the third term) 
significantly exceeds the \emph{perturbative contribution} (the first two 
terms) for phenomenologically interesting temperatures and remains 
quantitatively important even at asymptotically high temperatures. 

\vskip1ex
In principle it is possible to fix an appropriate gauge and calculate the 
longitudinal and transverse components of the gluon propagator or form factor 
nonperturbatively. 
In the spatial gluon correlation functions access to the magnetic gluons is 
permitted by the tranverse modes of the propagator, while access to the 
\emph{electric $A_0$ gluons} is permitted by its longitudinal modes. 
The first, pioneering studies were completed for the gauge group 
$\mr{SU}(2)$~\cite{Heller:1995qc, Heller:1997nqa, Karsch:1998tx, Cucchieri:2000cy, Cucchieri:2001tw}.  
The functional form of the nonperturbative inverse magnetic or electric 
screening lengths and their ratio, \ie $1/\lambda_M \sim g^2 T$ or 
$1/\lambda_E \sim g T$ and $\lambda_E/\lambda_M \sim g(T) $ 
respectively as predicted in the weak-coupling picture,  can be verified 
already at temperatures as low as $T\sim 2T_c$ in 
$\mr{SU}(2)$~\cite{Heller:1995qc, Heller:1997nqa, Karsch:1998tx}. 
The \emph{electric $A_0$ gluon} screening length $\lambda_E$ is indeed still 
dominated by large nonperturbative effects at temperatures as high as 
$T\sim 10^4 T_c$, and the singlet free energy in Landau gauge, \ie 
\Eqref{eq:CS}, in the $\mr{SU}(2)$ (pure gauge) Yang-Mills theory has the 
same asymptotic screening length, $\lambda_E$~\cite{Heller:1997nqa}. 
Further lattice studies comparing the four-dimensional theory and the 
dimensionally reduced effective field theory (EQCD) indicate that the 
individual gluon propagators are consistent between both formulations~\cite{Cucchieri:2001tw}.
The propagator of \emph{electric $A_0$ gluons} exhibits a strong gauge 
dependence in its UV part, but its decay is consistent with a 
gauge- and volume-independent pole mass. 
This is not true at all for the propagator of magnetic gluons, which is 
infrared suppressed and strongly volume dependent, thus excluding a 
pole mass as a possible cause of its exponential decay. 
For large momenta the magnetic gluon propagator becomes negative in Landau 
gauge, while being bounded in absolute value by its universally positive 
counterpart in maximal Abelian gauge~\cite{Cucchieri:2000cy}. 
Contrary to naive expectations from the weak-coupling picture it is the 
symmetric, namely, confining phase of EQCD that corresponds to the deconfined 
phase of the four-dimensional Yang-Mills theory~\cite{Karsch:1998tx}. 
There is very little sensitivity of the purely magnetic glue-balls on the 
adjoint scalar field representing the \emph{electric $A_0$ gluons}, namely, 
EQCD and three-dimensonal Yang-Mills theory have the same glue-ball spectrum. 
While the bound state masses of EQCD are compatible with a constituent model 
picture, there is no apparent connection between the bound states of the 
\emph{electric $A_0$ gluons} and the inverse screening length $1/\lambda_E$ 
of the electric gluon propagator. 
The qualitative features carry over from $\mr{SU}(2)$ to $\mr{SU}(3)$~\cite{Nakamura:2003pu}. 
Whereas the transverse form factor related to the magnetic gluons remains 
almost featureless through the phase transition, the longitudinal form factor 
exhibits the features of an order parameter (using Landau gauge) 
in $\mr{SU}(3)$ Yang-Mills theory~\cite{Silva:2013maa}. 
In particular, the overlap factor between the \emph{electric $A_0$ gluon} 
state and a massive quasi-particle state as well as the infrared mass scale 
associated with such a state expose unambiguous critical behavior. 
Both gluon screening lengths in the confined and deconfined phases are 
consistent with a running gluon mass picture that is natural in the 
Dyson-Schwinger equation framework. 
For temperatures above $T \gtrsim 400\,\mr{MeV}$ the 
\emph{electric $A_0$ gluon} screening length is largely consistent with 
the weak-coupling expectation, if a temperature independent nonperturbative 
contribution is permitted. 

\vskip1ex
The strongest correlations in the high temperature phase are not mediated 
by the exchange of individual gluons at asymptotic distances, but 
by the exchange of the lightest bound states available in each channel 
of EQCD, \ie the three-dimensional $ \mr{SU}(N_c) $ pure gauge theory coupled to the adjoint scalar representing the \emph{electric $A_0$ gluons}. 
This is similar to the nuclear force in the vacuum, which is mediated at 
the largest distances by the exchange of the lightest hadrons, \ie the pions. 
In the case of thermal QCD, this role is fulfilled by either the purely 
\emph{magnetic glue-balls} with masses at the confinement scale $ \sim g^2 T $, 
or by the bound states of \emph{electric $ A_0 $ gluons} (with individual masses 
$ \sim \md $ and the binding energy at the confinement scale $ \sim g^2T $). 
These states are classified by their quantum numbers $ J^{PC}_\mc{R} $, where 
$ J $ is spin, $ P $ is parity, $ C $ is charge conjugation, and $ \mc{R} $ 
is Euclidean time reflection. 
In particular, the even or odd sectors $ \mc{R}=\pm 1 $ under Euclidean time 
reflection have been referred to as \emph{magnetic} or \emph{electric} in an 
unfortunately misleading convention. 
The \emph{magnetic sector} ($ \mc{R}=+1 $) includes any bound states with 
even numbers of \emph{electric $ A_0 $ gluons} (including the purely 
\emph{magnetic glue-balls}), while the \emph{electric sector} ($ \mc{R}=-1 $) 
includes any bound states with odd numbers of \emph{electric $ A_0 $ gluons}.
\vskip1ex
For phenomenologically interesting temperatures, the na\"ive hierarchy is 
inverted in all channels (at least for spin $ J=0 $ or $ J=1 $), \ie 
bound states with $ 2n $ \emph{electric $ A_0 $ gluons} are found to be 
systematically lighter than glue-balls made from $ 2n $ \emph{magnetic gluons} 
(irrespective of the quantum numbers $ J^{PC}_\mc{R} $, or the numbers of the 
dynamical quark flavors)~\cite{Hart:2000ha}, see~\Figref{fig:EQCD spectrum}. 
In general, the mixing between the bound states with different 
\emph{electric $ A_0 $ gluon} content is rather weak. 
This inverted hierarchy is not particularly surprising, since the QCD 
coupling $ g(\nu) $ with $\nu \sim 2\pi T $ is typically larger than 
$ 1 $ for such temperatures. 
In the following we refer to the lightest bound state with given 
$ J^{PC}_\mc{R} $ simply as $ m(J^{PC}_\mc{R}) $ irrespective of the 
hierarchy. 
It has been suggested to define the inverse nonperturbative Debye mass 
$ \md|_\mr{NP} $ as the largest correlation length of any correlation 
function constructed from local, gauge-invariant operators that are odd 
under Euclidean time reflection~\cite{Arnold:1995bh}, \ie 
to define the Debye mass as $ \md|_\mr{NP} \equiv \min( m(J^{PC}_{-}) ) $. 
This resolves the issues associated with a perturbative definition of the 
Debye mass. 
The lightest bound state in the $ \mc{R}=-1 $ sector has been obtained in 
the dimensionally-reduced QCD for operators involving each one \emph{electric 
$ A_0 $ gluon} and one 
\emph{magnetic field strength}~\cite{Hart:2000ha}\footnote{
In the four-dimensional QCD language this bound state has the quantum 
numbers $ J^{PC}_\mc{R} = 0^{++}_- $. 
The quantum numbers $ J^{PC}_\mc{R} = 0^{-+}_- $ given for this state in 
the dimensionally-reduced QCD make use of a redefined parity that reflects 
only a single spatial direction~\cite{Hart:2000ha}.}, 
the minimal field content to achieve gauge invariance in the $ \mc{R} = -1 $ 
sector. 
\vskip1ex
The asymptotic screening masses of the Polyakov loop correlators are given 
in terms of the lightest bound state masses with which they mix. 
The Polyakov loop may be split into its real or imaginary parts 
$ L(\vec{x}) = \mr{Re}\,L(\vec{x}) +\ri\,\mr{Im}\,L(\vec{x}) $, which 
transform as even or odd under Euclidean time reflection, \ie they correspond 
to quantum numbers $ J^{PC}_\mc{R} = 0^{++}_+ $ or $J^{PC}_\mc{R} = 0^{+-}_- $. 
Thus, even if the correlation function of the imaginary part is separated, 
this channel is not suitable for studying the nonperturbative Debye mass,  
because the quantum numbers of $ \mr{Im}\,L(\vec{x}) $ are $ 0^{+-}_- $ 
instead of $ 0^{++}_- $ for the lowest state in the $ \mc{R}=-1 $ sector. 
On the one hand, the screening mass $ m(0^{++}_+) $ of the correlation 
function of the real part is either the mass of the lightest scalar bound 
state consisting of two \emph{electric $ A_0 $ gluons}, 
$ \sim 2 \md + E(2A_0) $ in the inverted na\"ive hierarchy at 
phenomenologically interesting temperatures (such that $ g \gtrsim 1 $), or 
the mass of the lightest magnetic glue-ball $ \sim g^2T $ in the na\"ive 
hierarchy at asymptotically high temperatures (such that $ g \ll 1 $). 
On the other hand, the screening mass $ m(0^{+-}_-) $ of the correlation 
function of the imaginary part is the mass of the lightest bound 
state consisting of three \emph{electric $ A_0 $ gluons}, 
$ \sim 3 \md + E(3A_0) $, 
or the mass of the bound state of one \emph{electric $ A_0 $ gluon} and two 
\emph{magnetic gluons}. 
Here, the binding energies of these states $ E(2A_0) $ and $ E(3A_0) $ are 
parametrically of order $ \sim g^2T $ (the confinement scale), whereas 
the dominant contribution to $ \md $ is parametrically of order $ \sim gT $. 
As the mass of the lightest ($ \mc{R}=+1 $) state is typically much smaller 
than the nonperturbative Debye mass or the mass of the lightest ($ \mc{R}=-1 $) state, \ie 
$ m(0^{++}_+) < m(0^{++}_-) \le m(0^{+-}_-) $, the free energy $ F_{q\bar q}^\mr{sub} $ 
obtained from the (full) Polyakov loop correlation function in the asymptotic 
screening regime is dominated by the contribution from the real parts of the 
Polyakov loops. 

\begin{figure*}\center
\includegraphics[width=8.6cm]{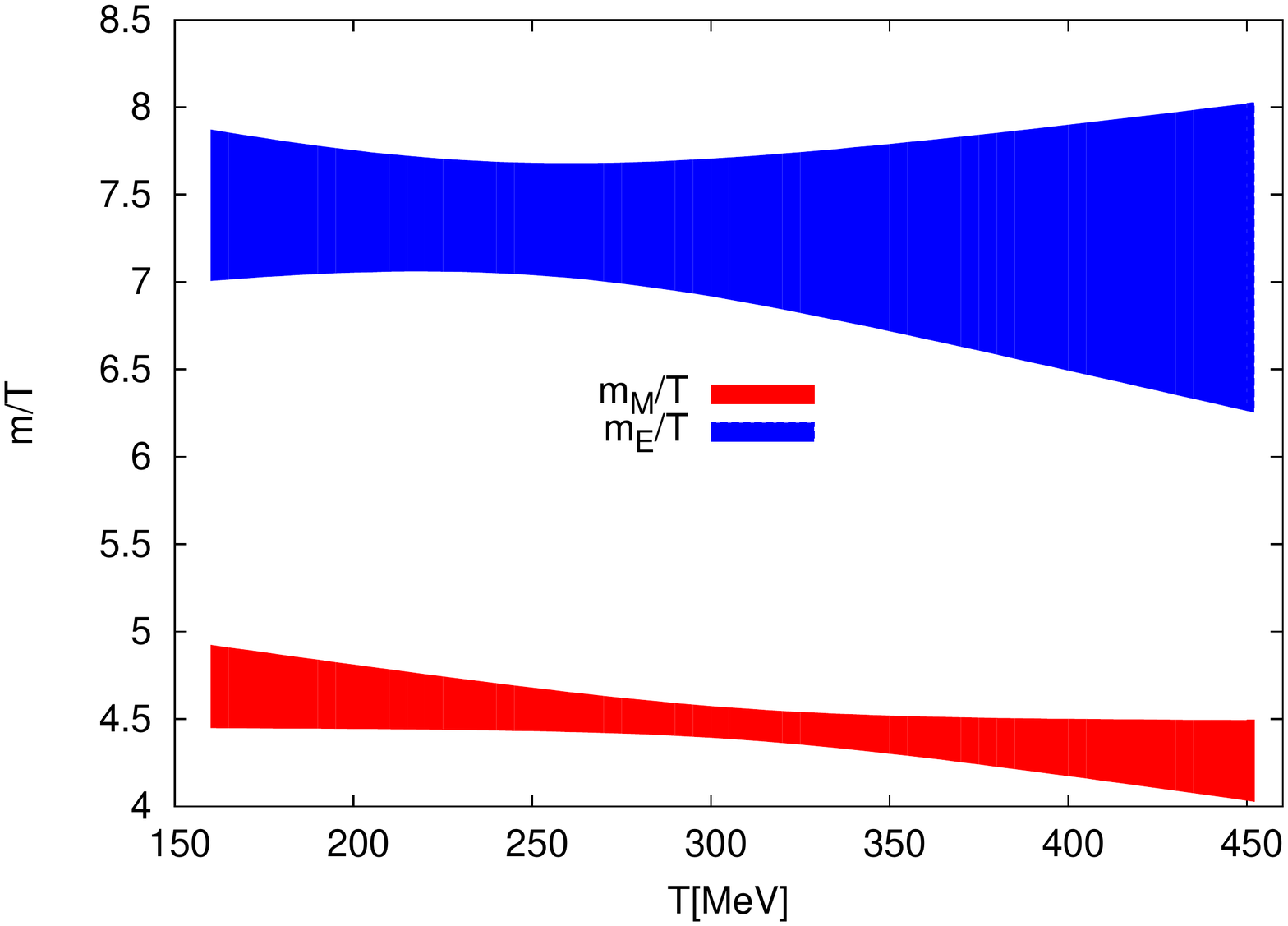}
\caption{
The continuum limit of the screening masses associated with the correlation 
function of the real (red) or imaginary parts (blue) of the Polyakov 
loops~\cite{Borsanyi:2015yka}. 
\label{fig:BW}
}
\end{figure*}

For temperatures up to $ T \lesssim 3 T_c $ the screening masses of the 
correlation functions of the real and imaginary parts of the Polyakov loop 
have been determined in (2+1)-flavor QCD lattice simulations at the physical 
point~\cite{Borsanyi:2015yka} using the stout-smeared staggered fermions and 
link smoothing techniques (4D hypercubic (HYP) smearing~\cite{Hasenfratz:2001hp}). 
Previous results from 2-flavor QCD lattice simulations using improved Wilson 
fermions and two unphysically large values of the sea quark 
mass~\cite{Maezawa:2010vj} are quantitatively similar, and verified, in 
particular, the consistency of the screening masses associated with the full 
Polyakov loop correlator and the real part correlator. 
The screening masses in the full QCD lattice calculation were found to be in 
fair agreement with the results obtained in the dimensionally-reduced QCD 
lattice simulations~\cite{Hart:2000ha}. 
For $T \approx 2 T_c $ the ratio between the corresponding masses is reported 
to be $ m(0^{+-}_-)/m(0^{++}_+) = 1.76(17) $ in the dimensionally-reduced QCD 
and $ m(0^{+-}_-)/m(0^{++}_+) = 1.63(8) $ in the QCD 
calculation~\cite{Borsanyi:2015yka}. 
The screening masses show a slightly decreasing trend for higher temperatures. 
(2+1)-flavor QCD lattice simulations using staggered (HISQ) quarks have shown 
that the asymptotic screening mass of $ F_{q\bar q}^\mr{sub} $ is only slightly 
larger than the corresponding expectation in a constituent model of two \emph{electric $A_0$ gluons}, $ 2\md|_\mr{NLO} $~\cite{Bazavov:2018wmo}, suggesting that the binding energy 
$ E(2A_0) \sim g^2T $ leads to a quantitative compensation of the nonperturbative 
dressing $ \sim g^2T $ of the two \emph{electric $ A_0 $ gluons} to a large extent.  

\begin{figure*}\center
\includegraphics[width=8.6cm]{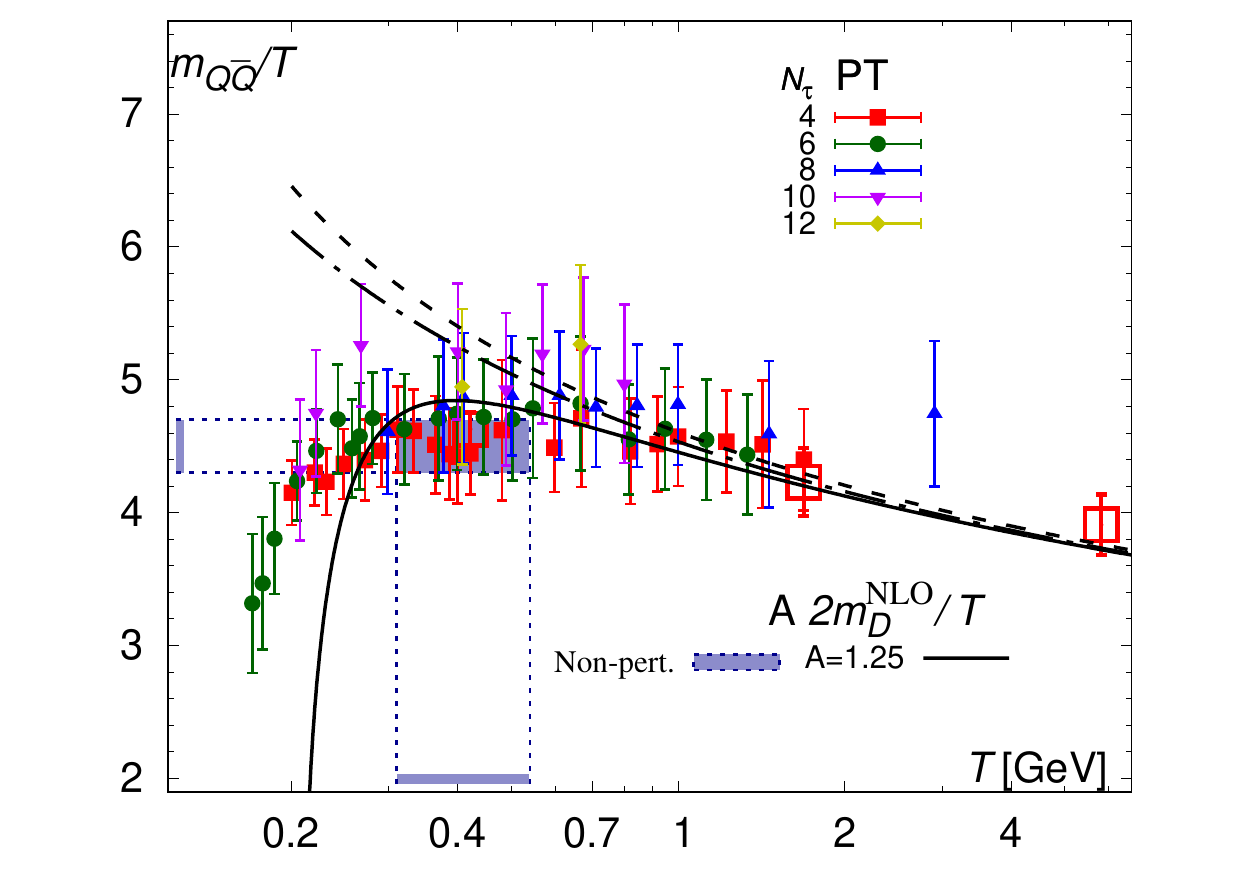}
\caption{
The screening mass associated with the free energy 
$ F_{q\bar q} $~\cite{Bazavov:2018wmo}. 
\label{fig:mqq}
}
\end{figure*}

\vskip1ex
The case of the color singlet correlation function in the asymptotic screening 
regime is still more intricate. 
First of all, the interaction that is dominant in the electric screening 
regime is mediated by the emission of one \emph{electric $ A_0 $ gluon}, which 
carries one unit of charge in the adjoint representation. 
Hence, due to its charge this exchanged \emph{electric $ A_0 $ gluon} cannot 
mix on its own with the bound states of the three-dimensional 
$ \mr{SU}(N_c) $ pure gauge theory, which have to transform in the trivial 
representation for any confining gauge theory. 
As a consequence, this mode of emission can only accumulate the 
nonperturbative dressing $ \sim g^2T$, and thus one would naively suppose 
that it should be screened with the nonperturbative Debye mass 
$ \md|_\mr{NP} $, or with some operator-dependent screening mass that 
receives contributions both $ \sim gT $ and $ \sim g^2T $ (due the definition 
of the nonperturbative Debye mass, operator-dependent screening masses for 
$ \mr{R}=-1$ cannot be smaller than $ \md|_\mr{NP} $). 
Whether or not this mode could propagate at asymptotically large distances 
may also depend on the details of the operators, \ie the gauge-fixed thermal 
Wilson line correlator or the path-dependent spatially-smeared cyclic Wilson 
loop.
In the former case, the Coulomb gauge gives rise to a 
\emph{dimension-two magnetic gluon condensate} $\Braket{ \bm{A}^2 }$. 
The \emph{electric $ A_0 $ gluon} may scatter on the \emph{magnetic gluons} 
of this condensate and acquire contributions from the confinement scale 
$ \sim g^2T $ of the three-dimensional $ \mr{SU}(N_c) $ pure gauge theory. 
In the latter case, the spatial Wilson lines involve spatially-extended 
operators, by which the emitted \emph{electric $ A_0 $ gluon} may couple 
to \emph{magnetic gluons} in a gauge-invariant multi-particle state of the 
three-dimensional $ \mr{SU}(N_c) $ pure gauge theory. 
For either correlation function, such states must have the quantum numbers 
$\mc{R}=-1$ and $P=+1$. 
Calling the energy of the lowest accessible state $ m(X^{+}_{-}) $, we note 
that it cannot have a lower energy than the lowest bound state with the same 
quantum numbers, namely $ m(X^{+}_{-}) \gtrsim \md|_\mr{NP} $.
Arguing from the rationale of the quantitative similarity between the cyclic 
Wilson loops with spatial smearing and the Wilson line correlators in Coulomb 
gauge in the asymptotic screening regime, see \Figref{fig:FW}, it appears as 
if these two operator-dependent singlet correlation functions 
may have access to the same set of screening lengths, which would suggest 
that the state $X^{+}_{-}$ could be in fact operator independent.
\vskip1ex
Yet this is not the whole story. 
The nonperturbative Debye mass is larger than $ m(0^{++}_+) $, \ie the mass 
of the lightest bound state with $ \mc{R}=+1 $ (whether this is a bound 
state of two \emph{electric $ A_0 $ gluons} or a \emph{magnetic glue-ball} 
is irrelevant in this regard). 
Just like the Polyakov loops, the untraced Wilson line operators in the 
singlet correlators making up \Eqsref{eq:cycWL} or \eqref{eq:CS} have 
separable real or imaginary parts, too, that each have well-defined behavior 
under Euclidean time reflection. 
In fact, the thermal Wilson lines also couple directly to two or three 
\emph{electric $ A_0 $ gluons} through their real or imaginary parts, respectively,
although these couplings are formally suppressed by two or four powers 
of the coupling $ g $. 
On the one hand, the singlet correlation function of the real parts of the 
thermal Wilson lines receives contributions from the $ \mc{R}=+1 $ bound 
states of the three-dimensional $ \mr{SU}(N_c) $ pure gauge theory.
The contribution from the emission of one \emph{electric $ A_0 $ gluon} is 
screened in the asymptotic regime with $ m(X^{+}_{-}) \gtrsim \md|_\mr{NP} $, 
while the contribution from the emission of two \emph{electric $ A_0 $ gluons} 
in a color singlet configuration mixes with the lightest bound state of 
the three-dimensional $ \mr{SU}(N_c) $ pure gauge theory with 
mass~$ m(0^{++}_+) $ satisfying $ m(0^{++}_+) < \md|_\mr{NP} $ for all 
thermal hierarchies. 
Hence, the latter contribution with screening length $1/m(0^{++}_+) $ has to 
dominate the singlet correlation function of the real parts eventually.
On the other hand, the singlet correlation function of the imaginary parts 
of the thermal Wilson lines, however, mixes the contribution from the 
emission of one \emph{electric $ A_0 $ gluon} with the contribution from 
the emission of three \emph{electric $ A_0 $ gluons}. 
Although the latter mixes with the $ 0^{+-}_- $ bound states of the 
three-dimensional $ \mr{SU}(N_c) $ pure gauge theory, all of these 
masses are larger than the nonperturbative Debye mass $ \md|_\mr{NP} $. 
Hence, in the end, either the potentially operator-dependent screening 
mass $ m(X^{+}_{-}) \gtrsim \md|_\mr{NP} $ or the mass of the lightest scalar 
bound state consisting of three \emph{electric $ A_0 $ gluons} in the 
three-dimensional $ \mr{SU}(N_c) $ pure gauge theory, $ \sim 3 \md + E(3A_0) $, 
may be the smallest inverse correlation length for the correlation function 
of the imaginary parts of the Wilson lines. 
Since the latter exchange mode is suppressed by four powers of $ g $, 
this channel may actually be well-suited to a determination of the 
operator-dependent screening mass $ m(X^{+}_{-}) \gtrsim \md|_\mr{NP} $ even 
with a possibly small mass difference between the $ X^{+}_{-} $ and 
$ 0^{+-}_- $ screening masses.

\begin{figure*}\center
\includegraphics[width=6.6cm]{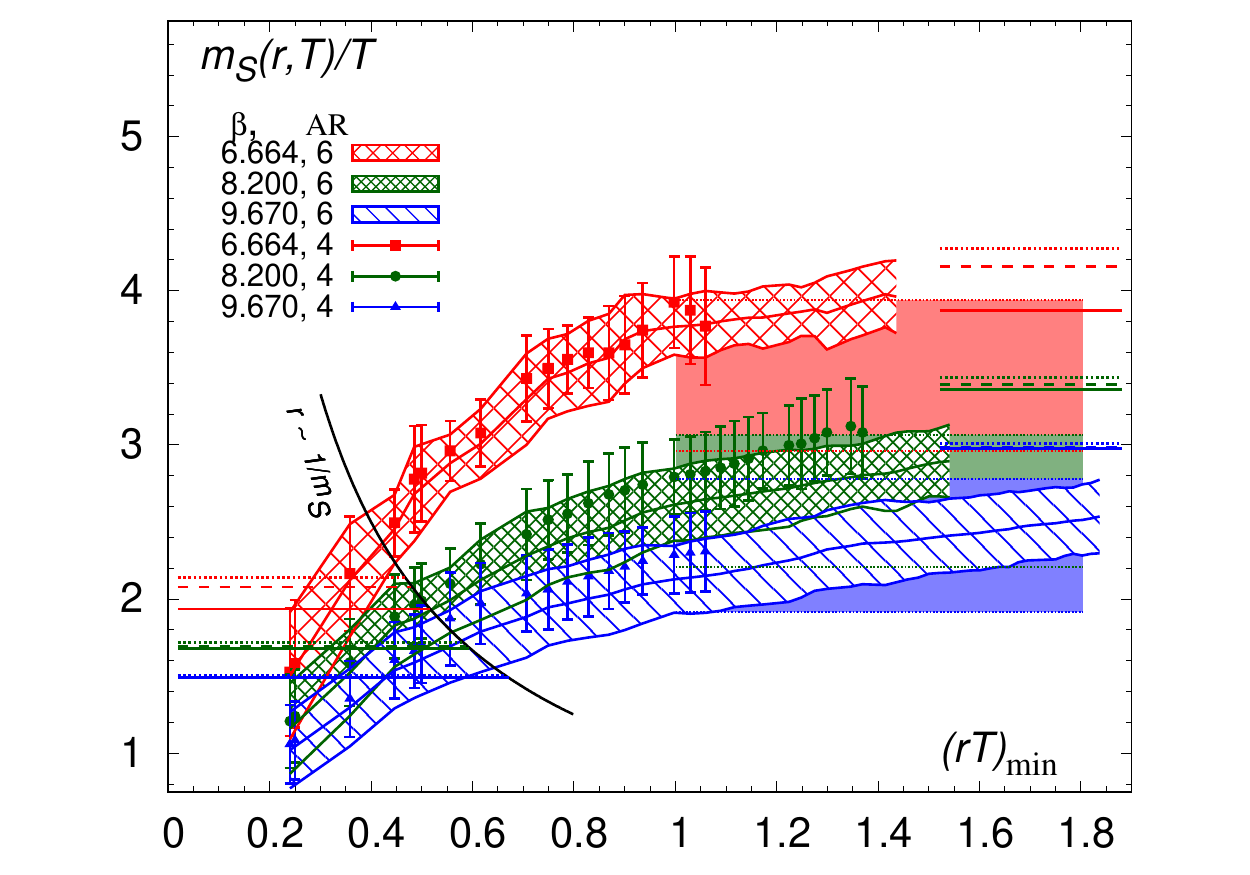}
\includegraphics[width=6.6cm]{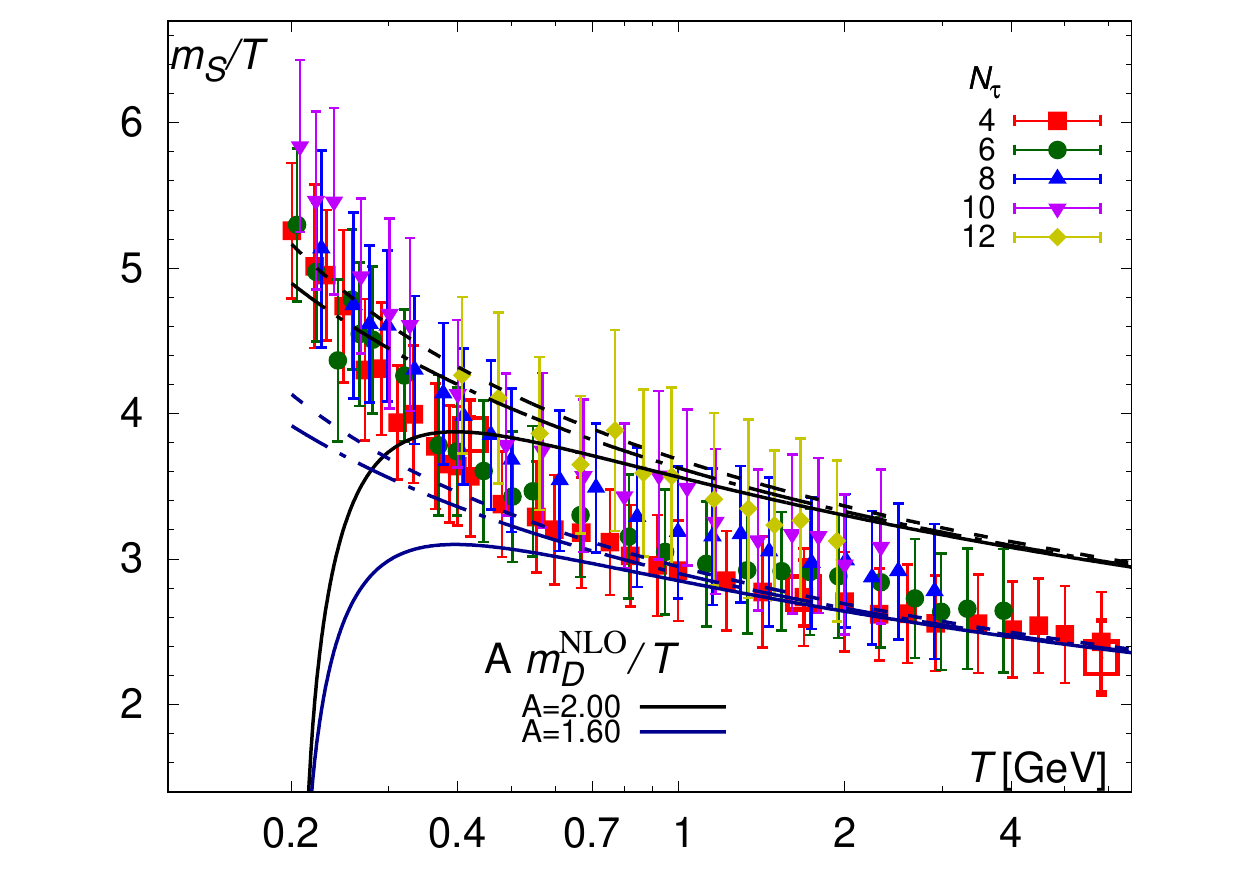}
\caption{
The screening mass associated with the singlet free energy 
$ F_S $~\cite{Bazavov:2018wmo}. 
(Left) The local screening mass associated with the singlet free 
energy defined in \Eqref{eq:CS} increases at larger distances. 
The horizontal lines to the left indicate the perturbative 
Debye mass $ \md|_\mr{NLO} $, and the horizontal lines to the right 
indicate twice the perturbative Debye mass, $ 2\md|_\mr{NLO} $. 
Available results are obtained with $ N_\tau = 4 $ and may be affected 
by substantial discretization effects. 
(Right) The screening mass associated with the singlet free energy 
has a similar temperature dependence than the perturbative Debye 
mass $ \md|_\mr{NLO} $.
\label{fig:ms}
}
\end{figure*}

\vskip1ex
(2+1)-flavor QCD lattice simulations using staggered (HISQ) quarks show that 
the local screening mass associated with $ F_S^\mr{sub} $ defined in terms 
of Wilson line correlation functions in Coulomb gauge, see \Eqref{eq:CS}, 
is only slightly larger than $ \md|_\mr{NLO} $ for 
$ r \sim 1/\md|_\mr{NLO} $~\cite{Bazavov:2018wmo}. 
This screening mass becomes systematically larger with increasing distances, 
and seems to saturate quite close to $ 2\md|_\mr{NLO} $ for $ r \gg 1/T $, see 
\Figref{fig:ms}
as in the case of the correlator of the real parts of two Polyakov loops.

This screening mass also exhibits a similar temperature dependence as 
$ \md|_\mr{NLO} $ for temperatures $ T > 300\,\mr{MeV} $. 
The unambiguous, quantitative analysis of the asymptotic screening 
of the singlet correlation function is still lacking and requires the use of 
link-smoothing techniques to overcome the severe signal-to-noise problem of 
static quark correlation functions, see \eg\mbox{Ref.}~\cite{Steinbeisser:2018sde} 
for preliminary results of ongoing work along these lines.
\vskip1ex
At temperatures that correspond to the vacuum phase, the asymptotic screening 
mass $ m(0^{++}_+) $ due to the lowest $ J^{PC}_\mc{R}=0^{++}_+ $ bound state 
of the dimensionally-reduced QCD smoothly connects to the screening mass due 
to the string breaking in the vacuum phase of QCD, \ie the energy difference 
between the static quark-antiquark energy and the mass of two static-light 
mesons. 
\vskip1ex
\subsubsection{Screening in different representations}


In the deconfined phase of $ \mr{SU}(N_c) $ gauge theory (with or without 
quarks) the notion of diquarks, which transform non-trivially under the 
gauge group, appears to be quite natural. 
In the vacuum, heavy-heavy diquarks seem to play an important role for the 
formation of heavy-light tetraquark systems, see \eg Ref.~\cite{Francis:2016hui}. 
In particular, it must be expected that  -- if diquarks exist at all as 
individual objects -- heavy-heavy diquarks are still quite strongly bound 
at temperatures slightly above $ T_c $. 
In the following we apply again the terminology for $ N_c=3 $, \ie use 
anti-triplet ($ \mb{\overline{3}} $) for the \emph{anti-fundamental} 
representation and sextet ($ \mb{6} $) for the 
$ \mb{(N_c^2-N_c)} $-dimensional representation arising from 
$ \mb{N_c} \times \mb{N_c} = \mb{\overline{N_c}} + \mb{(N_c^2-N_c)} $. 
The most simple objects related to the screening of the static diquarks 
are the Polyakov loops in other representations, which have been studied in 
the $\mr{SU}(3)$ pure gauge theory~\cite{Gupta:2007ax} and in the (2+1)-flavor 
QCD~\cite{Petreczky:2015yta,Bazavov:2016uvm}. 
Key results for single Polyakov loops in different representations are briefly discussed in \mbox{Sec.}~\ref{sec:ploop}. 
Correlation functions of Polyakov loops and thermal Wilson lines in 
suitable representations correspond to spatially extended diquarks. 

Thus, correlation functions of diquarks transforming in the anti-triplet or 
sextet representations can be studied with the lattice approach, \eg after 
fixing Coulomb gauge through 
\al{
 \label{eq:antitriplet}
 C_{\mb\overline{3}}(T,r) 
 &=
 \frac{1}{2N_c} \left\{
 \vphantom{ \Braket{
 		~\sum\limits_{\bm{x}}~L(\bm{x}) L(\bm{x}+\bm{r})~}-\Braket{~\sum\limits_{\bm{x}}~\tr
 		W(aN_\tau, \bm{x}; 0 \bm{x})  W(aN_\tau,\bm{x}+\bm{r}; 0 \bm{x}+\bm{r}) ~}^\mr{gf}}
 \Braket{
 	~\sum\limits_{\bm{x}}~L(\bm{x}) L(\bm{x}+\bm{r})~}\right.\nonumber\\
 &\left.-\Braket{~\sum\limits_{\bm{x}}~\tr
 W(aN_\tau, \bm{x}; 0 \bm{x})  W(aN_\tau,\bm{x}+\bm{r}; 0 \bm{x}+\bm{r}) ~}^\mr{gf}
 \right\}, \\
  \label{eq:sextet}
  C_{\mb 6}(T,r) 
 &=
 \frac{1}{4N_c} \left\{
 \vphantom{\Braket{~\sum\limits_{\bm{x}}~L(\bm{x}) L(\bm{x}+\bm{r})~}
 	+\Braket{~\sum\limits_{\bm{x}}~\tr
 		W(aN_\tau, \bm{x}; 0 \bm{x})  W(aN_\tau,\bm{x}+\bm{r}; 0 \bm{x}+\bm{r}) ~}^\mr{gf}}
 \Braket{~\sum\limits_{\bm{x}}~L(\bm{x}) L(\bm{x}+\bm{r})~}\right.\nonumber\\
 &\left.+\Braket{~\sum\limits_{\bm{x}}~\tr
 W(aN_\tau, \bm{x}; 0 \bm{x})  W(aN_\tau,\bm{x}+\bm{r}; 0 \bm{x}+\bm{r}) ~}^\mr{gf}
 \right\}.
} 
In particular, $ \mr{SU}(N_c) $ pure gauge theory, or 2-, or (2+1)-flavor 
QCD lattice studies using improved Wilson 
fermions~\cite{Maezawa:2011aa, Ejiri:2009hq, Maezawa:2007fc} found that the 
different channels with two thermal Wilson lines in the singlet, octet, 
anti-triplet and sextet representation satisfy the naive Casimir scaling 
with the Casimir factors
$ C_{\mb R}= \braket{ \sum_{c=1}^{N_c^2-1} t^c(\bm{x})t^c(\bm{x}+\bm{r}) }_{\mb R} $ 
for representations $ \mb{R} $ as 
\al{
 &C_{\mb 1} = -\frac{N_c^2-1}{2N_c},& 
 &C_{\mb 8} = \frac{1}{2N_c},& 
 &C_{\mb \overline{3}} = \frac{1}{N_c},& 
 &C_{\mb 6} = \frac{1-N_c}{N_c}
}
in the asymptotic regime to a good approximation for temperatures higher than $T \gtrsim 300\,\mr{MeV}$. 
2-flavor QCD lattice simulations~\cite{Doring:2007uh} with improved staggered 
fermions indicate consistent results. 
On the basis of the preceding discussion that the asymptotic screening length 
is determined in all of these cases by the inverse mass $ 1/m(0^{++}_+) $ of 
the same lightest scalar state, to which the real parts of the Wilson lines 
couple through the emission/absorption of two \emph{electric $ A_0 $ gluons} 
this is hardly surprising. 
\vskip1ex

\subsubsection{Screening at finite chemical potential}

Nonperturbative studies at finite chemical potential are challenging due to 
the sign problem associated with quark or baryon chemical potential in the 
lattice approach. 
One has to resort to either reweighting methods, or to the Taylor expansion 
in $ \mu/T $, or to analytical continuation of results obtained at imaginary 
chemical potential $ \mu_I =\ri \mu $.
Within their limited radii of applicability these approaches yield 
consistent results for the dependence of the free energies on the chemical 
potential~\cite{Takahashi:2013mja}. 
Available results are obtained with $ N_\tau = 4 $ and may be affected by 
significant discretization effects.

\begin{figure*}\center
  \includegraphics[width=0.5\textwidth]{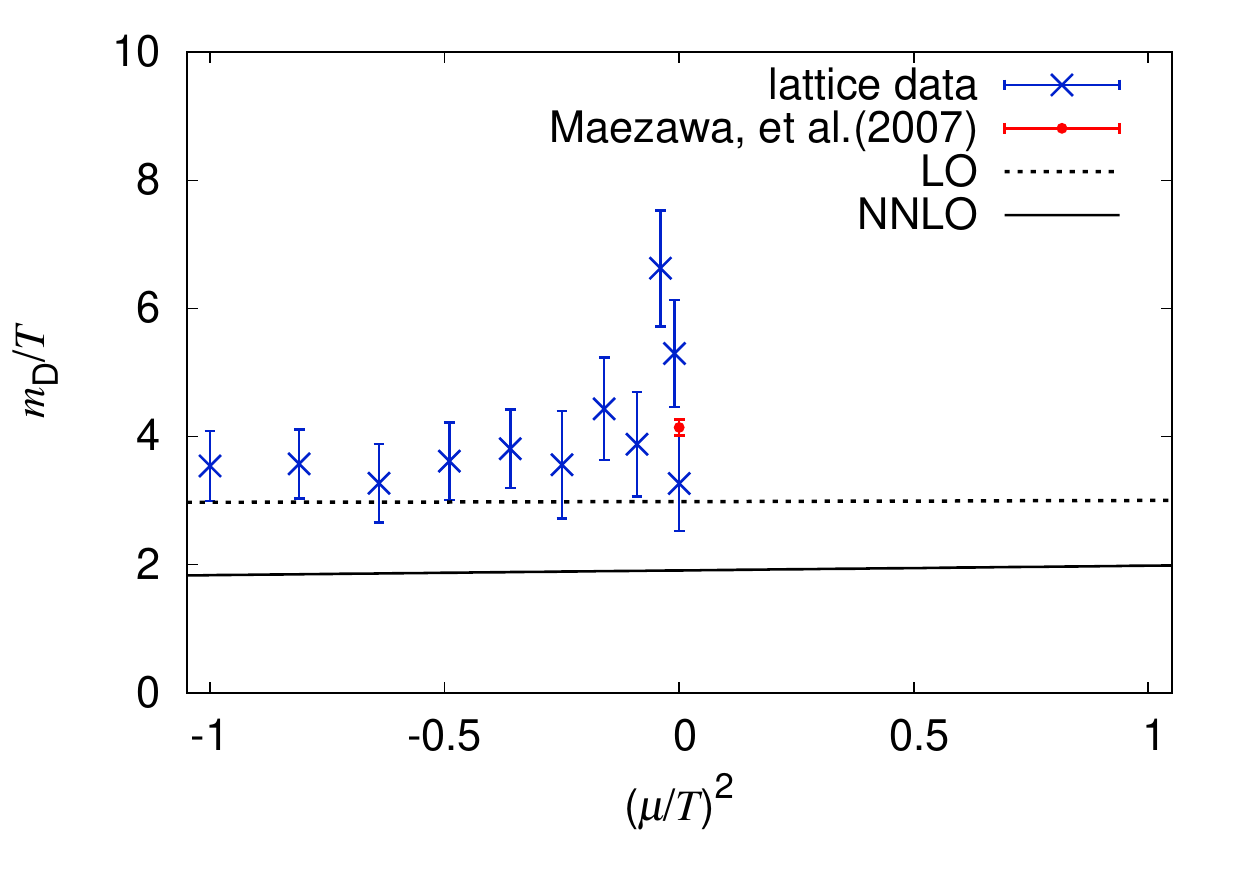}
\caption{
The screening mass associated with the singlet free energy 
$ F_S $ at imaginary chemical potential~\cite{Takahashi:2013mja}. 
Available results are obtained with $ N_\tau = 4 $.
\label{fig:mucb}
}
\end{figure*}

\vskip1ex
On the one hand, the singlet or octet correlation functions (as well as 
$ C_L $, see \Eqref{eq:fso}) are even functions of $ \mu/T $, while,  
on the other hand, the anti-triplet and sextet correlation functions, 
\Eqsref{eq:antitriplet} and \eqref{eq:sextet} contain 
the nontrivial odd contributions in the expansion in 
$ \mu/T $~\cite{Ejiri:2009hq}. 
The coefficients at higher orders in the Taylor expansion are suppressed 
by an order of magnitude against the $ \mu=0 $ result. 
The Taylor expansion coefficients associated with even powers $(\mu/T)^{2n}$ 
in the expansion of the asymptotic screening mass (\ie the mass of the 
lightest scalar $ m(0^{++}_+) $) are found to be positive, and about 
10\% of the $ \mu=0 $ result for temperatures 
$ T \approx 2T_c $, see \Figref{fig:mucb}.
\vskip1ex
At finite chemical potential charge conjugation and Euclidean time reflection 
$ \mc{R} $ cease to be good quantum numbers, since the number density operator 
in \Eqref{eq:muN} breaks the symmetries under charge conjugation and Euclidean 
time reflection. 
Hence, mixing between the bound states with even or odd numbers of 
\emph{electric $ A_0 $ gluons} may eventually become quite strong in the 
medium at finite density, which implies that the real or imaginary parts of 
the Polyakov loop cannot fluctuate independently. 
Yet the mass $ m(0^{+}_+) $ of the lightest scalar state has been found to 
be still significantly smaller than the masses of all other states in 
dimensionally-reduced QCD lattice simulations~\cite{Hart:2000ha} 
with an increase of only about 10\% at $ \mu/T = 1 $ and about 20\% at 
$ \mu/T = 2$.  
For this reason, the correlation functions that primarily couple to this 
state are expected to be only mildly affected. 
On the contrary, however, the screening lengths in the $ \mc{R} = -1 $ 
channels such $ J^{P}_\mc{R}=0^{+}_- $ are expected to increase quite 
dramatically towards the screening lengths associated with some scalar 
$ J^{P}_\mc{R}=0^{+}_+ $ channel. 

\begin{figure*}\center
\begin{minipage}{1.0\textwidth}
  \includegraphics[width=6.6cm]{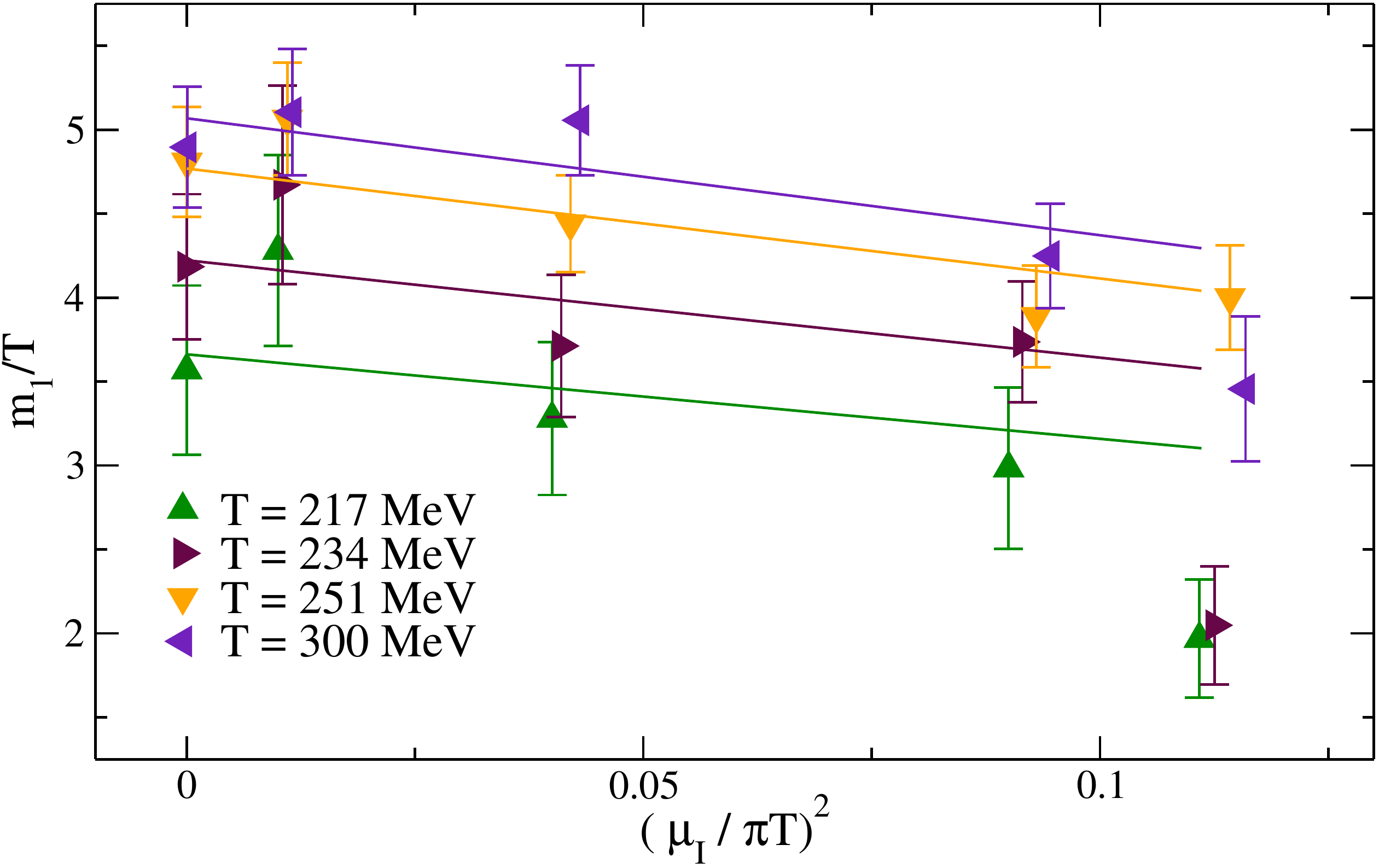}
  \includegraphics[width=6.6cm]{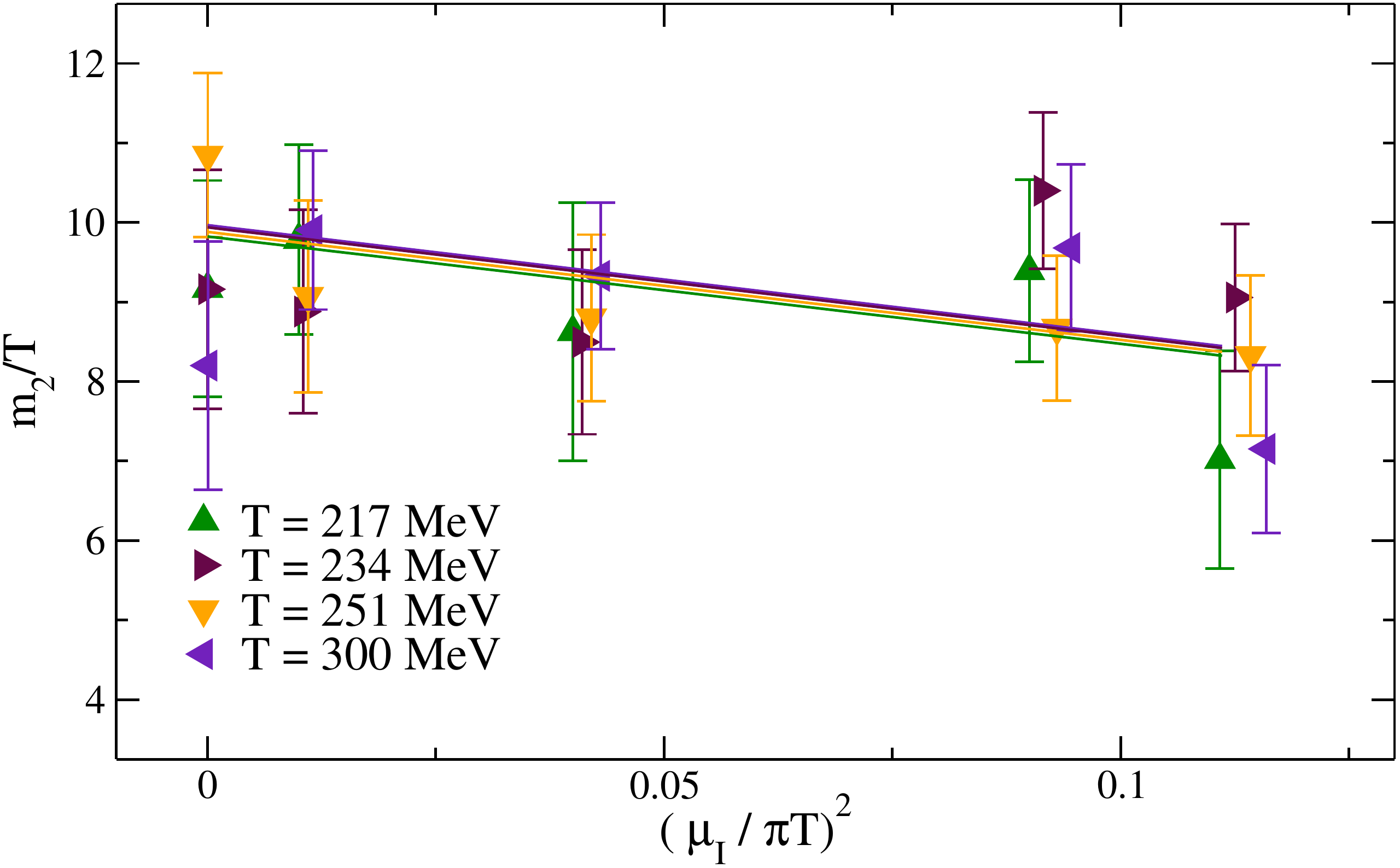}
\end{minipage}
\caption{
The screening masses $ m_1/T $ and $ m_2/T $ associated with the diagonalized 
components of the correlation matrix of the real and imaginary parts of the 
Polyakov loop at imaginary chemical potential~\cite{Andreoli:2017zie}. 
Available results are obtained with $ N_\tau = 8 $.
\label{fig:mu12}
}
\end{figure*}

\vskip1ex
The correlation functions of real and imaginary parts of the Polyakov 
loops are indeed mixed, \ie there is a nontrivial cross-correlator, and 
the $ 2 \times 2 $ correlation matrix has to be diagonalized. 
(2+1)-flavor QCD lattice simulations at imaginary chemical 
potential~\cite{Andreoli:2017zie} have verified this behavior. 
On the one hand, the smaller screening mass $ m_1/T $ in the diagonalized 
basis is found to be only marginally ($ \sim 10\% $) larger than for the 
correlation function of the real part, whereas, on the other hand, the 
larger screening mass $ m_2/T $ in the diagonalized basis increases by 
about 20\% compared to the $ \mu=0 $ result for temperatures sufficiently 
above the Roberge-Weiss transition up to $ T \lesssim 2T_c $. 
Note that the behavior in \Figref{fig:mu12} is the opposite, since the horizontal 
axis represents squared imaginary chemical potential. 
\vskip1ex

\subsubsection{Screening in external magnetic fields}

It has been known for some time that external magnetic fields modify the 
properties of the QCD crossover transition. 
In particular, the presence of the magnetic fields leads to the anisotropy 
of the quark-antiquark interaction for different orientations with respect 
to the magnetic field. 
For vanishing magnetic field the free energy $ F_{q\bar q} $ is for 
temperatures well below $ T_c $ and $ r \lesssim 1/\lMSb $ or for 
distances much smaller than the inverse temperature ($ r \ll 1/T $) up to 
the trivial change of normalization $ +T\ln(N_c^2) $ almost 
indistinguishable from the quark-antiquark static energy. 
In an external magnetic field the screening behavior of $ F_{q\bar q} $ 
is evident for temperatures much lower than $ T_c $, whereas the chiral 
condensate still does not show signs of inverse magnetic 
catalysis~\cite{Bonati:2016kxj}. 

\begin{figure*}\center
\begin{minipage}{1.0\textwidth}
  \includegraphics[width=6.6cm]{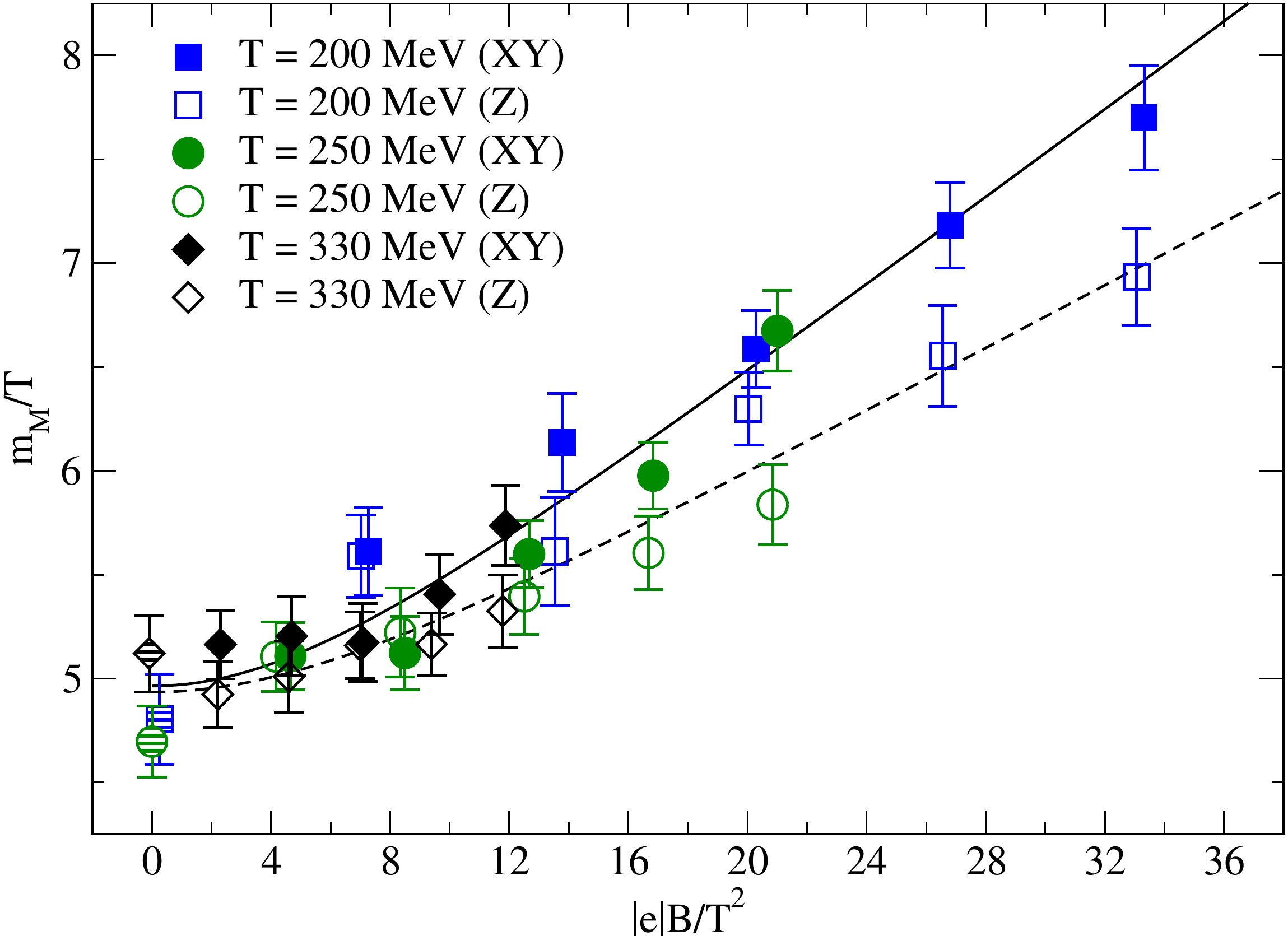}
  \includegraphics[width=6.6cm]{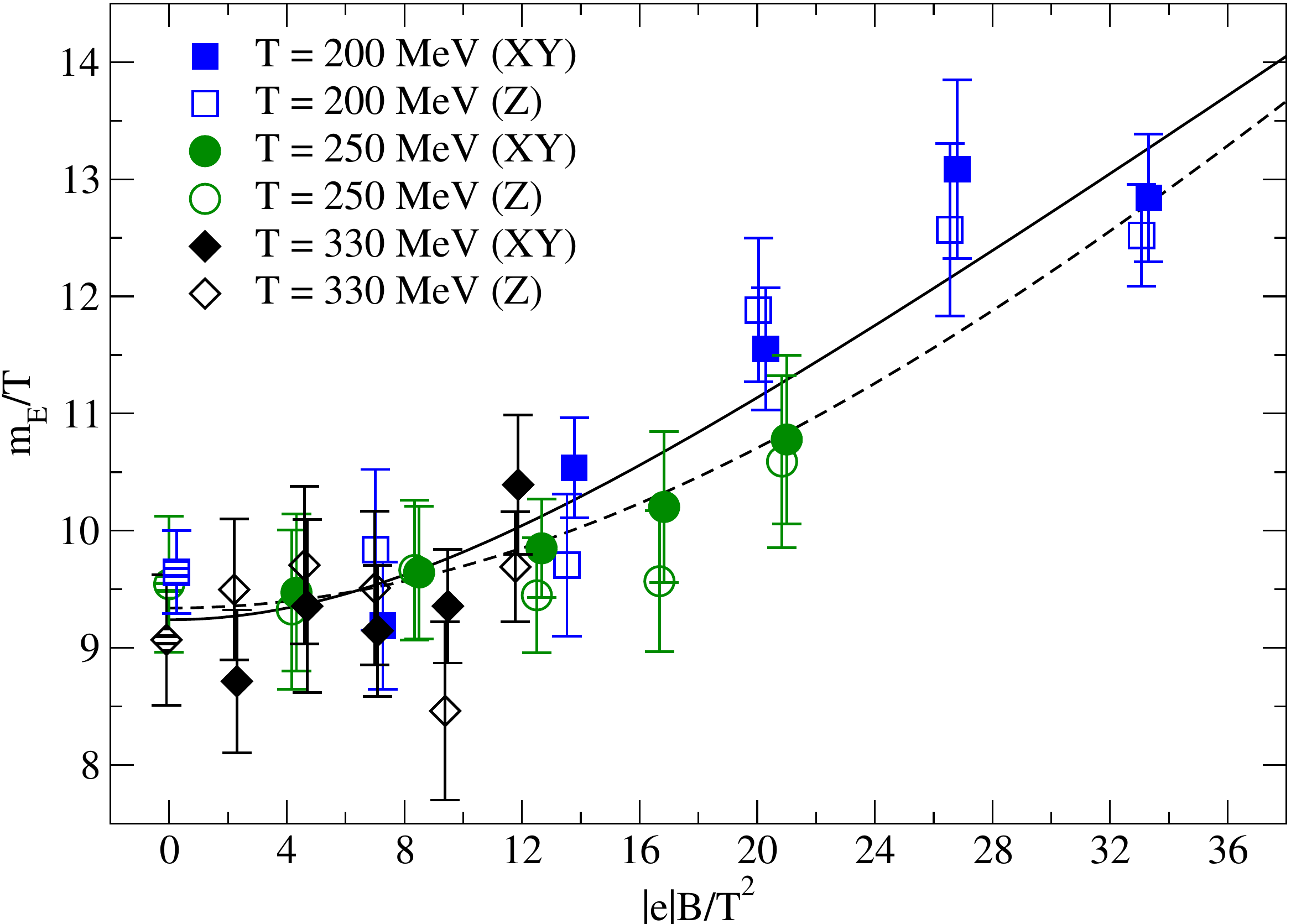}
\end{minipage}
\caption{
The screening masses $ m_m/T $ ($J^{PC}_\mc{R}=0^{++}_+$) and $ m_E/T $ 
($J^{PC}_\mc{R}=0^{+-}_-$) associated with the correlation function of 
the real or imaginary parts of the Polyakov loops in an external magnetic 
field~\cite{Bonati:2017uvz}. 
Available results are obtained with $ N_\tau = 8 $.
\label{fig:mag}
}
\end{figure*}

\vskip1ex
For this reason an influence of the magnetic fields on the thermal screening 
masses has to be expected. 
A calculation in (2+1)-flavor QCD lattice simulations~\cite{Bonati:2017uvz} 
indicated that the screening masses of the correlation function of the real 
and imaginary parts of the Polyakov loops are increased in strong magnetic 
fields. 
While the ratio between the screening masses in the $ J^{PC}_\mc{R} = 0^{+-}_- $ 
and  $ J^{PC}_\mc{R} = 0^{++}_+ $ is rather mildly affected, the individual 
masses increase quite strongly for a rising external magnetic field. 
The modified screening masses scale with $|e|B/T^2$, see \Figref{fig:mag}. 
Whereas the $ 0^{+-}_- $ channel is similarly affected for both alignments 
with regard to the external magnetic field, the $ 0^{++}_+ $ channel has to 
be aligned perpendicular to the external magnetic field for the largest 
modification. 

\section{Interplay between screening and dissociation}
\label{sec:dynamic}


\newcommand \ber {\begin{eqnarray}}
\newcommand \eer {\end{eqnarray}}
\newcommand \beq {\begin{equation}}
\newcommand \eeq {\end{equation}}
\newcommand \sgh {\sigma(\omega,\bm{p},T)}

\newcommand{\order}{\mathcal{O}}
\newcommand{\psib}{\bar{\psi}}
\newcommand{\chib}{\bar{\chi}}
\newcommand{\cM}{\mathcal{M}}
\newcommand{\op}{\psib (\Gamma \otimes t^a) \psi}
\newcommand{\Ns}{N_\sigma}
\newcommand{\Nt}{N_\tau}
\newcommand{\x}{\mathbf{x}}

\def \unitmatrix{1\!\!\!1}
\def \tc{\textcolor}


Shortly after the first conjectures of quark-gluon plasma as the state of 
nuclear matter at high temperature and the beginning of the era of heavy-ion 
collision experiments as tools for studies of the QCD phase diagram, the idea 
was brought up in \mbox{Ref.}~\cite{Matsui:1986dk} that color screening causes 
rearrangements of the in-medium bound states leading to the sequential 
melting of quarkonia.
The relative yields from the various in-medium quarkonia produced in heavy-ion 
collisions would then serve the role of an experimentally accessible probe for 
the temperature of the plasma. 
However, over the years this static picture has been replaced by a dynamical 
picture, in which both dissociation and recombination take place inside of 
the plasma, too. See \eg \mbox{Ref.}~\cite{Rothkopf:2019ipj} for a discussion 
of this paradigm change. 
These processes may be too rapid for permitting the restructuring of the bound 
states as demanded in the static picture and may even regenerate already 
depleted bound states. 
In this case, the in-medium quarkonia have to be considered as an open quantum 
system and treated in a real-time approach. 
While there are indeed indications that this is the case, the questions of 
color screening and dissociation in a realistic scenario cannot be treated 
separately. 
In the following, we will discuss the interplay between both for the static 
quarks that we have discussed so far and relax the infinite mass limit taking 
a look at relativistic heavy and light quarks as well.

\vskip1ex
We discuss the basic ideas about extracting real-time information from 
Euclidean lattice correlators through the spectral functions in 
subsection~\ref{sec:spectral}. 
We outline the key differences between how temporal and spatial meson 
correlators can provide information. 
After discussing the relatively simpler case of the real-time dynamics of 
static quark-antiquark pairs in subsection~\ref{sec:complex_F}, we turn our 
attention to heavy-heavy or heavy-light systems beyond the static limit in 
subsection~\ref{sec:mesons_h}. 
We close this discussion after comparing these results to the case of 
light-light systems in subsection~\ref{sec:mesons_l}.

\subsection{Euclidean correlation functions and spectral functions}
\label{sec:spectral}

Earlier we discussed how screening properties of the 
deconfined medium can be studied theoretically by looking at the 
response of the medium to insertion of static probe charges. 
We now turn to the discussion on screening properties of systems
with dynamical quarks.
Ultimately, we are interested in how the deconfined medium affects
the QCD spectrum, \textit{i.e.} various bound states and resonances 
composed of light ($u$, $d$, $s$) and heavy ($c$, $b$) dynamical
quarks.
We focus our discussion on mesons -- hadrons composed of quark and
antiquark.

\subsubsection{Temporal correlation functions}

The information on how the thermal medium, quark-gluon plasma modifies
hadrons and eventually leads to the their dissolution at high temperatures
is encoded in spectral functions. The latter are Fourier transforms
of the real-time correlation functions which are, however, not directly
accessible in lattice QCD. Instead, one can compute the Euclidean
correlation functions. A Euclidean temporal meson correlation function
projected to a given spatial momentum $\bm{p}$ has
the following form:
\begin{equation}
\label{eq:GtaupT}
G(\tau, \bm{p},T) = \int d^3x e^{i \bm{p}\cdot\bm{x}} 
\langle J_H(\tau, \bm{x}) J_H(0,
\bm{0}) \rangle \; ,
\end{equation}
where $J_H=\bar q \Gamma_H q$ is a meson operator and
$\Gamma_H=1,\gamma_5, \gamma_\mu, \gamma_5\gamma_\mu,\gamma_\mu\gamma_\nu$
projects onto a channel with given quantum numbers.

Taking into account the periodicity in the temporal direction
one can relate the Euclidean correlation functions to the spectral
functions $\rho(\omega,\bm{p},T)$:
\begin{eqnarray}
G(\tau, \bm{p},T) &=& \int_0^{\infty} d \omega
\rho(\omega,\bm{p},T) K(\omega, \tau, T), \label{eq.spect}\\
K(\omega, \tau, T) &=& \frac{\cosh(\omega(\tau-1/(2
	T)))}{\sinh(\omega/(2 T))}.
\label{eq.kernel}
\end{eqnarray}
The temperature dependence of $\rho(\omega,\bm{p},T)$ 
shows how the deconfined medium screens the interactions and at
what temperatures the mesons dissolve.

Ideally, one would like to calculate the spectral functions nonperturbatively
in lattice QCD. 
However, in the Euclidean lattice formalism Eq.~(\ref{eq.spect}) has then
to be considered as an integral equation from which the spectral function
$\rho(\omega,\bm{p},T)$ needs to be \textit{reconstructed}.
Eq.~(\ref{eq.spect}) poses a very ill-defined inverse problem.
There are several fundamental features that make
it particularly hard in practice.
First, the temporal extents of finite-temperature
lattices, where $G(\tau, \bm{p},T)$ is evaluated, are of 
$ \mc{O}(10) $ data points.
while the spectral function $\rho(\omega,\bm{p},T)$ has typically
a rich structure, requiring hundreds of frequency
points to be resolved. Right from the start the problem 
is very underdetermined.
Second, the kernel of the transformation $K(\omega, \tau, T)$ falls off
exponentially with the frequency $\omega$, suppressing the features of
the spectral function.
This exponential loss of information leads to very little sensitivity
of $G(\tau, \bm{p},T)$
to thermal modification of the spectral function. And, third, 
all correlation functions are 
determined from Monte Carlo sampling and, thus, feature significant
statistical fluctuations. 

If we restrict ourselves to a static quark-antiquark pair instead of a 
relativistic quark-antiquark pair, as we will be doing in 
subsection~\ref{sec:complex_F}, we will encounter a milder version of 
this inverse problem. 
In this case, the transformation kernel $K(\omega,\tau,T)$ simplifies to 
the Laplace kernel $\exp(-\omega\tau)$ and sheds its explicit temperature 
dependence.
Moreover, the Laplace kernel is not symmetric under $\tau \to 1/T-\tau$, 
such that the full range of the correlation function provides useful 
information. 

To deal with the inverse problem (\ref{eq.spect})
Bayesian methods, such as the Maximum Entropy Method,
are often employed~\cite{Jarrell:1996rrw}.
Despite almost two decades of effort
in the lattice QCD community starting with the pioneering
work of Refs.~\cite{Asakawa:2000tr,Asakawa:2003re,Datta:2003ww},
determination of spectral
functions with fully quantified uncertainties remains
an open problem. A recent comprehensive review
on the status of the field of calculating spectral
functions in perturbative approaches and
lattice QCD for heavy quarkonia can be
found in~\cite{Rothkopf:2019ipj}.

\subsubsection{Spatial correlation functions}

Given all the complications of extracting the information from the temporal
correlation functions, one could consider mesonic \textit{spatial}
correlation functions, as was first pointed out in
Refs.~\cite{DeTar:1987ar,DeTar:1987xb}:
\begin{equation}
G(z,T) = \int_0^{1/T} d \tau \int dx dy
\langle J_H(\tau,x,y,z) J_H(0,0,0,0) \rangle.
\end{equation}
They are related to the same spectral function as the temporal correlation
functions but in a slightly more involved way:
\begin{equation}
G(z,T)=\int_{0}^{\infty} \frac{2 d \omega}{\omega} \int_{-\infty}^{\infty} d p_z e^{i p_z z} \rho(\omega,p_z,T).
\label{eq:spatial}
\end{equation}
However, unlike for the temporal direction, the separation is not limited
to inverse temperature. Thus, the spatial correlation functions are
more sensitive to the in-medium modification effects,
since they can be studied at larger quark--antiquark separation.
Moreover, absence of the temperature-dependent kernel,
as apparent from comparing Eq.~(\ref{eq:spatial}) with (\ref{eq.spect}),
means that the temperature dependence of the spatial correlation function
comes entirely from the spectral function. Thus, one can directly study
the temperature dependence of $G(z,T)$ without the need of reconstructing
$\rho(\omega,p_z,T)$.
Deviations of the spatial correlation function at finite temperature from
its vacuum form directly signal in-medium modification of the corresponding
hadronic state.

At large distances the spatial correlation functions decay exponentially
\begin{equation}
G(z,T) \sim \exp(-M(T)z),
\end{equation}
where the fall-off is governed by the temperature-dependent
\textit{screening mass}, or inverse screening length, $M(T)$.
At small enough temperatures where
a well-defined mesonic bound state exists, the spectral function has
a $\delta$-like peak
\begin{equation}
\rho(\omega,p_z,T)\sim\delta(\omega^2-p_z^2-M_0^2)
\label{eq:rho_delta}
\end{equation}
and $M(T)$ coincides with the propagator pole mass $M_0$.

At high enough temperatures where the mesonic state is dissociated, the spatial
correlation function describes propagation of a free quark-antiquark pair.
In this case the screening mass is~\cite{Florkowski:1993bq}
\begin{equation}
M_{\rm free}=\sqrt{m_{q_1}^2+(\pi T)^2}+ \sqrt{m_{q_2}^2+(\pi T)^2} \; ,
\label{eq:Mfree}
\end{equation}
where $m_{q_1}$ and $m_{q_2}$ are the pole masses of the quark and 
antiquark that form the meson. 
As was shown in Ref.~\cite{Florkowski:1993bq}, and
was also argued earlier in Ref.~\cite{Eletsky:1988an} for the case of
massless quarks, the appearance of the lowest fermionic Matsubara
frequency mode $\pi T$ in the meson screening mass is a direct
consequence of the anti-periodic temporal boundary conditions for
fermions. Thus, a crossover from one limiting behavior,
Eq.~(\ref{eq:rho_delta}) to the other, Eq.~(\ref{eq:Mfree}), would
indicate that the mesonic state dissociates in the plasma and the
lowest-order contribution to the screening mass comes from two
independently propagating fermionic degrees of freedom.

\subsection{The complex static energy}
\label{sec:complex_F}

The quark-antiquark static energy $ E(T,r) $ in the thermal medium is 
modified, too. 
In particular, in the deconfined phase $ E(T,r) $ acquires a nonzero 
imaginary part and exhibits the color screening. 
However, contrary to the $ T=0 $ situation, lattice studies of the complex 
in-medium static energy $ E(T,r) $ are fraught with profound difficulties 
that we can illustrate only briefly in this review. 
\vskip1ex
In the following, we begin by juxtaposing the zero and finite temperature 
cases, beginning with the more simple situation at $ T = 0 $. 
In the vacuum the static energy $ E(r) $ can be defined in terms 
of the infinite time limit of the logarithm of the real-time Wilson loop 
$ W(t,r) $, \ie 
\al{\label{eq:Ereal}
 &E(r) = \lim\limits_{t \to \infty} { \frac{\ri}{t} \ln W(t,r)} ,&
 &W(t,r) 
 = \Braket{P \exp{\left\{ \ri \oint_{W(t,r)} dz^\mu A_\mu \right\}} },
}
or similarly in Euclidean space-time after $ t \to -\ri \tau $. 
As before, $P$ represents the path ordering. 
The Wilson loops have a spectral decomposition in the vacuum (using
Euclidean space-time)
\al{\label{eq:rhovac}
 &W(\tau,r) = \int\limits_{0}^\infty d\omega \rho_{W(r)}(\omega) e^{-\omega\tau},&
 & \rho_{W(r)}(\omega) = \sum\limits_{n=0}^\infty A_n \delta(E_n(r)-\omega),
}
where the spectral function $\rho_{W(r)}(\omega)$ is a weighted sum of delta 
functions that gives rise to the weighted sum of exponentials mentioned in \Eqref{eq:E vacuum}. 

\begin{figure*}\center
\includegraphics[width=8.6cm]{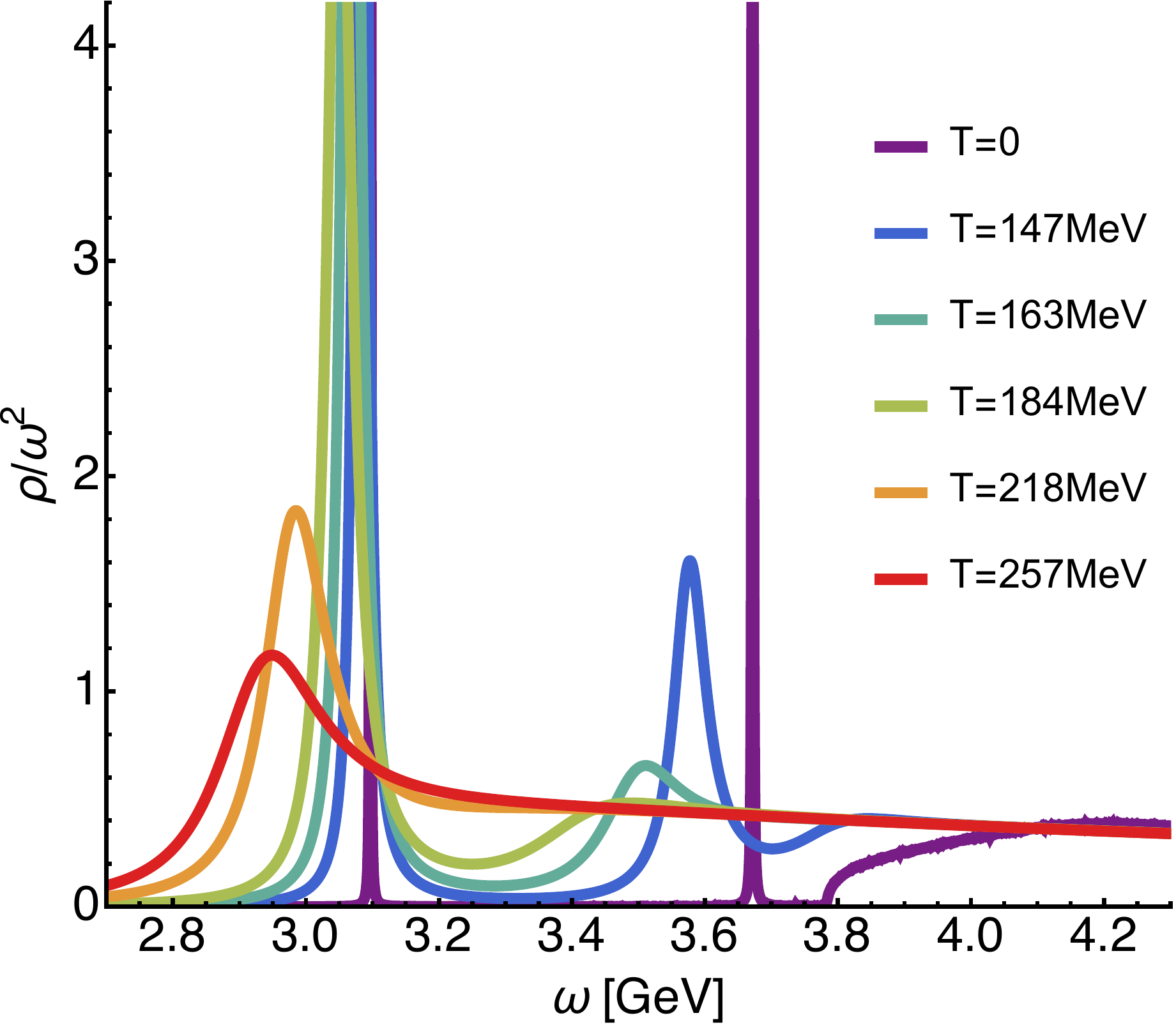}
\caption{
Charmonium in-medium spectral functions from a continuum corrected in-medium
heavy quark potential with AsqTad action~\cite{Burnier:2015tda}. 
\label{fig:rho}
}
\end{figure*}

\Eqref{eq:Ereal} can be directly evaluated on the lattice in the vacuum 
phase, see \Eqref{eq:E vacuum}, since the length of the Euclidean time 
direction can be as large as technically affordable such that only the 
first delta function in the sum in \Eqref{eq:rhovac} that is associated with 
the ground state contributes. 
If the Euclidean time direction is periodic with period $ aN_\tau$, then 
the Wilson loop must be evaluated at sufficiently large values of 
$ \tau < aN_\tau $ such that excited state contributions are quantitatively 
irrelevant or can be included in a robust fit. 
\vskip1ex
In the perturbative expansion $E(r)$ and $ F(r) = \del_r E(r) $ are known 
analytically to next-to-next-to-next-to-leading order, and at 
next-to-next-to-next-to-leading logarithmic accuracy~\cite{Brambilla:2009bi}. 
To leading order the static energy is given by the leading order singlet 
potential 
\al{ 
 E(r) 
 &= 
 -C_F \frac{\als}{r} = V_s(r).
}
Higher order contributions are proportional to the $ \mr{LO} $ result and 
involve higher powers of $ \als $, and --- starting at three loops --- 
$ \ln(\als) $. 
The latter are due to an interplay between the contribution from the 
singlet potential and from the ultrasoft contribution. 
The latter is due to transitions between color-singlet and -octet configurations on internal lines, starts at $ \mr{N^3LO} $, and includes contributions from 
coupling to the nonperturbative QCD condensates. 
With the exceptions of the ultrasoft contribution that is absent in Abelian 
gauge theories, this result is in all respects formally quite similar to the 
QED result.
Comparison to lattice simulations shows that the $ \mr{N^3LO} $ result 
describes the static energy very well up to 
$ r \lesssim 0.15\,\mr{fm} $~\cite{Bazavov:2019qoo}. 
\vskip1ex
Since there is no closed loop in the time direction, there cannot be any 
overlap with states involving (static) $ D $ mesons, and thus the 
screening of the static energy due to the string breaking in the vacuum 
cannot be resolved without explicitly including other operators that overlap 
with the pair of (static) $ D $ mesons and solving a generalized eigenvalue 
problem, see \mbox{Ref.}~\cite{Bulava:2019iut} for a recent calculation.
This is evidently different from the free energy in the vacuum, where the 
string breaking explicitly contributes. 
\vskip1ex
On the contrary, the spectral function of the finite temperature Wilson loops 
representing a static quark-antiquark pair in the thermal medium,
\al{
 &W(\tau,T,r) = \int\limits_{0}^\infty d\omega \rho_{W(T,r)}(\omega) e^{-\omega\tau},
}
is a sum of smeared delta peaks that may be shifted away from their zero 
temperature frequencies and a high frequency continuum that is due to 
the fully dissociated states. 
If there is a well-defined lowest peak that is clearly separable from the 
rest of the spectral function, then one can interpret its position as the 
real part of $E(T,r)$ and its width as the imaginary part of $E(T,r)$. 
However, as different peaks may merge with each other or into the 
encroaching high frequency continuum for increasing temperature, it is 
not completely obvious whether there actually are well defined peaks at 
all and if the notion of a finite temperature static energy $E(T,r)$ is 
physically appropriate at all. 
Moreover, the time direction is physically fixed to the inverse temperature 
$ aN_\tau=1/T $ in a lattice setup at finite temperature, and thus the limit 
$ \tau \to \infty $ in which the ground state could be isolated is completely 
out of reach in practical lattice simulations. 
For this reason, one has to deal with the excited state contamination, the 
early-time dynamics, and the bound state formation. 
It follows from very general considerations that the contribution from the 
complex static energy $E(T,r)$ to the spectral function takes the form of a 
skewed Breit-Wigner peak, where the skewing is due to the early time dynamics 
of the bound state formation~\cite{Burnier:2012az}. 
However, it is not a priori clear how much this spectral feature is 
distorted by the rest of the spectral function. 
The real-time static energy at finite temperature has been calculated to 
next-to-leading order~\cite{Laine:2006ns, Brambilla:2008cx} using the HTL 
approach. 
$E(T,r)$ has been found to exhibit an imaginary part in the electric 
screening regime $ r \sim 1/\md $,
\al{\label{eq:E electric screening}
 E(T,r) 
 &= -C_F\als(\nu) 
 \left\{ \frac{e^{-\md(\nu)r}}{r} + \md(\nu) 
 +\ri T \phi(r\md(\nu))    
 \right\}, \\
 \phi(x) 
 &=
 2\int\limits_{0}^{\infty} \frac{dz~z}{(z^2+1)^2} 
 \left\{ 1-\frac{\sin(zx)}{zx} \right \},
} 
where $ \phi(x) $ is a strictly monotonically increasing function. 
While the imaginary term appears already in the vacuum-like regime 
$r \ll 1/T$~\cite{Brambilla:2008cx}, the result is parametrically 
quite different as 
\al{\label{eq:E vacuum like}
 E(T,r) 
 &= -C_F\als(1/r) 
 +\left[ 
 \left\{ \#_1 \frac{\Delta V}{T} + \#_2 \frac{\md^2}{T^2} + \#_3 \frac{\md^3}{T^3} \right\}\right.\nonumber\\
 &+\left.\ri \left\{ \#_4 \frac{\Delta V^2}{T^2}+\#_5 \frac{\md^2}{T^2} \right\} 
 \right]  g^2 r^2 T^3,
} 
with coefficients $\#_i,~i=1,~\ldots,~5$ given in \mbox{Ref.}~\cite{Brambilla:2008cx}. 
It is understood that the origin of the imaginary part (in both regimes) 
is in part due to the dissipative scattering (Landau damping) of the emitted 
gluons with the various degrees of freedom in the thermal medium, and 
in part due to transitions between the color-singlet and -octet 
configurations of the static quark-antiquark pair. 
Beyond $ \mr{NLO} $ these two dynamical processes cannot be separated anymore. 
These weak-coupling results for two different regimes indicate that the 
notion of the finite temperature static energy $E(T,r)$ is at least justified 
for sufficiently small distances and sufficiently high temperatures such that 
the weak-coupling approach applies. 
Whether or not this concept is suitable at phenomenologically interesting 
temperatures and at distances relevant to the melting or survival of in-medium 
quarkonium bound states must be addressed using nonperturbative methods.
Whereas the real part $ \mr{Re}~E(T,r) $ agrees with the singlet free 
energy $ F_S(T,r) $ at order $g^3$ both in the vacuum-like regime 
$r \ll 1/T$ and in the electric screening regime $ r \sim 1/\md $, they 
differ at order $ g^4 $ due to different UV contributions.  
However, $ F_S(T,r) $ is screened already in the vacuum of full QCD (with sea 
quarks) due to the string breaking, whereas $ E(T,r) $ defined in terms of the 
large time limit of the Wilson loop cannot couple directly to states 
involving static $ D $ mesons\footnote{
This applies to a definition in terms of the large time limit of Wilson 
line correlators in Coulomb gauge as well.}. 
Hence, any screening mass associated with a nonperturbative analog of 
\Eqsref{eq:E electric screening} or \eqref{eq:E vacuum like} must necessarily vanish in the vacuum phase at $ T < T_c $. 
\vskip1ex
These different sources of information, \ie the dissociative imaginary part 
from the perturbative HTL result in \Eqref{eq:E electric screening}, and the real part from 
the zero temperature lattice calculation were used to construct a maximally 
binding, minimally dissociative model potential with 
ad-hoc exponential screening in \mbox{Ref.}~\cite{Petreczky:2010tk}.
It could be shown that the dominant source for in-medium quarkonium 
melting is the dissociation even in such a simplistic model.  
\vskip1ex
In order to calculate $ E(T,r) $ directly, the real-time formalism is 
required, since the imaginary part arises due to dynamical real-time 
processes. 
However, the real-time formalism introduces a sign problem in the QCD path 
integral, and thus prevents the application of the importance sampling 
in the Markov Chain Monte Carlo algorithm employed in lattice simulations. 
Nevertheless, both the real-time and the imaginary-time Wilson loops 
are related to the same underlying spectral function 
$ \rho_{W(T,r)}(\omega) $ through analyticity,
\al{\label{eq:rhoreal}
 W(t,T,r) 
 &= \int\limits_{-\infty}^{+\infty} d\omega e^{-\ri \omega t} 
 \rho_{W(T,r)}(\omega) &&\text{real time}, \\
 \label{eq:rhoimag}
 W(\tau,T,r) 
 &= \int\limits_{-\infty}^{+\infty} d\omega e^{-\omega \tau} 
 \rho_{W(T,r)}(\omega) &&\text{imaginary time},
}
where $ W(\tau,T,r) $ is directly calculable on the lattice, 
at most with $ \mc{O}(10) $ data points for a thermal correlation function.  
However, any reliable reconstruction of narrow spectral features and some 
continuum requires covering hundreds of frequency values. 
This makes the inverse problem of reconstructing 
$ \rho_{W(T,r)}(\omega) $ from the imaginary-time Wilson loop 
$ W(\tau,T,r) $ ill-posed, 
and solutions to it are beyond the scope of this review. 
One has to resort to model assumptions on the form of the spectral function, 
or Bayesian analysis that incorporates prior knowledge. 
Even a brief discussion of the various Bayesian techniques that have been 
developed for solving this problems exceeds the scope of this review, 
see~\cite{Burnier:2013nla, Rothkopf:2019dzu, Rothkopf:2019ipj} for an 
overview of the current state of the art. 
Upon assuming that the spectral function $ \rho_{W(T,r)}(\omega) $ 
has been determined using a nonperturbative lattice simulation, \ie 
see \Figref{fig:rho} for results obtained from (2+1)-flavor lattice 
simulations using the AsqTad action, the complex, real-time static 
energy follows from its lowest peak structure. 

\begin{figure*}\center
\includegraphics[width=8.6cm]{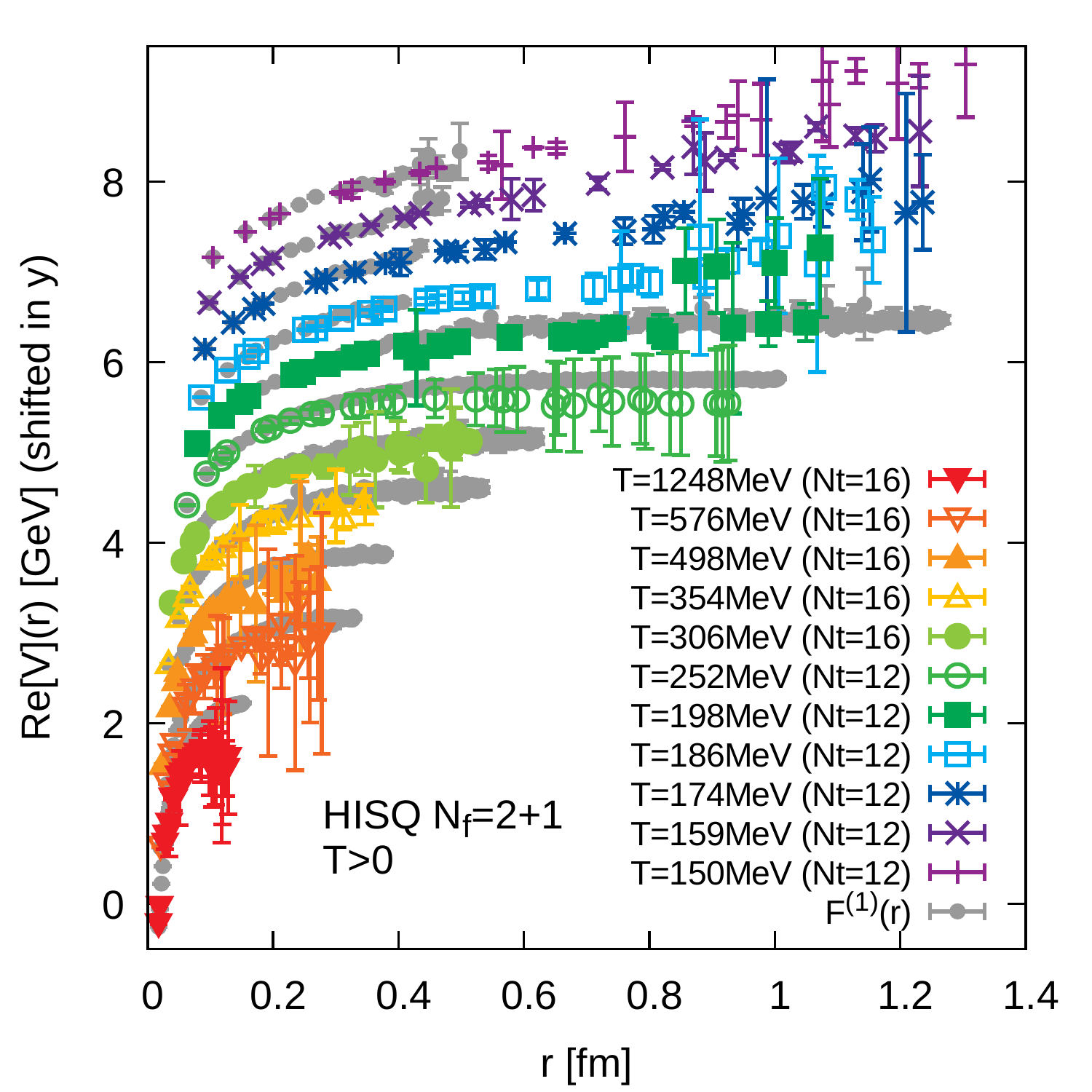}
\caption{
The real part of the potential obtained from Pad\'e reconstructed spectral 
functions of the Wilson line correlation function in Coulomb gauge on 
$ 48^3 \times 12 $, and $ 64^3 \times 16 $ lattices in (2+1)-flavor QCD 
with HISQ action~\cite{Petreczky:2018xuh}. 
The values are shifted by hand in y-direction for better readability from 
the lowest temperature $T = 151\,\mr{MeV} $ on top to the highest temperature 
$ T = 1248\,\mr{MeV} $ at the bottom. 
The gray data represent the color singlet free energy in Coulomb gauge 
calculated on the same lattices.
\label{fig:ReV}
}
\end{figure*}

\begin{figure*}\center
\includegraphics[width=6.6cm]{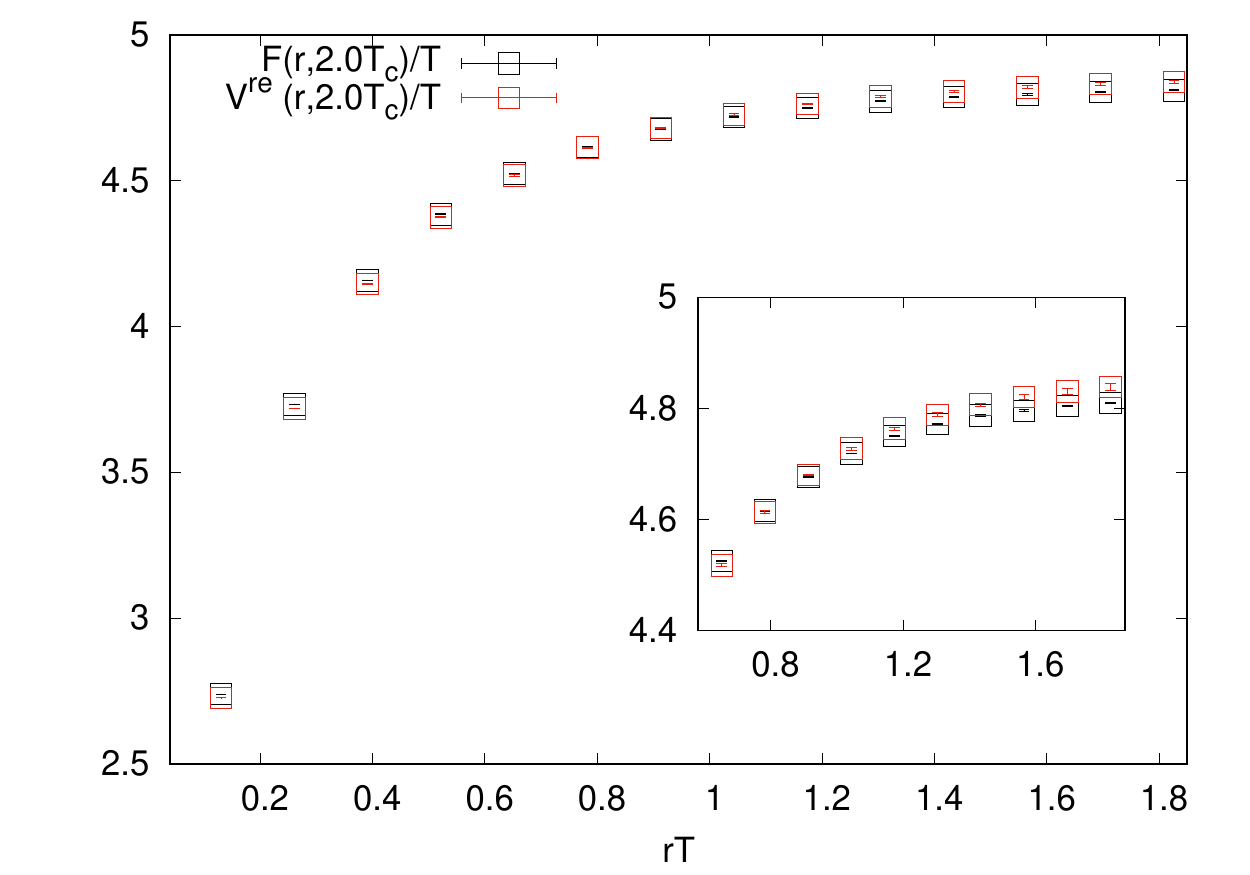}
\includegraphics[width=6.8cm]{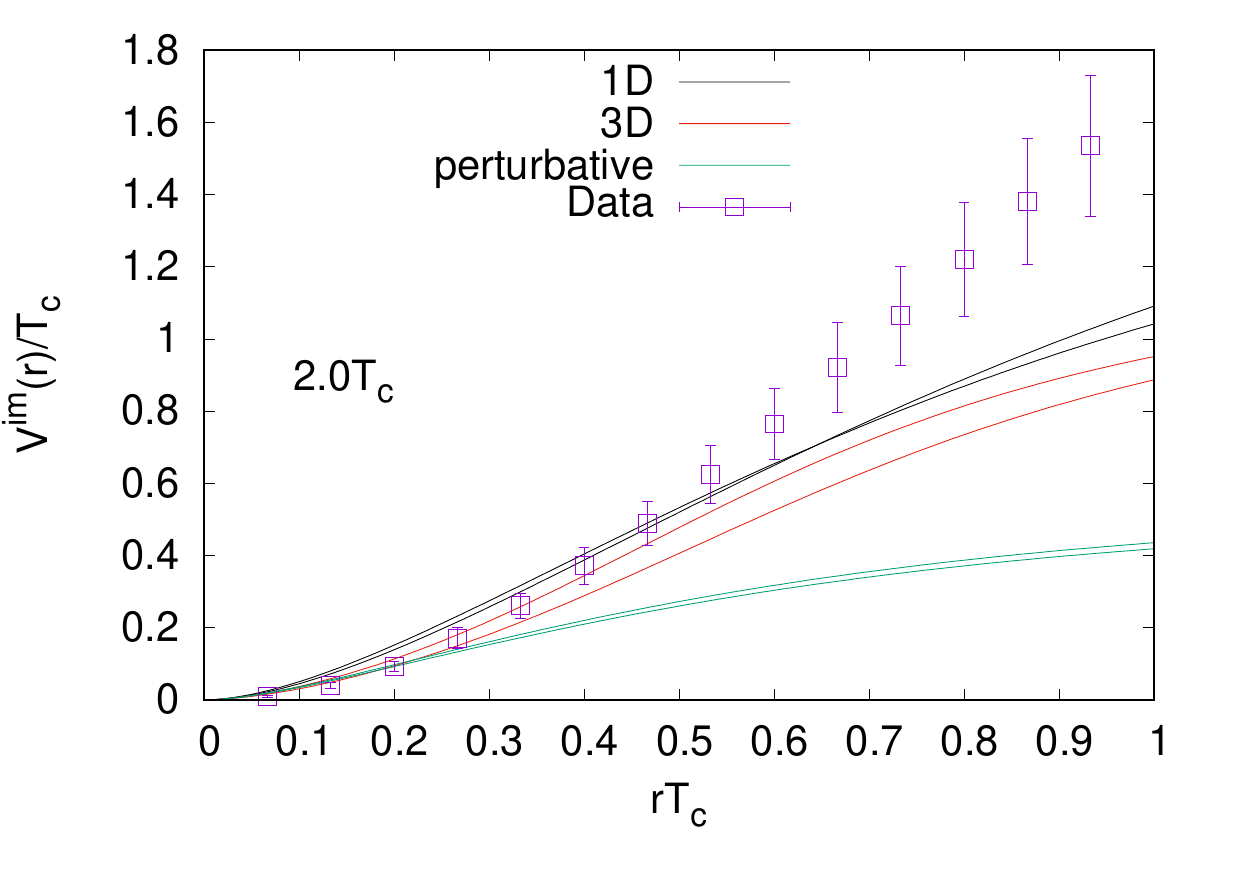}
\caption{
(Left) Detailed comparison of $ \mr{Re}~E(T,r) $ and $ F_S $ in $ \mr{SU}(3) $ 
pure gauge theory indicate small differences between the quantities that become 
larger in the asymptotic screening regime. 
(Right) $ \mr{Im}~E(T,r) $ obtained in $ \mr{SU}(3) $ pure gauge theory lattice 
simulations increases faster than the HTL result, and also faster than the 
imaginary parts obtained with various forms of medium permittivity, 
see~\cite{Bala:2019cqu} for details. 
\label{fig:su3V}
}
\end{figure*}

\vskip1ex
$ \mr{SU}(3) $ pure gauge theory~\cite{Burnier:2015nsa} or (2+1)-flavor QCD 
lattice simulations using the AsqTad action~\cite{Burnier:2014ssa} have been 
employed to calculate \Eqref{eq:rhoimag}, while tackling the inverse problem 
through the Bayesian reconstruction method~\cite{Burnier:2013nla}. 
The solution of the inverse problem in the analysis of more precise correlators 
from (2+1)-flavor QCD lattice simulations using the HISQ action proved to be 
more difficult due to the smaller statistical errors, and has led to somewhat 
unclear, preliminary results so far, whether by using rescaled results from 
HTL calculations~\cite{Bazavov:2014kva}, fits to the moments of the lattice 
correlators~\cite{Petreczky:2017aiz}, or by using the analytic continuation 
of Pad\'e fits or the BR method~\cite{Petreczky:2018xuh}, see \Figref{fig:ReV} 
for the current state of the art result from (2+1)-flavor QCD lattice 
simulations with the HISQ action.
In particular, a reliable determination of $ \mr{Im}~E(T,r) $ in QCD has proved 
elusive so far. 
Very recently exciting new results in $ \mr{SU}(N_c) $ pure gauge theory 
indicate that an elegant solution to the inverse problem may have finally been 
identified~\cite{Bala:2019cqu}, see~\Figref{fig:su3V}.

\vskip1ex
All present results indicate that at the level of the somewhat large systematic 
uncertainties the relation between $ \mr{Re}~E(T,r) $ and 
$ F_S(T,r) $ at order $g^3$ seems to be approximately realized in the 
nonperturbative calculation with relatively mild differences. 
Furthermore, the screening mass parameter that has been extracted from the real 
part of the static energy $ \mr{Re}~E(T,R) $ seems to be consistent with 
going to zero at $ T_c $ and in the vacuum. 
For this reason, an increase of the difference at larger distances has to be 
expected even at the lowest temperatures.

\subsection{Meson screening masses for heavy and heavy-light mesons}
\label{sec:mesons_h}

To illustrate the utility of Eq.~(\ref{eq:spatial}) we start the discussion with
the spatial Euclidean correlation functions as calculated on the
lattice.
A ratio of the Euclidean correlation function at non-zero temperature to its
zero-temperature counterpart directly probes thermal modification of the
spectral function. Ref.~\cite{Bazavov:2014cta} calculated the correlation functions
in various channels using staggered fermions up to temperature of 250~MeV which
is about $1.6T_c$ in 2+1 flavor QCD with the physical light quark masses.

A complication with staggered fermions is that mesonic correlation functions
contain contributions 
from excitations with opposite parity eigenvalues, where one causes a non-oscillating 
and the other causes an oscillating contribution:
\begin{equation}
 G(z) = A_{NO}^2 \left( e^{-M_{NO}z} + e^{-M_{NO}(N_\sigma - z)} \right)
-(-1)^z A_{O}^2 \left( e^{-M_{O}z} + e^{-M_{O}(N_\sigma - z)} \right).
\label{eq:fitGz}
\end{equation}
These contributions need to be separated before comparing to zero-temperature
results. This is possible by constructing effective mass correlators as described
in Ref.~\cite{Bazavov:2014cta}.

\begin{figure*}\center
	\includegraphics[width=0.49\textwidth]{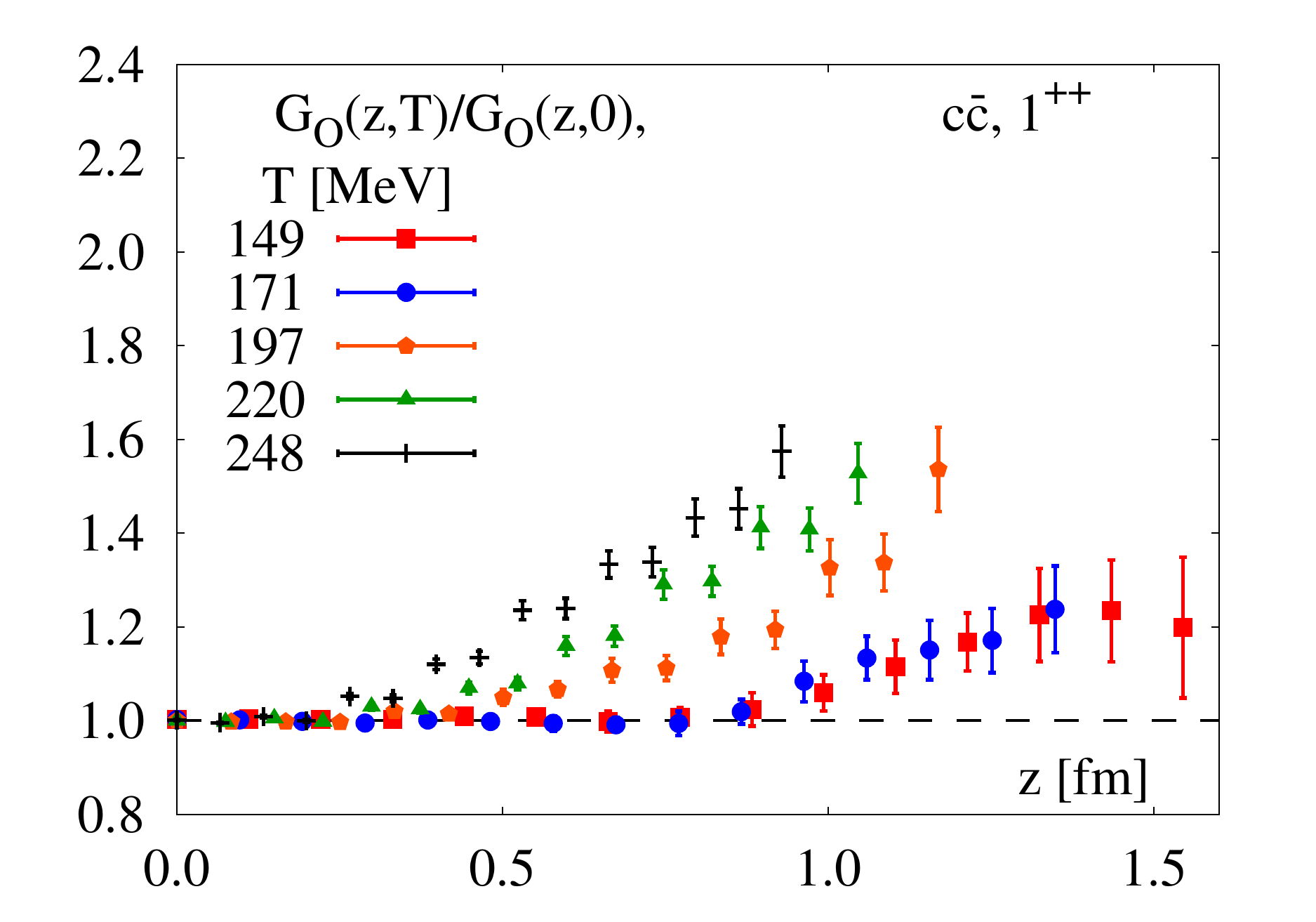}
	\includegraphics[width=0.49\textwidth]{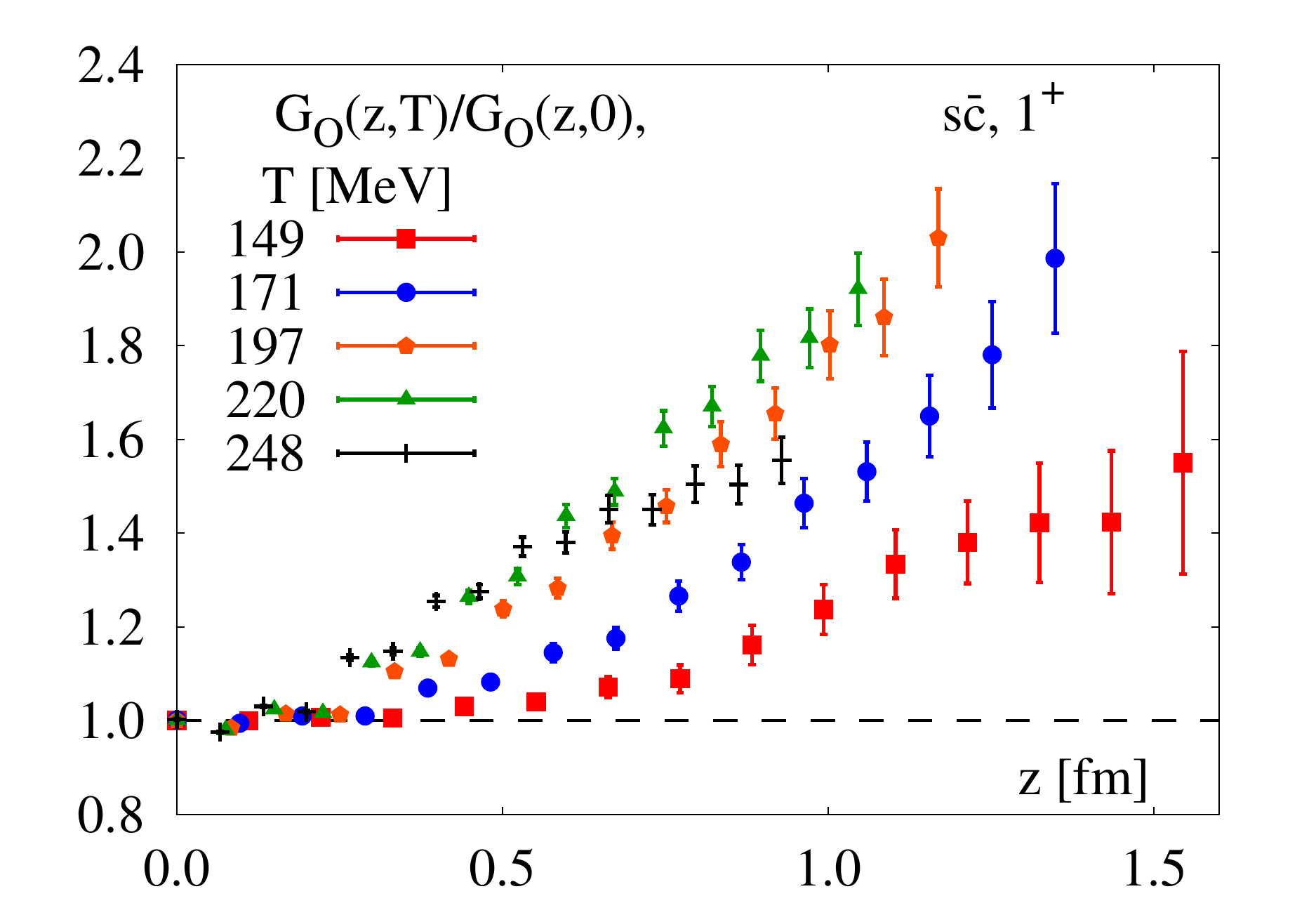}
	\caption{
		The oscillating (positive parity) parts of the axial-vector ($1^{++}$ or $1^+$) 
		correlation functions for the $c\bar c$ (left) and $s\bar c$ (right)
		sectors at different temperatures normalized by the zero-temperature results. 
		\label{fig:spcorO}
	}
\end{figure*}

From the point of view of the static picture of color screening it is to be 
expected that axial-vector mesons --- being in a P-wave, and thus being 
larger --- are dissociated already at lower temperatures than the 
corresponding vector mesons. 
The tightly bound pseudoscalars are expected to behave similarly to the vector mesons, whereas the scalar mesons ought to exhibit a pattern more akin to the axial-vector states.
Moreover, due to the smaller quark masses involved in open heavy-flavor mesons leading to a larger size of the bound states, the thermal modification is expected to be more pronounced and at lower temperatures even in the static picture. 
Since dynamical processes certainly enhance these trends, any observation 
of these patterns does not lead to a statement whether the static color 
screening or the dynamic dissociation and recombination is the dominant 
cause of the thermal modification. 

On the one hand, the ratio of the positive parity contribution to the 
spatial Euclidean axial-vector correlation function at finite temperature 
to the one at zero temperature is shown in Fig.~\ref{fig:spcorO} (left) 
for $c\bar c$ and Fig.~\ref{fig:spcorO} (right) for $s\bar c$ mesons. 
At zero temperature the ground states in these channels are, respectively, $\chi_{c1}$ or $D_{s1}$ mesons. 
For the ratios in the $\chi_{c1}$ channel significant thermal modifications 
of the ground state are seen at temperatures below $T\sim 200\,\mr{MeV}$ only at 
large distances $z \gtrsim 1\,\mr{fm}$, while these modifications \
occur at much smaller distances for temperatures above $T\sim 200\,\mr{MeV}$.
This is consistent with the analyses attempting to reconstruct the spectral 
functions from the temporal correlation by solving the inverse problem,
\Eqref{eq.spect}~\cite{Rothkopf:2019ipj}. 
In contrast, the temperature dependence resolved in these ratios 
is more complicated for the open-charm state $D_{s1}$, with a deviation of 
about 20\% from the zero-temperature value already in the crossover region, 
\ie at $T\sim 150$~MeV. 
At even higher temperatures, the increase of the ratio stalls below  $T\sim 250\,\mr{MeV}$. 
The case of the scalar channels --- the respective zero temperature ground states being $\chi_{c0}$ or $D_{s0}^\ast$ mesons --- is quantitatively similar. 
On the other hand, the vector meson channels with $J/\psi$ or $D_s^\ast$ 
mesons, or the pseudoscalar meson channels with $\eta_c$ 
or $D_s$ mesons as the respective ground states, exhibit less pronounced features.
For both $c\bar c$ channels significant thermal modifications 
of the ground states are seen at temperatures above $T\sim 200\,\mr{MeV}$ with a decreasing slope in $z$, but do not show any clearly non-monotonic $z$- or temperature-dependence. 
For both open-charm channels, a deviation of about 15 to 20\% from 
the zero-temperature value is present already at $T\sim 170\,\mr{MeV}$. 
Therefore, these ratios confirm the intuition of sequential melting based on the static picture of color screening. 
A recent analysis of the charm-quark susceptibilities~\cite{Bazavov:2014yba} 
is also consistent with this observation suggesting that open-charm
states start to melt at temperatures around the chiral crossover temperature
$T_c=156.5$~MeV.

\begin{figure*}\center
	\includegraphics[width=0.49\textwidth]{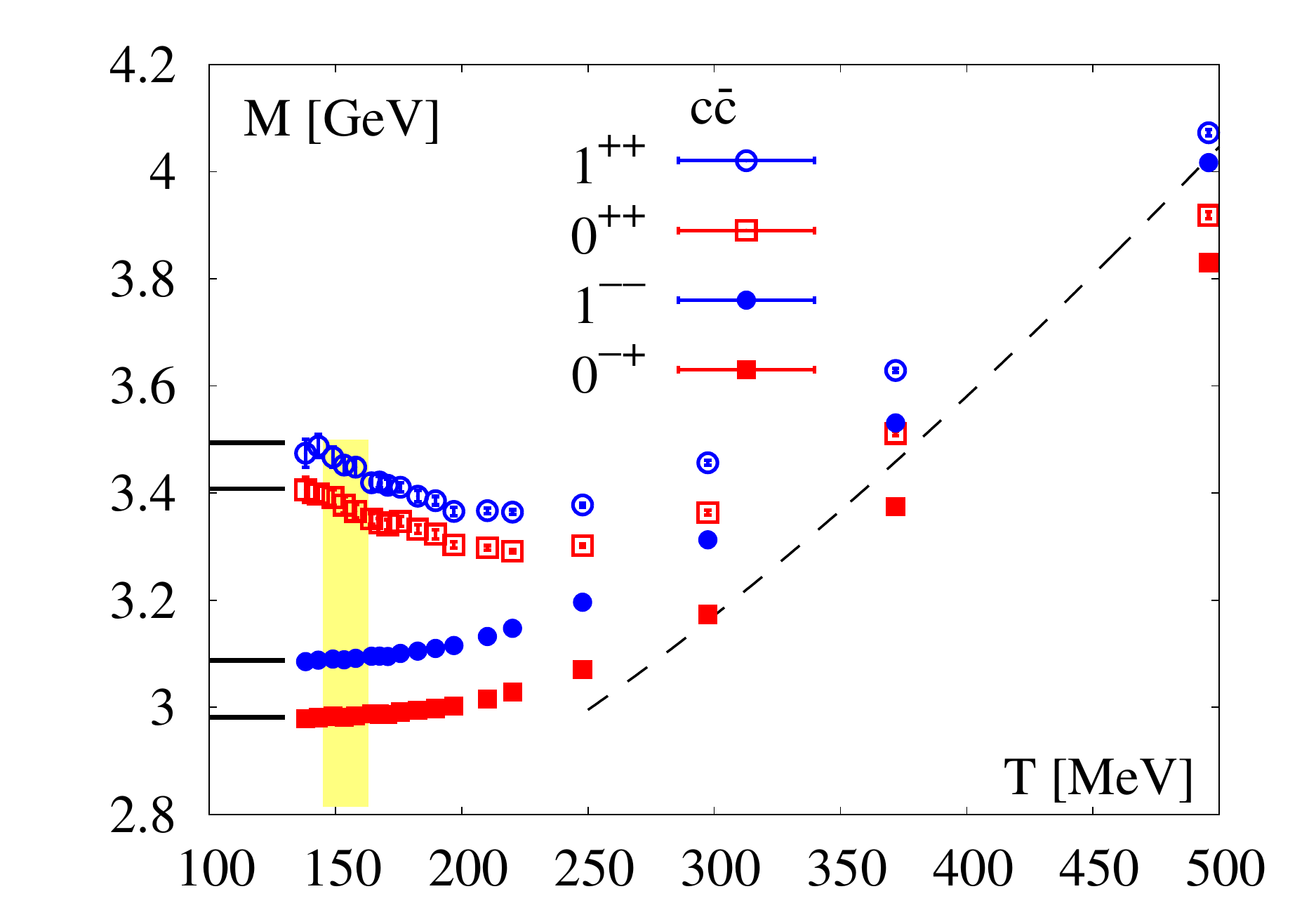}
	\includegraphics[width=0.49\textwidth]{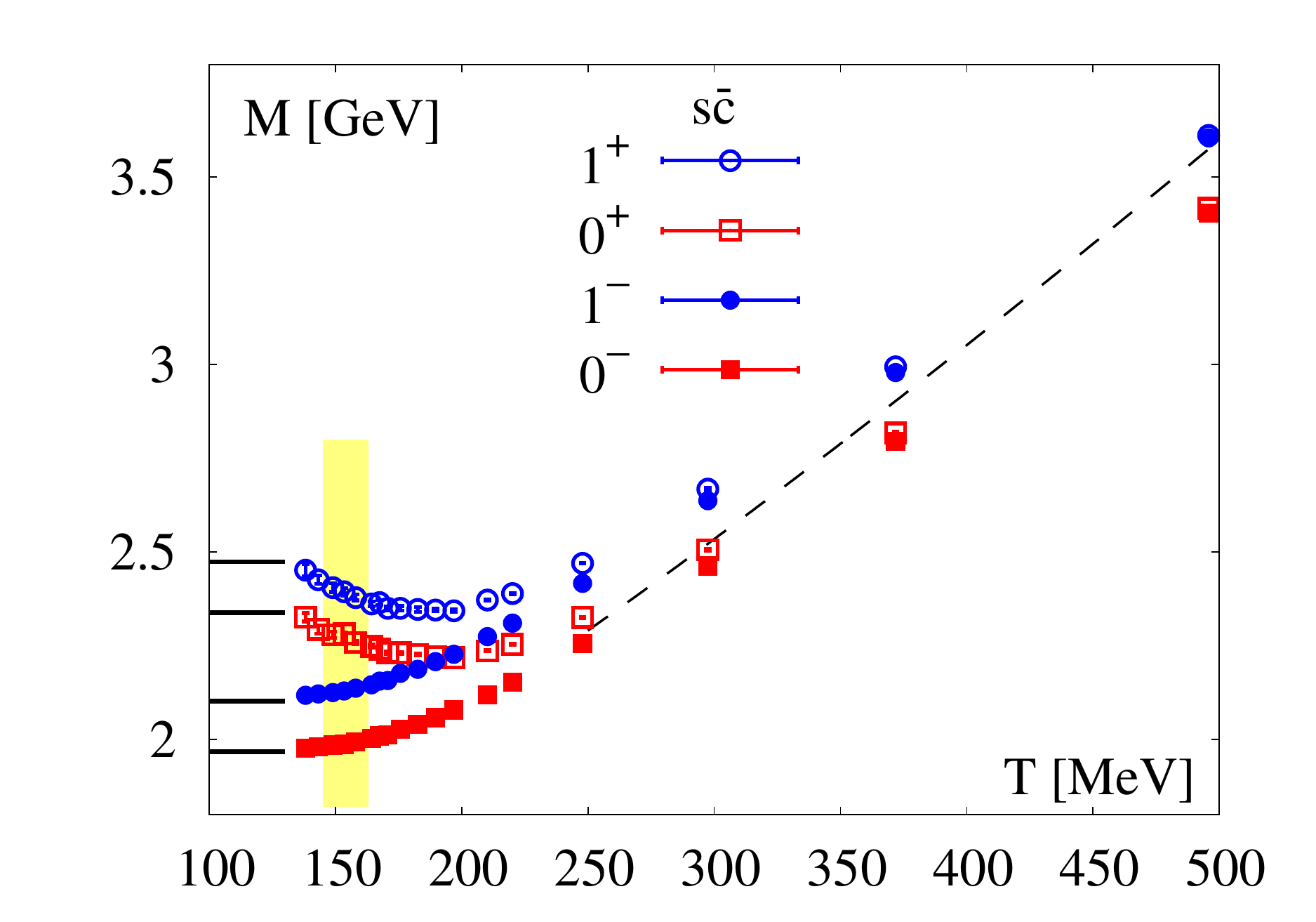}
	\caption{
		The screening masses for different channels in the $c\bar c$ (left)
		and $s\bar c$ (right) sectors as functions of temperature.
		The solid horizontal lines are the zero-temperature masses of the 
		corresponding ground-state mesons and the dashed lines indicate the
		free theory result. 
		\label{fig:spcorM}
	}
\end{figure*}

To further assess thermal modification effects, one can extract the screening
masses in the corresponding channels by fitting the long-distance behavior of
the correlation functions, \Eqref{eq:fitGz}. The temperature dependence of the 
screening masses for all four (axial vector, scalar, vector and pseudo-scalar)
channels for the $c\bar c$ and $s\bar c$ mesons is shown in
Fig.~\ref{fig:spcorM}. Three qualitatively distinct regions can be identified:
the low temperature region, where the screening masses are close to the corresponding vacuum
masses (horizontal solid lines),
the intermediate temperature region, where there are about 10--15\%
changes in the values of the screening masses with respect to the corresponding vacuum
masses and the high temperature region, where the screening masses approach
the free theory result (dashed lines).
The onset of the high temperature behavior in the $c\bar c$ sector starts at 
$T>300$~MeV, and in the $s\bar c$ sector earlier, at about $T=250$~MeV.
This matches the previous observation in the ratios of the correlation functions
that thermal modifications depend on the quark content of the states and appear
earlier for the states containing 
quarks with lower masses.

The behavior of the screening masses corresponding to the negative and positive parity
states in the low and intermediate temperature regions is qualitatively different.
The screening masses of the negative parity states increase with temperature
from their vacuum values monotonically. Those of the positive parity states first
decrease, with the decrease beginning close to the chiral crossover region.
In the intermediate temperature region the trend reverses and the masses increase
to eventually follow the high-temperature asymptotics. Moreover, in the intermediate
region the ordering of the screening masses changes and the masses of the opposite
parity partners approach each other to become degenerate at sufficiently high
temperatures. While this is less apparent in the $c\bar c$ sector, since it requires
temperatures on 
the order of the charm mass, it is evident in the $s\bar c$ sector, where
the masses of the pseudoscalar and scalar or vector and axial-vector states become
degenerate above 350~MeV. In the high temperature region 
the pseudoscalar screening masses stay below the vector screening masses,
as was observed in earlier lattice~\cite{Cheng:2010fe}
and Dyson-Schwinger formalism calculations~\cite{Wang:2013wk}.

\begin{figure*}\center
	\includegraphics[width=8.6cm]{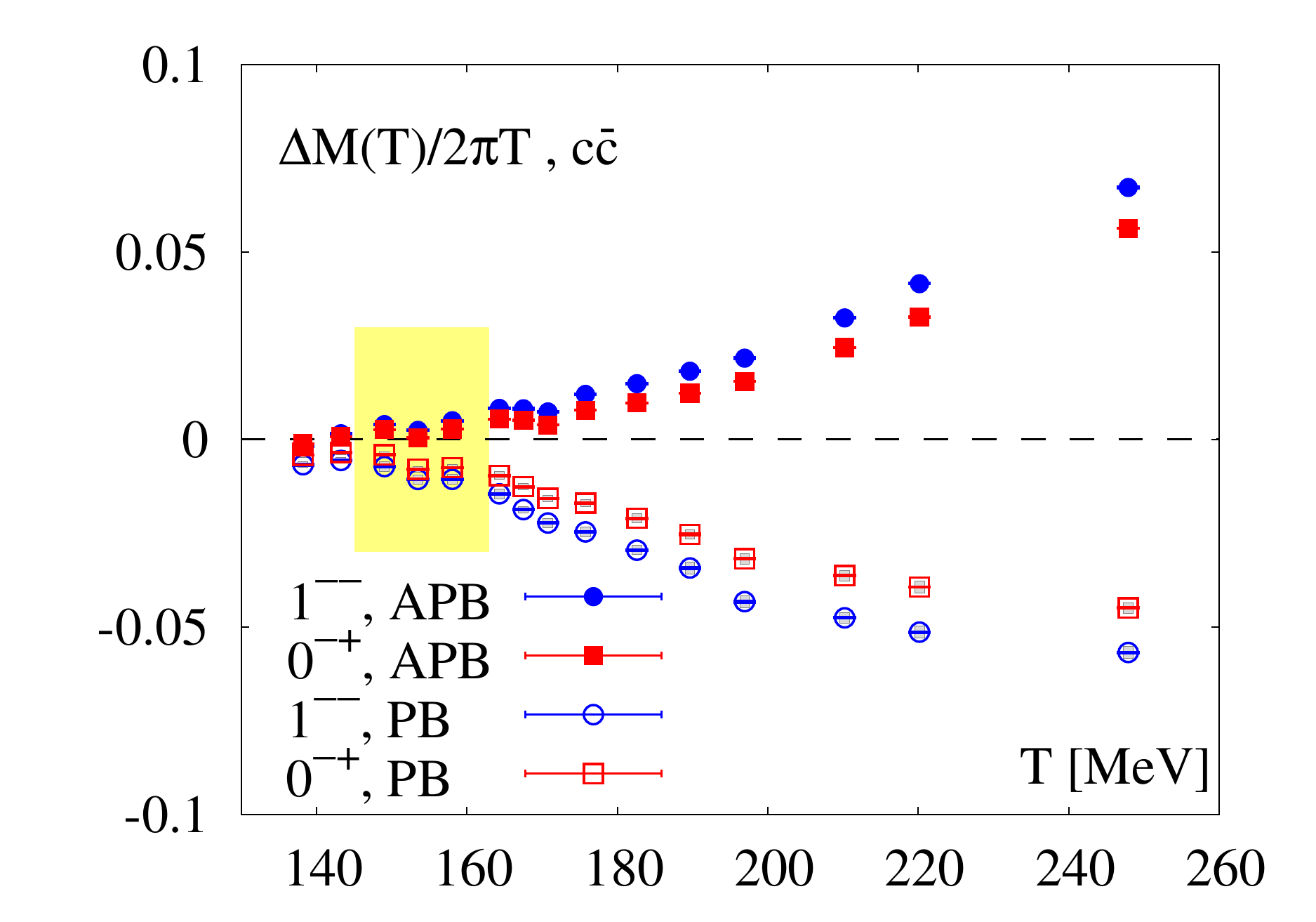}
	\caption{
		The differences in the screening masses for the pseudoscalar and vector
		charmonia states and the vacuum masses, Eq.~(\ref{eq:dM}),
		for anti-periodic (filled symbols) and periodic (open symbols)
		temporal boundary conditions for fermions.
		\label{fig:dMcc_scr}
	}
\end{figure*}

Following earlier work~\cite{Boyd:1994np,Mukherjee:2008tr,Karsch:2012na}
Ref.~\cite{Bazavov:2014cta} also considered sensitivity of the charmonia screening
masses to the temporal boundary conditions. At finite temperature the boundary
conditions in the temporal direction must be periodic for the bosonic and anti-periodic
for the fermionic fields. However, on the lattice the Euclidean correlation functions,
\Eqref{eq:GtaupT}, can be also measured with artificially imposed periodic temporal boundary
conditions for fermions. 
The vacuum masses of stable mesons are insensitive to the boundary conditions.
At very high temperatures where the bound states
dissolve and the two quarks propagate independently,
the mesonic screening masses
approach twice the value of the lowest Matsubara frequency, as evident from
\Eqref{eq:Mfree}, which is $2\pi T$ due to the anti-periodic boundary conditions
on fermions. In contrast, if periodic temporal boundary conditions for fermions are imposed,
the screening masses should become vanishingly small at very high temperatures.
Thus sensitivity of correlation functions and in turn the screening masses to the
boundary conditions helps one to judge if at a given temperature quarks still constitute
a bosonic bound state, or if thermal modifications uncover its fermionic structure.

The charmonium screening masses in the pseudo-scalar
and vector channels calculated using anti-periodic and periodic
boundary conditions are shown in Fig.~\ref{fig:dMcc_scr}. The difference between the
screening mass $M(T)$ and its vacuum value $M_0$
\begin{equation}
\label{eq:dM}
\Delta M(T)=M(T)-M_0
\end{equation}
is normalized with $2\pi T$, since one expects that at asymptotically high temperature the
quadratic difference between the screening masses with the two types of boundary conditions
approaches $(2\pi T)^2$. Although the screening masses become sensitive to the modification
of the boundary conditions already in the chiral crossover region, the sensitivity
is small up to about 170~MeV. The difference gets larger with increasing temperature and the overall
picture supports melting of the $\eta_c$ and $J/\psi$ states above 200~MeV.

\subsection{Meson screening masses for light mesons}
\label{sec:mesons_l}

\begin{figure*}\center
	\includegraphics[width=0.33\textwidth]{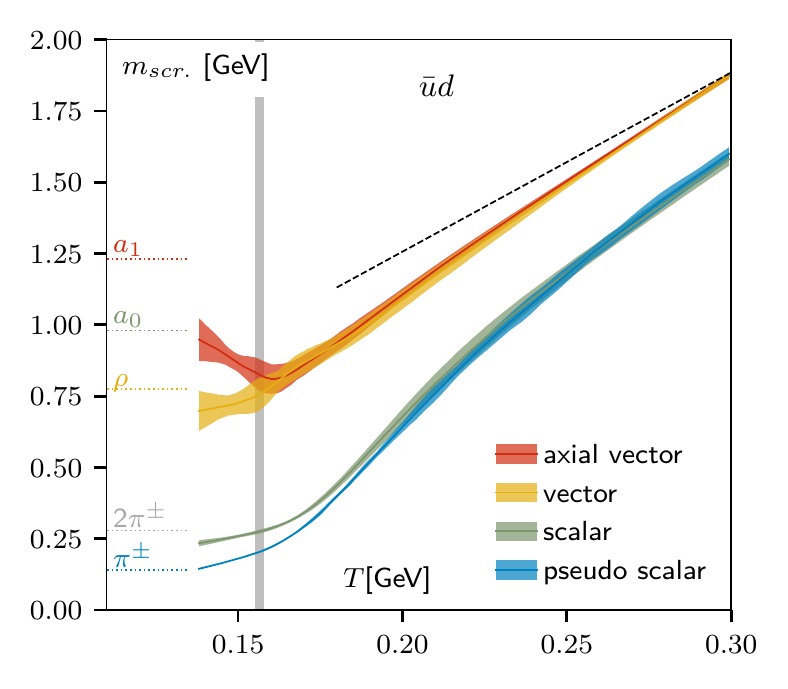}\hfill
	\includegraphics[width=0.33\textwidth]{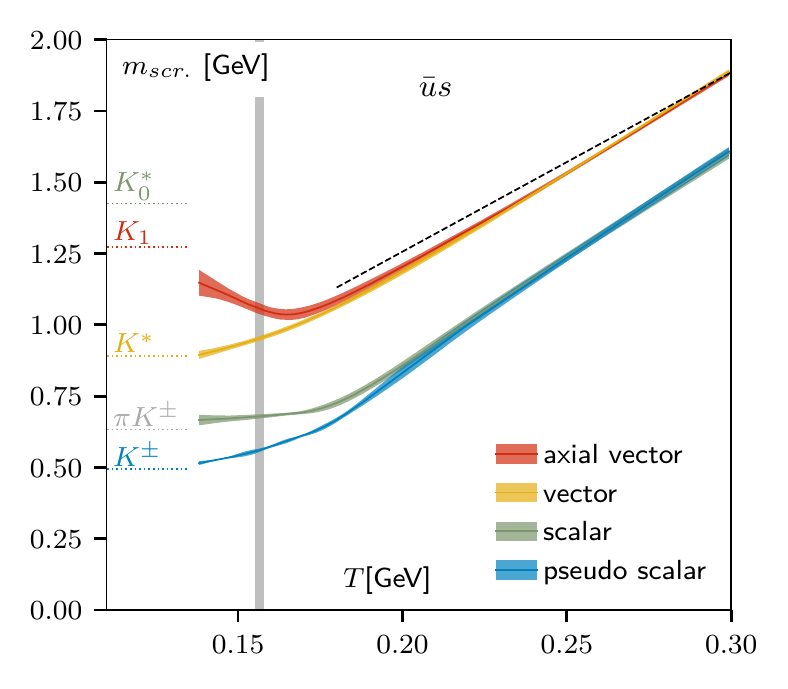}\hfill
	\includegraphics[width=0.33\textwidth]{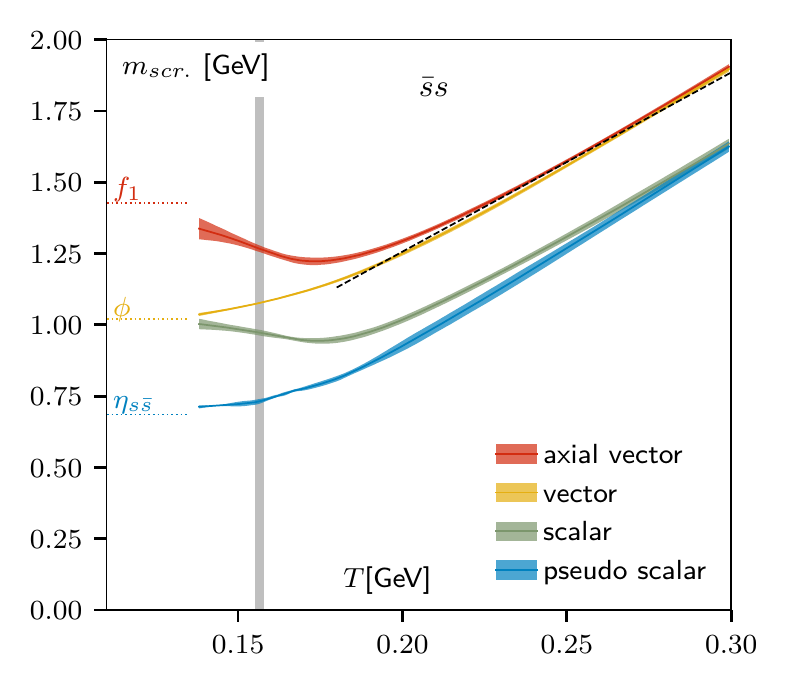}
	\caption{
		The continuum extrapolated mesonic
		screening masses for different channels in the $\bar u d$ (left),
		$\bar u s$ (middle) and $\bar s s$ (right)
		sectors as functions of temperature.
		The solid horizontal lines are the zero-temperature masses of the 
		corresponding ground-state mesons or two-particle states (see text)
		and the dashed lines indicate the
		free theory result.  
		\label{fig:sc_lowT}
	}
\end{figure*}

The observables that we discussed in the previous sections, related to static
or heavy quarks act as external probes of the deconfined medium since their mass
scales are well separated from the temperature scales of the transition.
It is natural to ask how the color screening properties of quark-gluon plasma affect
the states in the QCD spectrum composed of light and strange quarks such as 
$\pi$, $K$, $\rho$, etc. The situation there is more complicated since the dynamics
of the transition is driven by the dynamics of the light quarks and their composites.
For instance, the Hadron Resonance Gas (HRG)
model~\cite{Dashen:1969ep,Venugopalan:1992hy,Hagedorn:1983wk,Fiore:1984yu,Hagedorn:1984hz}
approximates the
partition function of QCD by an ideal gas of stable particles and resonances.
This approximation works surprisingly well up to temperatures of about 140--150~MeV,
and, of course, breaks down close to the chiral crossover. The dominant contribution
into the HRG partition function comes from the lightest states, \textit{e.g.} the
expansion for observables with zero strangeness and baryon number starts with the pion,
for non-zero strangeness with kaon and so on.
Light degrees of freedom are also closely related to the fundamental symmetries of QCD
such as the $\mr{SU_L}(2)\times \mr{SU_R(2})$ chiral symmetry and the anomalous axial
$\mr{U_A}(1)$ symmetry. While the chiral symmetry is completely restored at the chiral
phase transition temperature $T^0_c$ (in the chiral limit) and is smoothly restored within
a narrow temperature range around the chiral crossover temperature $T_c$ in QCD
with physical light quark masses as indicated by the melting of the chiral condensate,
the fate of the $\mr{U_A}(1)$ is not completely clear at present. Most studies agree that
it gets effectively restored in the high temperature phase in 
QCD. However, some differ in
if it happens at the same temperature as the chiral symmetry restoration or at some
higher temperature.

The large distance behavior of the spatial correlation functions
defined in Eq.~(\ref{eq:spatial}) is sensitive to various patterns of chiral symmetry
restoration~\cite{DeTar:1987ar,DeTar:1987xb}.
A recent analysis of the screening masses for mesons composed of light and strange
quarks was performed in Ref.~\cite{Bazavov:2019www} with the HISQ action.
The continuum extrapolated mesonic screening masses for the four channels 
are shown in Fig.~\ref{fig:sc_lowT} for the light-light, light-strange and strange-strange
mesons. We should first note that extraction of the scalar state poses difficulties
with staggered fermions when the quark masses are light. Due to the taste exchange
interactions there are unphysical contributions in the scalar channel that allow the
scalar state composed of $u$ and $d$ quarks
decay into two pions at finite lattice spacing~\cite{Prelovsek:2005rf},
while this decay does not occur in nature due to parity, isospin and
$G$-parity conservation. For this reason the scalar screening mass in
the left panel of Fig.~\ref{fig:sc_lowT} approaches the energy of the two-pion
state instead of the true scalar ground state ($a_0(980)$ or $a_0(1450)$)
or the allowed $\pi\eta$ two-particle state.
This problem could be resolved if the continuum limit is taken for
the spatial correlation function (\ref{eq:fitGz}) first and then the screening
mass is extracted from the continuum correlator. This approach is however difficult
due to the oscillating terms in Eq.~(\ref{eq:fitGz}), and Ref.~\cite{Bazavov:2019www}
resorted to extracting the screening masses from correlators at finite lattice
spacing and then taking the continuum limit for the screening masses.
In the $\bar u s$ channel the situation is better since the decay to $K\pi$
occurs in nature. The continuum limit of the scalar screening mass
extracted from the $\bar u s$ correlator at finite lattice spacing approaches the
$K\pi$ state as indicated in the middle panel of Fig.~\ref{fig:sc_lowT}.

The overall trends in Fig.~\ref{fig:sc_lowT} are similar to the ones observed
for heavy-heavy and heavy-light states discussed in Sec.~\ref{sec:mesons_h}.
Thermal modifications happen at lower temperatures for the states with lower
quark masses. Due to the restoration of chiral symmetry one expects that
the vector ($\rho$) and axial vector ($a_1$) screening masses become
degenerate. As can be seen from the left panel of Fig.~\ref{fig:sc_lowT},
the axial vector screening mass decreases 
significantly (it is already about 20\% below the
vacuum value at the lowest temperature available in the calculation) while
the vector mass slightly increases and the two indeed become degenerate 
at the chiral crossover temperature. For the states involving the strange
quark in the middle and right panel of  Fig.~\ref{fig:sc_lowT}
the degeneracy of the axial vector and vector screening
masses happens at higher temperature.

Restoration of the $U_A(1)$ symmetry is signaled by degeneracy of
the scalar and pseudoscalar screening masses. In the $\bar u d$ sector
it is observed at temperature about 200~MeV, however, one has to be careful
with its interpretation due to the unphysical effects in the scalar channel.
Discussion of the technical subtleties is beyond the scope of this review and
we refer the reader to \cite{Bazavov:2019www} where another measure such as a
difference of continuum extrapolated integrated scalar and pseudoscalar correlators 
was constructed to estimate the temperature of the $U_A(1)$ symmetry restoration.
The numerical evidence points out to the restoration temperature $T\sim200$~MeV,
in general, consistent with the degeneracy of the screening masses observed
in the left panel of Fig.~\ref{fig:sc_lowT}.

\begin{figure*}\center
	\includegraphics[width=0.33\textwidth]{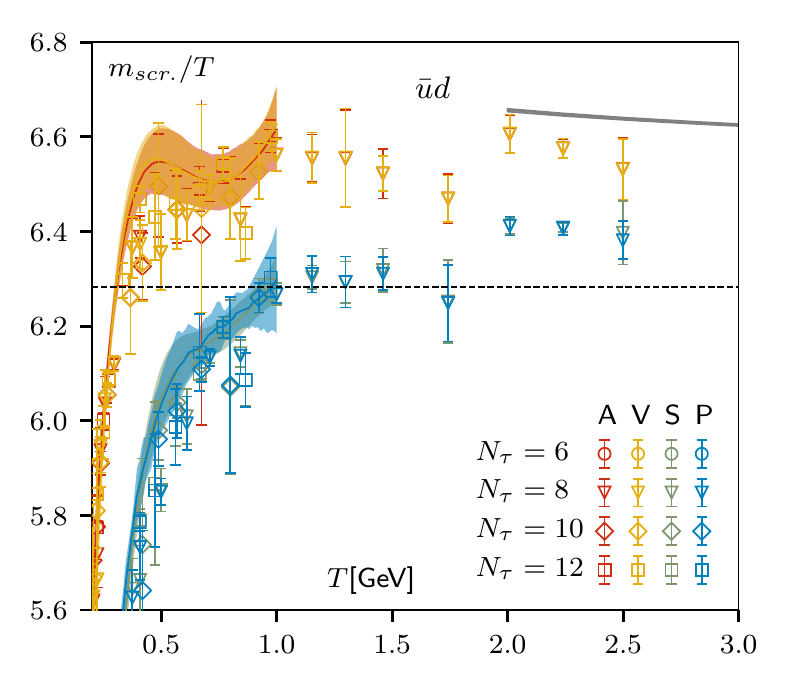}\hfill
	\includegraphics[width=0.33\textwidth]{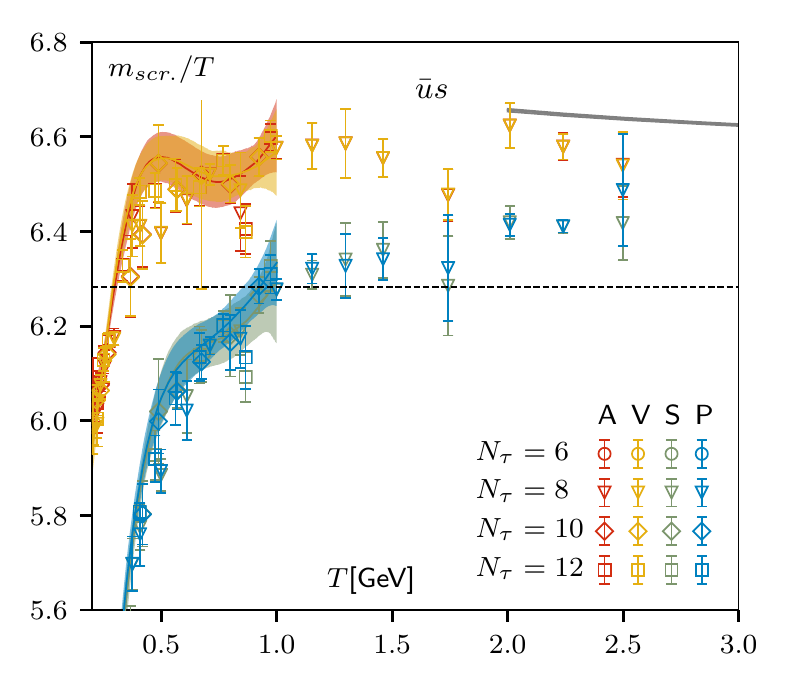}\hfill
	\includegraphics[width=0.33\textwidth]{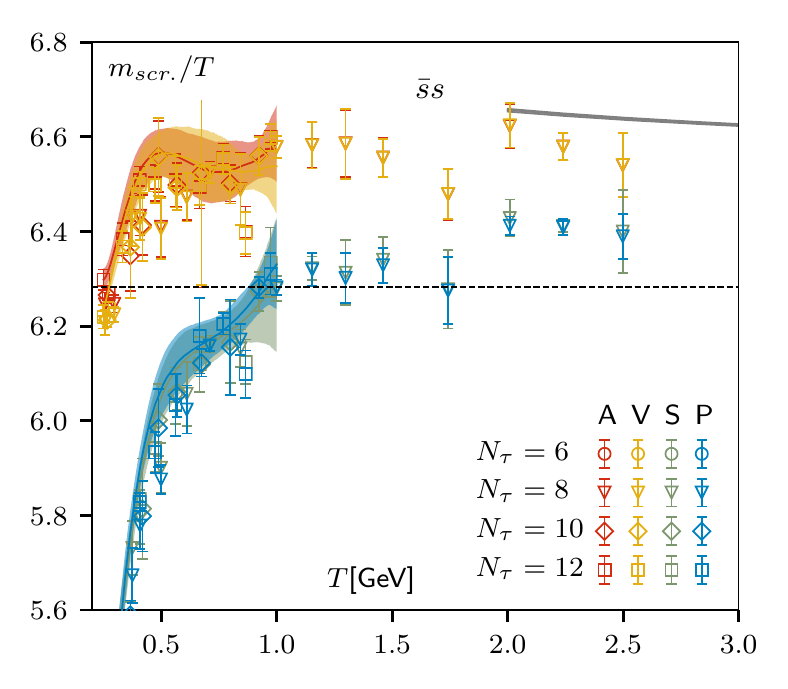}
	\caption{
		The mesonic screening masses normalized by the temperature
		for different channels in the $\bar u d$ (left),
		$\bar u s$ (middle) and $\bar s s$ (right)
		sectors as functions of temperature shown for higher temperatures
		in the deconfined phase, compare with Fig.~\ref{fig:sc_lowT}.
		The dashed horizontal line indicates the free theory result, $2\pi$,
		and the solid line the EQCD correction (see text).
		\label{fig:sc_highT}
	}
\end{figure*}

The temperature dependence of the screening masses becomes
qualitatively consistent
with the free theory behavior at temperatures above 300~MeV.
Around that temperature 
the vector and axial vector masses are numerically consistent with
the free theory behavior, the scalar and pseudoscalar masses 
are below by 10-20\%.
As full degeneracy of the screening masses is expected at infinite
temperature, Ref.~\cite{Bazavov:2019www} followed the temperature
dependence of the screening masses up to $T~\sim2.5$~GeV. The screening
masses normalized by the temperature in the high temperature region are shown
in Fig.~\ref{fig:sc_highT}. It is argued that above $T\sim1$~GeV
the cutoff effects are
small and the calculation is performed only on $N_\tau=8$ lattices above
that temperature.

Perturbatively the correction to the free theory screening mass
can be calculated in electrostatic QCD (EQCQ)~\cite{Braaten:1995jr}.
Ref.~\cite{Laine:2003bd} evaluated this correction which turns out to be
independent of the spin. 
Its value is positive and is qualitatively 
consistent with the lattice results, as shown by the solid lines in
Fig.~\ref{fig:sc_highT}.
As can be seen from the figure, the vector and axial vector screening masses
overshoot the free theory result at $T\sim400$~MeV and stay approximately
constant reasonably close to the weak-coupling EQCD result.
The scalar and pseudoscalar screening masses increase past the
free theory value at $T\sim1$~GeV and in the observed temperature range stay
significantly below the weak-coupling result. It hints that higher-order,
spin-dependent corrections~\cite{Koch:1992nx,Shuryak:1993kg}
may be important since the the EQCD
coupling $g_E^2$ is not small in this temperature range. Moreover, since
it decreases logarithmically, only at significantly higher temperatures
(by orders of magnitude) one may expect the screening masses to
approach the free theory value $2\pi T$.

\section{Summary}
\label{sec:summary}
\emph{}
  In this review paper we discussed the color screening in 
the quark-gluon plasma as it is studied using lattice QCD. 
We presented a brief overview of the field-theoretical foundations 
and summarically contrasted specific phenomena associated with the color 
screening in QCD with their counterparts in QED or cold nuclear matter. 
We reviewed in pure gauge theory and in full QCD the Polyakov loop 
and its various correlation functions, which are the primary observables 
by which color screening is still being studied on the lattice and in other approaches. 
Going beyond static limit we reviewed the status of dynamic, spatial meson 
screening correlation functions for heavy-heavy, heavy-light, and 
light-light flavors and spin $0$ or $1$ states. 
In particular, we reviewed the behavior and phenomena from the upper end 
of the confined phase at low temperature all the way up to phenomena in 
the weakly-coupled quark-gluon plasma at high temperature. 
To date, all known color screening phenomena tend to become fairly compatible 
with the weak-coupling picture for $T \gtrsim 300\,\mr{MeV}$, which largely coincides with the QCD scale $\lMSb$. 

We highlighted the role of the Polyakov loop in pure gauge theory as the 
order parameter of the deconfinement transition and its continued relevance 
in full QCD, where its renormalization is required. 
In full QCD the renormalization scheme independent static quark entropy 
shift that signals deconfinement helps with better understanding of 
the transition regime at $T \sim T_c$ (with $T_c=156.5(1.5)\,\mr{MeV}$). 
Whether the coincidence of chiral symmetry restoration and deconfinement, 
which has been observed down to almost physical quark masses, holds even in 
the chiral limit is one of the open questions regarding the Polyakov loop.

We scrutinized the different regimes of static, spatial screening 
correlation functions in various channels, and how the corresponding 
screening behavior changes with the separation of the quark and antiquark, relating these to the picture sketched by the dimensionally-reduced QCD and 
the hard thermal loop QCD. 
In particular, we juxtaposed direct lattice QCD calculations with results 
from the weak-coupling approach, where those were available and applicable.  
Eventually, direct lattice QCD simulations quantitatively confirm these 
ideas, establish the existence of a vacuum-like regime, a dissociation regime, 
an electric screening regime, and an asymptotic screening regime, where finally 
the nonperturbative physics becomes dominant, and in part explain the success 
of weak-coupling descriptions of color screening phenomena. 
We reviewed the existence of the qualitatively different screening patterns 
at different values of the external control parameters temperature $T$, 
chemical potential $\mu/T$ and magnetic field $\vert{eB}\vert/T$.
We indicated open issues regarding these static screening correlation functions.

Lastly, we addressed the interplay of the real-time dynamical processes 
and the color screening in quark-gluon plasma, as it plays out in the complex 
static energy at finite temperature, which has to be obtained from lattice QCD 
after solving the inverse problem of reconstructing the spectral function from 
the Euclidean correlation functions. 
Finally passing on from the static limit to the dynamical heavy and light 
quarks, we discussed meson correlation functions and repeated the arguments 
for using spatial meson screening correlation functions. 
We reviewed studies of these with a wide variety of flavor contents down from 
the charm- to the average light-quark considering hidden and open flavor.  
The naive expectation of the earlier modification of correlation functions 
involving lower quark masses is quantitatively confirmed and the weak-coupling 
like behavior is postponed to higher temperatures for larger quark masses, 
while the approximate degeneracy between parity partners sets in a bit later. 
Indications of the degeneracy of scalar and pseudoscalar states that signals 
the effective $\mr{U_A}(1)$ restoration is consistent with about 
$T \sim 200\,\mr{MeV}$. 
Open heavy flavor mesons are generally modified already slightly above the 
QCD crossover transition. 
\vskip1ex
Taken together, all of these observations largely support the sequential melting picture and suggest that in-medium quark-antiquark systems can be understood in terms of the weak-coupling picture for $T \gtrsim 300\,\mr{MeV}$.

\section*{Acknowledgments}

We would like to thank the HotQCD and TUMQCD collaborations for the productive work, and 
F.~Karsch, A.~Lahiri and P.~Petreczky for valuable discussions, a careful reading and comments on the manuscript. 
This work was in part supported by the U.S. Department of Energy, Office of Science, Office 
of Nuclear Physics and Office of Advanced Scientific Computing Research within 
the framework of Scientific Discovery through Advance Computing (SciDAC) award 
``Computing the Properties of Matter with Leadership Computing Resources''
and the U.S. National Science Foundation under the award PHY-1812332.

\bibliographystyle{unsrt}
\bibliography{ref}

\end{document}